\documentstyle[twoside,epsf,rotate]{article}

\catcode`\@=11
\long\def\@makefntext#1{
\protect\noindent \hbox to 3.2pt {\hskip-.9pt
$^{{\eightrm\@thefnmark}}$\hfil}#1\hfill}		%CAN BE USED

\def\thefootnote{\fnsymbol{footnote}}
\def\@makefnmark{\hbox to 0pt{$^{\@thefnmark}$\hss}}	%ORIGINAL

\def\ps@myheadings{\let\@mkboth\@gobbletwo
\def\@oddhead{\hbox{}
\rightmark\hfil\eightrm\thepage}
\def\@oddfoot{}\def\@evenhead{\eightrm\thepage\hfil
\leftmark\hbox{}}\def\@evenfoot{}
\def\sectionmark##1{}\def\subsectionmark##1{}}

\oddsidemargin=\evensidemargin
\renewcommand{\thefootnote}{\fnsymbol{footnote}}

\newcounter{sectionc}
\newcounter{subsectionc}
\newcounter{subsubsectionc}
\renewcommand{\section}[1] {\vspace{12pt}\addtocounter{sectionc}{1}
\setcounter{subsectionc}{0}\setcounter{subsubsectionc}{0}\noindent
	{\tenbf\thesectionc. #1}\par\vspace{5pt}}
\renewcommand{\subsection}[1] {\vspace{12pt}
\addtocounter{subsectionc}{1}\setcounter{subsubsectionc}{0}\noindent
	{\bf\thesectionc.\thesubsectionc.
        {\kern1pt \bfit #1}}\par\vspace{5pt}}
\renewcommand{\subsubsection}[1] {\vspace{12pt}
\addtocounter{subsubsectionc}{1}\noindent
        {\tenrm\thesectionc.\thesubsectionc.\thesubsubsectionc.
	{\kern1pt \tenit #1}}\par\vspace{5pt}}

\newcounter{appendixc}
\newcounter{subappendixc}[appendixc]
\newcounter{subsubappendixc}[subappendixc]
\renewcommand{\thesubappendixc}{\Alph{appendixc}.
        \arabic{subappendixc}}
\renewcommand{\thesubsubappendixc}{\Alph{appendixc}.
        \arabic{subappendixc}.\arabic{subsubappendixc}}

\renewcommand{\appendix}[1] {\vspace{12pt}
        \refstepcounter{appendixc}
        \setcounter{figure}{0}
        \setcounter{table}{0}
        \setcounter{lemma}{0}
        \setcounter{theorem}{0}
        \setcounter{corollary}{0}
        \setcounter{definition}{0}
        \setcounter{equation}{0}
        \renewcommand{\thefigure}{\Alph{appendixc}.\arabic{figure}}
        \renewcommand{\thetable}{\Alph{appendixc}.\arabic{table}}
        \renewcommand{\theappendixc}{\Alph{appendixc}}
        \renewcommand{\thelemma}{\Alph{appendixc}.\arabic{lemma}}
        \renewcommand{\thetheorem}{\Alph{appendixc}.\arabic{theorem}}
        \renewcommand{\thedefinition}{\Alph{appendixc}.
         \arabic{definition}}
        \renewcommand{\thecorollary}{\Alph{appendixc}.
         \arabic{corollary}}
        \renewcommand{\theequation}{\Alph{appendixc}.
         \arabic{equation}}
        \noindent{\tenbf Appendix \theappendixc #1}\par\vspace{5pt}}
\newcommand{\subappendix}[1] {\vspace{12pt}
        \refstepcounter{subappendixc}
        \noindent{\bf Appendix \thesubappendixc. {\kern1pt \bfit #1}}
	\par\vspace{5pt}}
\newcommand{\subsubappendix}[1] {\vspace{12pt}
        \refstepcounter{subsubappendixc}
        \noindent{\rm Appendix \thesubsubappendixc.
        {\kern1pt \tenit #1}}\par\vspace{5pt}}

\newcommand{\textlineskip}{\baselineskip=13pt}
\newcommand{\smalllineskip}{\baselineskip=10pt}

\def\eightcirc{
\begin{picture}(0,0)
\put(4.4,1.8){\circle{6.5}}
\end{picture}}
\def\eightcopyright{\eightcirc\kern2.7pt\hbox{\eightrm c}}

\newcommand{\copyrightheading}[1]
  {\vspace*{-2.5cm}\smalllineskip{\flushleft
  {\footnotesize International Journal of Modern Physics A, #1}\\
  {\footnotesize $\eightcopyright$\, World Scientific Publishing
   Company}\\
  }}

\newcommand{\publisher}[2]{{\begin{center}\footnotesize\smalllineskip
	Received #1\\
	Revised #2
	\end{center}
	}}

\def\abstracts#1#2#3{{
	\centering{\begin{minipage}{4.5in}\baselineskip=10pt
        \footnotesize
	\parindent=0pt #1\par
	\parindent=15pt #2\par
	\parindent=15pt #3
	\end{minipage}}\par}}

\renewenvironment{thebibliography}[1]
	{\frenchspacing
	 \ninerm\baselineskip=11pt
	 \begin{list}{\arabic{enumi}.}
	{\usecounter{enumi}\setlength{\parsep}{0pt}
	 \setlength{\leftmargin 12.7pt}{\rightmargin 0pt}
	 \setlength{\itemsep}{0pt} \settowidth
	{\labelwidth}{#1.}\sloppy}}{\end{list}}

\newcounter{itemlistc}
\newcounter{romanlistc}
\newcounter{alphlistc}
\newcounter{arabiclistc}

\newcommand{\fcaption}[1]{
        \refstepcounter{figure}
        \setbox\@tempboxa = \hbox{\footnotesize Fig.~\thefigure. #1}
        \ifdim \wd\@tempboxa > 5in
           {\begin{center}
        \parbox{5in}{\footnotesize\smalllineskip Fig.~\thefigure. #1}
            \end{center}}
        \else
             {\begin{center}
             {\footnotesize Fig.~\thefigure. #1}
              \end{center}}
        \fi}

\newcommand{\tcaption}[1]{
        \refstepcounter{table}
        \setbox\@tempboxa = \hbox{\footnotesize Table~\thetable. #1}
        \ifdim \wd\@tempboxa > 5in
           {\begin{center}
        \parbox{5in}{\footnotesize\smalllineskip Table~\thetable. #1}
            \end{center}}
        \else
             {\begin{center}
             {\footnotesize Table~\thetable. #1}
              \end{center}}
        \fi}

\def\@citex[#1]#2{\if@filesw\immediate\write\@auxout
	{\string\citation{#2}}\fi
\def\@citea{}\@cite{\@for\@citeb:=#2\do
	{\@citea\def\@citea{,}\@ifundefined
	{b@\@citeb}{{\bf ?}\@warning
	{Citation `\@citeb' on page \thepage \space undefined}}
	{\csname b@\@citeb\endcsname}}}{#1}}

\newif\if@cghi
\def\cite{\@cghitrue\@ifnextchar [{\@tempswatrue
	\@citex}{\@tempswafalse\@citex[]}}
\def\citelow{\@cghifalse\@ifnextchar [{\@tempswatrue
	\@citex}{\@tempswafalse\@citex[]}}
\def\@cite#1#2{{$\null^{#1}$\if@tempswa\typeout
	{IJCGA warning: optional citation argument
	ignored: `#2'} \fi}}

\def\pmb#1{\setbox0=\hbox{#1}
	\kern-.025em\copy0\kern-\wd0
	\kern.05em\copy0\kern-\wd0
	\kern-.025em\raise.0433em\box0}

\def\fnt#1#2{\footnotetext{\kern-.3em
	{$^{\mbox{\scriptsize #1}}$}{#2}}}

\def\fpage#1{\begingroup
\voffset=.3in
\thispagestyle{empty}\begin{table}[b]\centerline{\footnotesize #1}
	\end{table}\endgroup}

\def\runninghead#1#2{\pagestyle{myheadings}
\markboth{{\protect\footnotesize\it{\quad #1}}\hfill}
{\hfill{\protect\footnotesize\it{#2\quad}}}}
\font\tenrm=cmr10
\font\tenit=cmti10
\font\tenbf=cmbx10
\font\bfit=cmbxti10 at 10pt
\font\ninerm=cmr9

\font\eightrm=cmr8

\def\qed{\hbox{${\vcenter{\vbox{		%HOLLOW SQUARE
   \hrule height 0.4pt\hbox{\vrule width 0.4pt height 6pt
   \kern5pt\vrule width 0.4pt}\hrule height 0.4pt}}}$}}

\renewcommand{\thefootnote}{\fnsymbol{footnote}}
\input epsf
\global\arraycolsep=2pt
\def\spose#1{\hbox to 0pt{#1\hss}}
\def\lsim{\mathrel{\spose{\lower 3pt\hbox{$\mathchar"218$}}
 \raise 2.0pt\hbox{$\mathchar"13C$}}}
\def\gsim{\mathrel{\spose{\lower 3pt\hbox{$\mathchar"218$}}
 \raise 2.0pt\hbox{$\mathchar"13E$}}}

\begin{document}

%%% title page %%%%%%%%%%%%%

\thispagestyle{empty}

\rightline{TTP96-58}
\rightline{hep-ph/9612446}
\rightline{December 1996}
\vspace*{1.5truecm}
\bigskip
\boldmath
\begin{center}
{\Large {\bf CP Violation and the Role of Electroweak}}
\end{center}
\vspace*{0.2truecm}
\begin{center}
{\Large{\bf Penguins in Non-leptonic $B$ Decays}}\\
\end{center}
\unboldmath
\vspace*{1.3truecm}
\smallskip
\begin{center}
{\large{\sc Robert Fleischer}}\footnote{Internet: {\tt
rf@ttpux1.physik.uni-karlsruhe.de}}\\
\vspace{0.2cm}
{\sl Institut f\"{u}r Theoretische Teilchenphysik}\\
{\sl Universit\"{a}t Karlsruhe}\\
{\sl D--76128 Karlsruhe, Germany}
\vspace*{1.4cm}
\end{center}
\begin{abstract}
\noindent
The phenomenon of CP violation in the $B$ system and strategies 
for extracting CKM phases are reviewed. We focus both on general 
aspects and on some recent developments including CP-violating 
asymmetries in $B_d$ decays, the $B_s$ system in light of a possible 
width difference $\Delta\Gamma_s$, charged $B$ decays, and $SU(3)$ 
relations among certain transition amplitudes. In order to describe 
the relevant non-leptonic $B$ decays, low energy effective Hamiltonians 
calculated beyond the leading logarithmic approximation are used. 
Special emphasis is given to the role of electroweak penguin operators 
in such transitions. These effects are analyzed both within a general 
framework and more specifically in view of the theoretical cleanliness 
of methods to determine CKM phases. Strategies for obtaining insights 
into the world of electroweak penguins are discussed.
\end{abstract}

\vspace*{1.7cm}

\begin{center}
{\small To appear in {\it International Journal of Modern Physics} {\bf A}}
\end{center}

\newpage

%%% end title page %%%%%%%%%%%%%

\normalsize\textlineskip
\thispagestyle{empty}
\setcounter{page}{1}

\copyrightheading{}			%{Vol. 0, No. 0 (1993) 000--000}

\vspace*{0.88truein}

\fpage{1}

\centerline{\bf CP VIOLATION AND THE ROLE OF ELECTROWEAK}
\centerline{\bf PENGUINS IN NON-LEPTONIC B DECAYS}
\vspace*{0.37truein}

\centerline{\footnotesize ROBERT FLEISCHER}
\vspace*{0.015truein}
\centerline{\footnotesize\it Institut f\"{u}r Theoretische Teilchenphysik,
Universit\"{a}t Karlsruhe}
\baselineskip=10pt
\centerline{\footnotesize\it D--76128 Karlsruhe, Germany}

\vspace*{0.225truein}
\publisher{(received date)}{(revised date)}

\vspace*{0.21truein}
\abstracts{The phenomenon of CP violation in the $B$ system and 
strategies for extracting CKM phases are reviewed. We focus both on 
general aspects and on some recent developments including CP-violating 
asymmetries in $B_d$ decays, the $B_s$ system in light of a possible 
width difference $\Delta\Gamma_s$, charged $B$ decays, and $SU(3)$ 
relations among certain transition amplitudes. In order to describe 
the relevant non-leptonic $B$ decays, low energy effective Hamiltonians 
calculated beyond the leading logarithmic approximation are used. 
Special emphasis is given to the role of electroweak penguin operators 
in such transitions. These effects are analyzed both within a general 
framework and more specifically in view of the theoretical cleanliness 
of methods to determine CKM phases. Strategies for obtaining insights 
into the world of electroweak penguins are discussed.}{}{}
\vspace*{1pt}\textlineskip

\textheight=7.8truein
\setcounter{footnote}{0}
\renewcommand{\thefootnote}{\alph{footnote}}

\runninghead{Setting the Scene} {Setting the Scene}
\section{Setting the Scene}
\noindent
Although the experimental discovery of CP violation by Christenson, Cronin,
Fitch and Turlay\cite{ccft} goes back to the year 1964, the non-conservation 
of the CP symmetry still remains one of the unsolved mysteries in particle 
physics.

\runninghead{Setting the Scene} {CP Violation in the $K$-System}
\subsection{CP Violation in the $K$-System}
\noindent
So far CP violation has been observed only within the neutral $K$-meson 
system, where it is described by two complex quantities called 
$\varepsilon$ and $\varepsilon'$ which are defined by the following 
ratios of decay amplitudes:
\begin{equation}\label{defs-eps}
\frac{A(K_{\mbox{{\scriptsize L}}}\to\pi^+\pi^-)}{A(K_{\mbox{{\scriptsize 
S}}}
\to\pi^+\pi^-)}=\varepsilon+\varepsilon',\quad
\frac{A(K_{\mbox{{\scriptsize L}}}\to\pi^0\pi^0)}{A(K_{\mbox{{\scriptsize 
S}}}
\to\pi^0\pi^0)}=\varepsilon-2\varepsilon'.
\end{equation}
While $\varepsilon=(2.26\pm0.02)\cdot e^{i\frac{\pi}{4}}\cdot10^{-3}$ 
parametrizes ``indirect'' CP violation originating from the fact that 
the mass eigenstates of the neutral $K$-meson system are not eigenstates
of the CP operator, the quantity Re$(\varepsilon'/\varepsilon)$ provides
a measure of ``direct'' CP violation in $K\to\pi\pi$ transitions. 
Unfortunately the experimental situation concerning Re$(\varepsilon'/
\varepsilon)$, which has been subject of very involved experiments performed 
both at CERN and Fermilab by the the NA31 and E731 collaborations, 
respectively, is unclear at present. Whereas NA31 
finds\cite{NA31} Re$(\varepsilon'/\varepsilon)=(23\pm7)\cdot10^{-4}$ 
indicating already direct CP violation, the 
result\cite{E731} Re$(\varepsilon'/\varepsilon)=
(7.4\pm5.9)\cdot10^{-4}$ of the Fermilab experiment E731 provides no 
unambiguous evidence for a non-zero effect. In about two years this 
situation is hopefully clarified by the improved measurements at the 
$10^{-4}$ level of these two collaborations as well as by the KLOE 
experiment\cite{KLOE} at DA$\Phi$NE. 

Theoretical analyses\cite{buras-ichep96} of Re$(\varepsilon'/\varepsilon)$ 
are very difficult and suffer from large hadronic uncertainties. They are,
however, consistent with present experimental data. Because of this 
rather unfortunate theoretical situation, the measurement 
of a non-vanishing value of Re$(\varepsilon'/\varepsilon)$ will not provide 
a powerful quantitative test of our theoretical description of CP 
violation. Consequently the major goal of a possible future observation of 
Re$(\varepsilon'/\varepsilon)\not=0$ would be the exclusion of ``superweak''
theories\cite{superweak} of CP violation predicting a vanishing value of 
that quantity. 

\runninghead{Setting the Scene} {The Standard Model 
Description of CP Violation}
\subsection{The Standard Model Description of CP Violation}
\noindent
At present the observed CP-violating effects arising in the neutral
$K$-meson system can be described successfully by the Standard Model 
(SM) of electroweak interactions\cite{sm}. Within that framework CP 
violation is closely related to the quark-mixing-matrix -- the 
Cabibbo--Kobayashi--Maskawa matrix\cite{cab,km} (CKM matrix) -- connecting 
the electroweak eigenstates $(d',s',b')$ of the $d$-, $s$- and $b$-quarks 
with their mass eigenstates $(d,s,b)$ through the following unitary 
transformation:
\begin{equation}\label{ckm}
\left(\begin{array}{c}
d'\\
s'\\
b'
\end{array}\right)=\left(\begin{array}{ccc}
V_{ud}&V_{us}&V_{ub}\\
V_{cd}&V_{cs}&V_{cb}\\
V_{td}&V_{ts}&V_{tb}
\end{array}\right)\cdot
\left(\begin{array}{c}
d\\
s\\
b
\end{array}\right)\equiv\hat V_{\mbox{{\scriptsize CKM}}}\cdot
\left(\begin{array}{c}
d\\
s\\
b
\end{array}\right).
\end{equation}
The elements of the CKM matrix describe charged-current couplings as
can be seen easily by expressing the non-leptonic charged-current interaction 
Lagrangian
\begin{equation}\label{cc-lag1}
{\cal L}_{\mbox{{\scriptsize int}}}^{\mbox{{\scriptsize CC}}}=
-\frac{g_2}{\sqrt{2}}\left(\begin{array}{ccc}\bar
u_{\mbox{{\scriptsize L}}},& \bar c_{\mbox{{\scriptsize L}}},&
\bar t_{\mbox{{\scriptsize L}}}\end{array}\right)\gamma^\mu\left(
\begin{array}{c}
d'_{\mbox{{\scriptsize L}}}\\
s'_{\mbox{{\scriptsize L}}}\\
b'_{\mbox{{\scriptsize L}}}
\end{array}\right)W_\mu^\dagger\quad+\quad h.c.
\end{equation}
in terms of the electroweak eigenstates (\ref{ckm}):
\begin{equation}\label{cc-lag2}
{\cal L}_{\mbox{{\scriptsize int}}}^{\mbox{{\scriptsize CC}}}=
-\frac{g_2}{\sqrt{2}}\left(\begin{array}{ccc}\bar
u_{\mbox{{\scriptsize L}}},& \bar c_{\mbox{{\scriptsize L}}},&
\bar t_{\mbox{{\scriptsize L}}}\end{array}\right)\gamma^\mu\,\hat
V_{\mbox{{\scriptsize CKM}}}
\left(
\begin{array}{c}
d_{\mbox{{\scriptsize L}}}\\
s_{\mbox{{\scriptsize L}}}\\
b_{\mbox{{\scriptsize L}}}
\end{array}\right)W_\mu^\dagger\quad+\quad h.c.,
\end{equation}
where the gauge coupling $g_2$ is related to the gauge group 
$SU(2)_{\mbox{{\scriptsize L}}}$ and the $W_\mu^{(\dagger)}$ field 
corresponds to the charged $W$-bosons. Since neutrinos are massless within 
the SM, the analogue of the CKM matrix in the leptonic sector is equal to 
the unit matrix. Furthermore, since the CKM matrix is unitary in
flavor-space, flavor changing neutral current (FCNC) processes are absent
at tree-level within the SM. Therefore the unitarity of the CKM matrix is 
the basic requirement of the ``GIM-mechanism'' describing that  
feature\cite{gim}.

The elements of the CKM matrix are fundamental parameters of the SM and have 
to be extracted from experimental data. Whereas a single real parameter -- 
the Cabibbo angle $\Theta_{\mbox{{\scriptsize C}}}$ -- suffices to 
parametrize the CKM matrix in the case of two fermion generations\cite{cab}, 
three generalized Cabibbo-type angles and a single {\it complex phase} are 
needed in the three generation case\cite{km}. This complex phase is the 
origin of CP violation within the SM. Concerning phenomenological 
applications, the following parametrization of the CKM matrix, which 
exhibits nicely the hierarchical structure of its elements, is particularly 
useful:
\begin{equation}\label{wolf1}
\hat V_{\mbox{{\scriptsize CKM}}} =\left(\begin{array}{ccc}
1-\frac{1}{2}\lambda^2 & \lambda & A\lambda^3(\rho-i\,\eta) \\
-\lambda & 1-\frac{1}{2}\lambda^2 & A\lambda^2\\
A\lambda^3(1-\rho-i\,\eta) & -A\lambda^2 & 1
\end{array}\right)+\,{\cal O}(\lambda^4).
\end{equation}
The basic idea of that parametrization, which is due to 
Wolfenstein\cite{wolf}, is a phenomenological expansion of the CKM 
matrix in powers of the small quantity $\lambda\equiv|V_{us}|=
\sin\theta_{\mbox{{\scriptsize C}}}\approx0.22$. A treatment of the 
neglected higher order terms can be found e.g.\ in Refs.\cite{wolf,blo}.

Since at present only a single CP-violating observable, i.e.\ $\varepsilon$, 
has to be fitted, many different ``non-standard'' model descriptions of CP 
violation are imaginable. Since $\varepsilon'/\varepsilon$ is also not in 
a good shape to give an additional stringent constraint, the $K$-meson 
system by itself cannot provide a powerful test of the CP-violating sector 
of the SM.

\runninghead{Setting the Scene}{The Unitarity Triangle}
\subsection{The Unitarity Triangle}
\begin{figure}[t]
\centerline{
\rotate[r]{
\epsfysize=6truecm
\epsffile{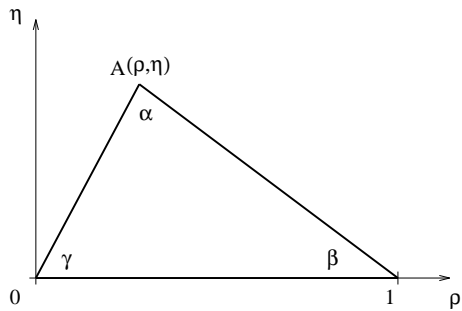}}}
\caption{The unitarity triangle of the CKM matrix in the $(\rho,\eta)$
plane.}\label{figUT}
\end{figure}

\noindent
As we will work out in detail in this review, the $B$-meson system 
represents
a very fertile ground for testing the SM description of CP violation. 
Concerning such tests, the central target is the ``unitarity triangle'' 
which is a graphical illustration of the fact that the CKM matrix is 
unitary\cite{ut}. The unitarity of the CKM matrix is expressed by
\begin{equation}
\hat V_{\mbox{{\scriptsize CKM}}}^{\,\,\dagger}\cdot\hat 
V_{\mbox{{\scriptsize CKM}}}=
\hat 1=\hat V_{\mbox{{\scriptsize CKM}}}\cdot\hat V_{\mbox{{\scriptsize 
CKM}}}^{\,\,\dagger}
\end{equation}
and leads to a set of twelve equations, where six equations are related to 
the normalization of the columns and rows of the CKM matrix, and the 
remaining six equations describe the orthogonality of different columns 
and rows, respectively. The orthogonality relations are of particular 
interest since they can be represented as {\it triangles} in the complex 
plane\cite{akl}. It can be shown that all of these triangles have the same 
area\cite{ut}, however, only in two of them all three sides are of comparable 
magnitude ${\cal O}(\lambda^3)$, while in the others one side is suppressed 
relative to the remaining ones by ${\cal O}(\lambda^2)$ or
${\cal O}(\lambda^4)$. The latter four triangles are therefore rather 
squashed ones and hence play a minor phenomenological role. A closer look 
at the two non-squashed triangles shows that they agree at leading order
in the Wolfenstein expansion so that one actually has to deal with a
single triangle -- {\it the} unitarity triangle (UT) of the CKM matrix --
that is described by
\begin{equation}\label{UT}
V_{ub}^\ast-\lambda V_{cb}^\ast+V_{td}=0.
\end{equation}
Here terms of ${\cal O}(\lambda^5)$ have been neglected. Expressing 
(\ref{UT}) in terms of the Wolfenstein parameters\cite{wolf} and 
rescaling all sides of the corresponding triangle by $A\lambda^3$ gives
\begin{equation}
(\rho+i\,\eta)-1+(1-\rho-i\,\eta)=0.
\end{equation}
Consequently the UT can be represented in the complex $(\rho,\eta)$ plane 
as has been shown in Fig.~\ref{figUT}. Defining the UT more strictly through
\begin{equation}
V_{ud}V_{ub}^\ast+V_{cd}V_{cb}^\ast+V_{td}V_{tb}^\ast=0,
\end{equation}
which is the exact CKM phase convention independent definition, the 
upper corner $A$ of the triangle depicted in that figure receives 
corrections of ${\cal O}(\lambda^2)$. In
Ref.\cite{blo} it was pointed out that these corrections can be
included straightforwardly by replacing $\rho\to\bar\rho\equiv\rho(1-
\lambda^2/2)$ and $\eta\to\bar\eta\equiv\eta(1-\lambda^2/2)$. To an 
accuracy of 3\% we have $\bar\rho=\rho$, $\bar\eta=\eta$ and as far as the 
phenomenological applications discussed in this review are concerned 
these corrections are inessential. 

The Wolfenstein parametrization (\ref{wolf1}) can be modified as follows
to make the dependence on the angles $\beta$ and $\gamma$ of the UT
explicit:
\begin{equation}\label{wolf2}
\hat V_{\mbox{{\scriptsize CKM}}} =\left(\begin{array}{ccc}
1-\frac{1}{2}\lambda^2 & \lambda & A\lambda^3 R_b\, e^{-i\gamma} \\
-\lambda & 1-\frac{1}{2}\lambda^2 & A\lambda^2\\
A\lambda^3R_t\,e^{-i\beta} & -A\lambda^2 & 1
\end{array}\right)+\,{\cal O}(\lambda^4),
\end{equation}
where
\begin{equation}\label{RbRt}
A\equiv\frac{1}{\lambda^2}\left|V_{cb}\right|,\,\,
R_b\equiv\frac{1}{\lambda}\left|\frac{V_{ub}}{V_{cb}}\right|=
\sqrt{\rho^2+\eta^2},\,\,
R_t\equiv\frac{1}{\lambda}\left|\frac{V_{td}}{V_{cb}}\right|=
\sqrt{(1-\rho)^2+\eta^2}.
\end{equation}
The presently allowed ranges\cite{al} for these parameters are 
$A=0.810\pm0.058$, $R_b=0.363\pm0.073$ and $R_t={\cal O}(1)$. 
The status of $R_t$ and strategies to fix this CKM factor have been 
summarized recently in Ref.\cite{buras-ichep96}. Note that the 
$3^{\mbox{{\scriptsize rd}}}$ angle $\alpha$ of the UT can be obtained 
straightforwardly through the relation
\begin{equation}
\alpha+\beta+\gamma=180^\circ.
\end{equation}

At present the UT can only be constrained indirectly through experimental 
data from CP-violating effects in the neutral $K$-meson system, 
$B^0_d-\overline{B^0_d}$ mixing, and from certain tree decays measuring 
$|V_{cb}|$ and $|V_{ub}|/|V_{cb}|$. Such analyses have been performed
by many authors and can be found e.g.\ in 
Refs.\cite{buras-ichep96,blo,al,bbl-rev}. It should, however, be possible 
to determine the three angles $\alpha$, $\beta$ and $\gamma$ of the UT 
independently in a {\it direct} way at future $B$ physics 
facilities\cite{babar}$^-$\cite{LHC-B} by measuring CP-violating effects 
in $B$ decays. Obviously one of the most exciting 
questions related to these measurements is whether the results for $\alpha$, 
$\beta$, $\gamma$ will agree one day or not. The latter possibility 
would signal ``New Physics''\,\cite{newphys} beyond the SM.

\runninghead{Setting the Scene} {Outline of the Review}
\subsection{Outline of the Review}
\noindent
In view of these experiments starting at the end of this millennium it 
is mandatory for theorists working on $B$ physics to search for decays 
that should allow interesting insights both into the mechanism of CP 
violation and into the structure of electroweak interactions in general. 
A review of such studies is the subject of the present article (for a
very compact version see Ref.\cite{dubna96}) which is organized as follows:

Since non-leptonic $B$-meson decays play the central role in respect to
CP violation and extracting angles of the UT, let us have a closer
look at these transitions in Section~2. A very useful tool to analyze
such decays are low energy effective Hamiltonians evaluated in
renormalization group improved perturbation theory. The general structure 
of these Hamiltonians consisting of perturbatively calculable Wilson
coefficient functions and local four-quark operators is presented 
in that section, and the problems caused by renormalization scheme 
dependences arising beyond the leading logarithmic approximation as well 
as their cancellation in the physical transition amplitudes are discussed. 

Section~3 is devoted to CP violation in non-leptonic $B$-meson decays and
reviews strategies for extracting the angles of the UT. Both general aspects, 
a careful discussion of the ``benchmark modes'' to determine $\alpha$, 
$\beta$ and $\gamma$, some recent developments including CP-violating 
asymmetries in $B_d$ decays, the $B_s$ system in light of a possible width 
difference $\Delta\Gamma_s$, charged $B$ decays, and relations among certain 
non-leptonic $B$ decay amplitudes are discussed. 

In Sections~4 and 5 we shall focus on electroweak penguin effects in 
non-leptonic $B$ decays and in strategies for extracting CKM phases,
respectively. This issue led to considerable interest in the recent 
literature. Na\"\i vely one would expect that electroweak 
penguins should only play a minor role since the ratio 
$\alpha/\alpha_s={\cal O}(10^{-2})$ of the QED and QCD couplings is very 
small. However, because of the large top-quark mass, electroweak penguins 
may nevertheless become important and may even compete with QCD penguins. 
These effects are discussed within a general framework in 
Section~4. There we will see that some non-leptonic $B$ decays are 
affected significantly by electroweak penguins and that a few of them 
should even be dominated by these contributions. The question 
to what extent the usual strategies for extracting angles of the 
UT are affected by the presence of electroweak penguins is addressed in 
Section~5. There also methods for obtaining experimental insights 
into the world of electroweak penguins are discussed. 

Finally in Section~6 a brief summary and some concluding remarks are
given. 

\runninghead{Non-leptonic $B$ Decays and Low Energy Effective Hamiltonians} 
{Non-leptonic $B$ Decays and Low Energy Effective Hamiltonians}
\section{Non-leptonic $B$ Decays and Low Energy Effective 
Hamiltonians}
\noindent
The subject of this section is an introduction to a very useful tool to deal
with non-leptonic $B$ decays: low energy effective Hamiltonians. Since the 
evaluation of these operators beyond the leading logarithmic approximation
has been reviewed in great detail by Buchalla, Buras and Lautenbacher in 
a recent paper\cite{bbl-rev}, only the general structure of these 
Hamiltonians is discussed here. For the technicalities the reader is 
referred to Ref.\cite{bbl-rev}. Before turning to these Hamiltonians, 
let us classify briefly non-leptonic $B$ decays in the following subsection.

\runninghead{Non-leptonic $B$ Decays and Low Energy Effective Hamiltonians} 
{Classification of Non-leptonic $B$ Decays}
\subsection{Classification of Non-leptonic $B$ Decays}
\noindent
Non-leptonic $B$ decays are caused by $b$-quark transitions of 
the type $b\to q_1\,\overline{q}_2\,q_3$ with $q_1\in\{d,s\}$ and 
$q_2,q_3\in\{u,d,c,s\}$ and can be divided into three classes:
\begin{itemize}
\item[i)]$q_2=q_3\in\{u,c\}$: both tree and penguin diagrams contribute.
\item[ii)]$q_2=q_3\in\{d,s\}$: only penguin diagrams contribute.
\item[iii)]$q_2\not=q_3\in\{u,c\}$: only tree diagrams contribute.
\end{itemize}
\begin{figure}
\centerline{
\epsfxsize=8.5truecm
\epsffile{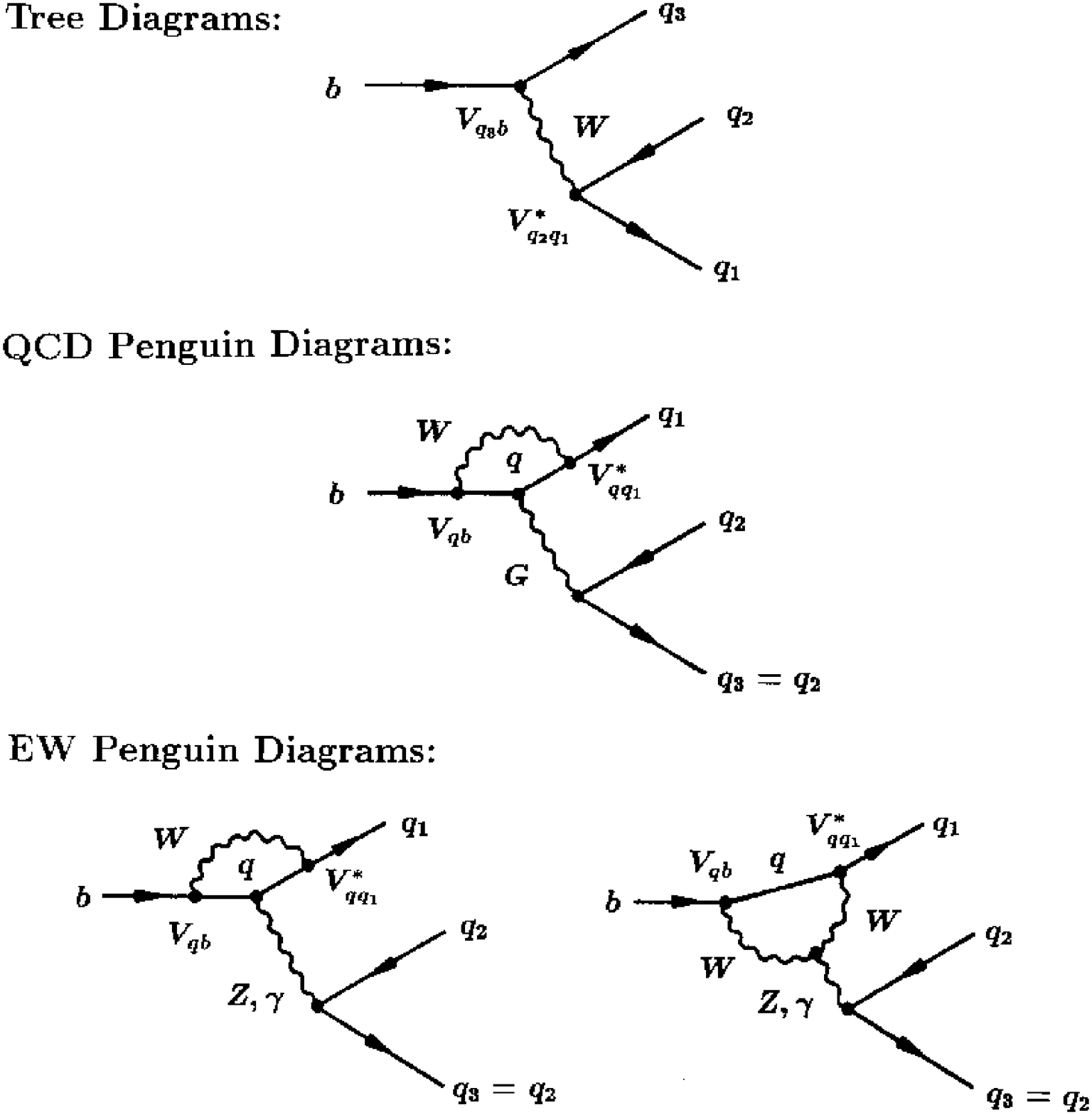}}
\caption{Lowest order contributions to non-leptonic $b$-quark
decays $(q\in\{u,c,t\})$.}\label{feyndiags}
\end{figure}
The corresponding lowest order Feynman diagrams are shown in 
Fig.~\ref{feyndiags}. There are two types of penguin topologies: 
{\it gluonic} (QCD) and {\it electroweak} (EW) penguins originating 
from strong and electroweak interactions, respectively. Such 
penguin diagrams play also an important role in the $K$-meson system. 
The corresponding operators were introduced there by Vainshtein, Zakharov 
and Shifman\cite{vzs}.

Concerning CP violation, decay classes i) and ii) are very promising. 
These modes, which are usually referred to as $|\Delta B|=1$, $\Delta 
C=\Delta U=0$ transitions, will hence play the major role in Section~3. 
Since we shall analyze such transitions by using low energy effective 
Hamiltonians 
calculated in renormalization group improved perturbation theory, let us 
have a closer look at these operators in the following subsection. 
Decays belonging to class iii) allow in some cases clean extractions of 
the angle $\gamma$ of the UT without any hadronic uncertainties and are 
therefore also very important. The structure of their low energy effective 
Hamiltonians can be obtained straightforwardly from the $|\Delta B|=1$, 
$\Delta C=\Delta U=0$ case.

\runninghead{Non-leptonic $B$ Decays and Low Energy Effective Hamiltonians} 
{Low Energy Effective Hamiltonians}
\subsection{Low Energy Effective Hamiltonians}
\noindent
In order to evaluate low energy effective Hamiltonians, one makes use of
the operator product expansion\cite{ope} (OPE) yielding transition matrix
elements of the structure
\begin{equation}\label{opetme}
\langle f|{\cal H}_{\mbox{{\scriptsize eff}}}|i\rangle\propto\sum_{k}
\langle f|Q_{k}(\mu)|i\rangle\,C_{k}(\mu),
\end{equation}
where $\mu$ denotes an appropriate renormalization scale. The OPE allows
one to separate the ``long-distance'' contributions to that decay
amplitude from the ``short-distance'' parts. Whereas the former
pieces are related to non-perturbative hadronic matrix elements 
$\langle f|Q_{k}(\mu)|i\rangle$, the latter ones are described by 
perturbatively calculable Wilson coefficient functions $C_k(\mu)$. 

In the case of $|\Delta B|=1$, $\Delta C=\Delta U=0$ transitions we have
\begin{equation}
{\cal H}_{\mbox{{\scriptsize eff}}}={\cal H}_{\mbox{{\scriptsize eff}}}
(\Delta B=-1)+{\cal H}_{\mbox{{\scriptsize eff}}}(\Delta B=-1)^\dagger
\end{equation}
with
\begin{equation}\label{LEham}
{\cal H}_{\mbox{{\scriptsize eff}}}(\Delta B=-1)=\frac{G_{\mbox{{\scriptsize
F}}}}{\sqrt{2}}\left[\sum\limits_{j=u,c}V_{jq}^\ast V_{jb}\left\{\sum
\limits_{k=1}^2Q_k^{jq}\,C_k(\mu)+\sum\limits_{k=3}^{10}Q_k^{q}\,C_k(\mu)
\right\}\right].
\end{equation}
Here $G_{\mbox{{\scriptsize F}}}$ denotes the Fermi constant, the 
renormalization scale $\mu$ is of ${\cal O}(m_b)$, the flavor label 
$q\in\{d,s\}$ corresponds to $b\to d$ and $b\to s$ transitions, respectively, 
and $Q_k^{jq}$ are four-quark operators that can be divided into three 
categories:
\begin{itemize}
\item[i)]current-current operators:
\begin{equation}\label{cc-def}
\begin{array}{rcl}
Q_{1}^{jq}&=&(\bar{q}_{\alpha}j_{\beta})_{\mbox{{\scriptsize V--A}}}
(\bar{j}_{\beta}b_{\alpha})_{\mbox{{\scriptsize V--A}}}\\
Q_{2}^{jq}&=&(\bar{q}_\alpha j_\alpha)_{\mbox{{\scriptsize 
V--A}}}(\bar{j}_\beta b_\beta)_{\mbox{{\scriptsize V--A}}}.
\end{array}
\end{equation}
\item[ii)]QCD penguin operators:
\begin{equation}\label{qcd-def}
\begin{array}{rcl}
Q_{3}^q&=&(\bar{q}_\alpha b_\alpha)_{\mbox{{\scriptsize V--A}}}\sum_{q'}
(\bar{q}'_\beta q'_\beta)_{\mbox{{\scriptsize V--A}}}\\
Q_{4}^q&=&(\bar{q}_{\alpha}b_{\beta})_{\mbox{{\scriptsize V--A}}}
\sum_{q'}(\bar{q}'_{\beta}q'_{\alpha})_{\mbox{{\scriptsize V--A}}}\\
Q_{5}^q&=&(\bar{q}_\alpha b_\alpha)_{\mbox{{\scriptsize V--A}}}\sum_{q'}
(\bar{q}'_\beta q'_\beta)_{\mbox{{\scriptsize V+A}}}\\
Q_{6}^q&=&(\bar{q}_{\alpha}b_{\beta})_{\mbox{{\scriptsize V--A}}}
\sum_{q'}(\bar{q}'_{\beta}q'_{\alpha})_{\mbox{{\scriptsize V+A}}}.
\end{array}
\end{equation}
\item[iii)]EW penguin operators:
\begin{equation}\label{ew-def}
\begin{array}{rcl}
Q_{7}^q&=&\frac{3}{2}(\bar{q}_\alpha b_\alpha)_{\mbox{{\scriptsize V--A}}}
\sum_{q'}e_{q'}(\bar{q}'_\beta q'_\beta)_{\mbox{{\scriptsize V+A}}}\\
Q_{8}^q&=&\frac{3}{2}(\bar{q}_{\alpha}b_{\beta})_{\mbox{{\scriptsize V--A}}}
\sum_{q'}e_{q'}(\bar{q}_{\beta}'q'_{\alpha})_{\mbox{{\scriptsize V+A}}}\\
Q_{9}^q&=&\frac{3}{2}(\bar{q}_\alpha b_\alpha)_{\mbox{{\scriptsize V--A}}}
\sum_{q'}e_{q'}(\bar{q}'_\beta q'_\beta)_{\mbox{{\scriptsize V--A}}}\\
Q_{10}^q&=&\frac{3}{2}(\bar{q}_{\alpha}b_{\beta})_{\mbox{{\scriptsize V--A}}}
\sum_{q'}e_{q'}(\bar{q}'_{\beta}q'_{\alpha})_{\mbox{{\scriptsize V--A}}}.
\end{array}
\end{equation}
\end{itemize}
Here $\alpha$ and $\beta$ denote $SU(3)_{\mbox{{\scriptsize C}}}$ 
color indices, V$\pm$A refers to the Lorentz structures $\gamma_\mu(1\pm
\gamma_5)$, respectively, $q'$ runs over the quark flavors being 
active at the scale \mbox{$\mu={\cal O}(m_{b})$}, i.e.\ 
$q'\in\{u,d,c,s\}$, and $e_{q'}$ are the corresponding electrical quark 
charges. The current-current, QCD and EW penguin operators are 
related to the tree, QCD and EW penguin processes depicted in
Fig.~\ref{feyndiags}. 

In the case of transitions belonging to class iii), only 
current-current operators contribute. The 
structure of the corresponding low energy effective Hamiltonians is 
completely analogous to (\ref{LEham}). We have simply to replace both 
the CKM factors $V_{jq}^\ast V_{jb}$ and the flavor contents of the 
current-current operators (\ref{cc-def}) straightforwardly, and have 
to omit the sum over penguin operators. We shall come back to the
resulting Hamiltonians\cite{bbl-rev,acmp,bw} in our discussion of $B_s$ 
decays originating from $\bar b\to\bar uc\bar s$ ($b\to c\bar u s$) 
quark-level transitions that is presented in 3.4.5.

The Wilson coefficient functions $C_k(\mu)$ can be calculated in
renormalization group improved perturbation theory. Within that 
framework the Wilson coefficients are evolved from a scale of the
order of the $W$-boson mass $M_W$ down to $\mu={\cal O}(m_b)$ by solving 
the renormalization group equations. The use of the renormalization group 
technique allows one to sum up large logarithms $\log(M_W/\mu)$. 
In the leading logarithmic approximation (LO)
terms of the type $(\alpha_s\log(M_W/\mu))^n$ are summed, in the 
next-to-leading logarithmic approximation (NLO) also terms 
$(\alpha_s)^n(\log(M_W/\mu))^{n-1}$ are summed, and so on. That 
procedure has been described extensively in an excellent recent 
review\cite{bbl-rev}, where all technicalities can be found. Let us 
therefore not go into details except one important feature discussed in the 
following subsection.  

\runninghead{Non-leptonic $B$ Decays and Low Energy Effective Hamiltonians} 
{Renormalization Scheme Dependences}
\subsection{Renormalization Scheme Dependences}
\noindent
Beyond LO problems arise from renormalization scheme dependences which
are reflected by the fact that the Wilson coefficient functions $C_k(\mu)$
depend both on the form of the operator basis specified in 
(\ref{cc-def})-(\ref{ew-def}) and on the scheme to renormalize the matrix 
elements of the corresponding operators\cite{buras-nlo}. In order to 
study the cancellation of these scheme dependences explicitly, it is 
convenient to introduce the following {\it renormalization scheme 
independent} Wilson coefficient functions\cite{buras-nlo}:
\begin{equation}
\overline{\vec C}(\mu)=\left[\hat 1+\frac{\alpha_{s}(\mu)}{4\pi}\hat
r^{T}_{s}+\frac{\alpha}{4\pi}\hat r^{T}_{e}\right]\cdot
\vec C(\mu).
\end{equation}
Here the scheme dependence of $\vec C(\mu)$ is cancelled through the
one of the scheme dependent matrices $\hat r^{T}_{s}$ and $\hat
r^{T}_{e}$. Using this parametrization we find
\begin{equation}\label{e626}
\vec Q^{T}\cdot \vec C(\mu)=\vec Q^{T}\cdot\left[\hat 1-
\frac{\alpha_{s}(\mu)}{4\pi}\hat r^{T}_{s}-\frac{\alpha}{4\pi}\hat
r^{T}_{e}\right]\cdot\overline{\vec C}(\mu),
\end{equation}
where the elements of the column vector $\vec Q$ are given by the
operators $Q_{1},\ldots,Q_{10}$ (flavor labels are suppressed 
in the following discussion to make the expressions more transparent). 
Taking into account one-loop QCD and QED matrix elements of the operators 
$Q_{k}$, which define matrices $\hat m_{s}(\mu)$ and $\hat m_{e}(\mu)$ 
through
\begin{equation}\label{e627}
\left\langle\vec Q^{T}(\mu)\right\rangle=\left\langle\vec Q^{T}
\right\rangle_{0}\cdot
\left[\hat 1+\frac{\alpha_{s}(\mu)}{4\pi}\hat
m^{T}_{s}(\mu)+\frac{\alpha}{4\pi}\hat m^{T}_{e}(\mu)\right],
\end{equation}
yields
\begin{eqnarray}
\lefteqn{\left\langle\vec Q^{T}(\mu)\cdot\vec C(\mu)\right
\rangle}\label{e628}\\
&&=\left\langle\vec Q^{T}\right\rangle_{0}\cdot\left[\hat 1+
\frac{\alpha_{s}(\mu)}{4\pi}\left(\hat m_{s}(\mu)-\hat r_{s}\right)^{T}+
\frac{\alpha}{4\pi}\left(\hat m_{e}(\mu)-\hat r_{e}\right)^{T}\right]
\cdot\overline{\vec C}(\mu),\nonumber
\end{eqnarray}
where terms of ${\cal O}(\alpha_{s}(\mu)^{2})$, ${\cal O}(\alpha\,
\alpha_{s}(\mu))$ and ${\cal O}(\alpha^{2})$ have been neglected and the
components of the vector $\langle\vec Q\rangle_{0}$ denote 
the tree level matrix elements of the operators $Q_{1},\ldots,Q_{10}$. Since 
the matrices $\hat r_{s,e}$ are special cases of the matrices 
$\hat m_{s,e}$ (see Ref.\cite{buras-nlo}), the renormalization scheme 
dependences of these matrices cancel in (\ref{e628}). Therefore the matrix 
element given in that expression is {\it renormalization scheme independent}.
Since penguin contributions play a central role in this review,  
the penguin sector of the matrix element (\ref{e628}) will be of 
particular interest:
\begin{eqnarray}
\left\langle\vec Q^{T}(\mu)\cdot\vec
C(\mu)\right\rangle^{\mbox{{\scriptsize pen}}}&=&\sum\limits_{k=3}^{6}\langle
Q_{k}\rangle_{0}\biggr[\overline{C}_{k}(\mu)+\frac{\alpha_{s}(\mu)}{4\pi}
\left(\hat m_{s}(\mu)-\hat r_{s}\right)_{2k}\overline{C}_{2}(\mu)
\biggr]\label{e629}\\
&&+\sum\limits_{k=7}^{10}\langle
Q_{k}\rangle_{0}\biggl[\overline{C}_{k}(\mu)+\frac{\alpha}{4\pi}
\sum\limits_{j=1}^{2}\left(\hat
m_{e}(\mu)-\hat r_{e}\right)_{jk}\overline{C}_{j}(\mu)\biggr].\nonumber
\end{eqnarray}
Here one-loop matrix elements of penguin operators have been neglected 
as in Ref.\cite{rf1}.  Moreover it has been taken into account that the
current-current operator $Q_{1}$ does not mix with QCD penguin
operators at the one-loop level because of its color-structure.
\begin{figure}
\centerline{
\epsfxsize=6.5truecm
\epsffile{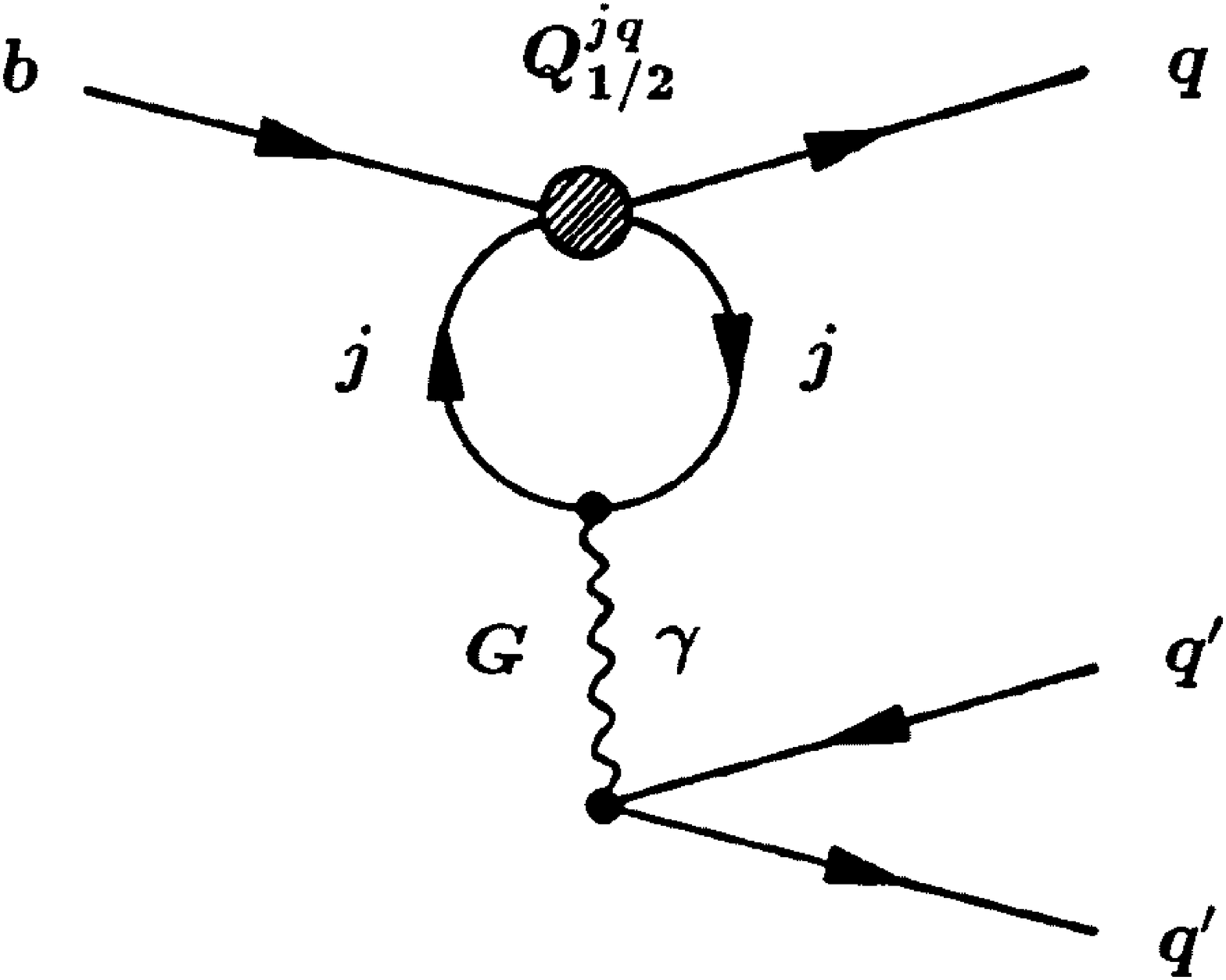}}
\caption{One-loop QCD and QED time-like penguin matrix elements of the
current-current operators $Q_{1/2}^{jq}$.}\label{pen-me}
\end{figure}

Calculating both the one-loop QCD and QED time-like penguin matrix
elements of the current-current operators $Q_{1/2}$ shown in
Fig.~\ref{pen-me}, one finds\cite{rf1,rfewp1} the following non-vanishing 
elements $(\hat m_{s/e}(\mu))_{jk}$~($j\in\{1,2\}$,~$k\in\{3,\ldots,10\}$) 
of the matrices $\hat m_{s}(\mu)$ and $\hat m_{e}(\mu)$:
\begin{eqnarray}
(\hat m_{s}(\mu))_{23}&=&(\hat m_{s}(\mu))_{25}=
\,\,\,\,\,\frac{1}{6}\left[\frac{2}{3}\kappa+G(m,k,\mu)\right]\nonumber\\
(\hat m_{s}(\mu))_{24}&=&(\hat m_{s}(\mu))_{26}=
-\frac{1}{2}\left[\frac{2}{3}\kappa+G(m,k,\mu)\right]\nonumber\\
(\hat m_{e}(\mu))_{17}&=&(\hat m_{e}(\mu))_{19}=
-\frac{4}{3}\left[\frac{2}{3}\kappa+G(m,k,\mu)\right]\label{e630}\\
(\hat m_{e}(\mu))_{27}&=&(\hat m_{e}(\mu))_{29}=
-\frac{4}{9}\left[\frac{2}{3}\kappa+G(m,k,\mu)\right]\,.\nonumber
\end{eqnarray}
Here $\kappa$ parametrizes the renormalization scheme dependences and
distinguishes between different mass-independent renormalization schemes. 
The function $G(m,k,\mu)$ is defined by
\begin{equation}\label{e633}
G(m,k,\mu)=-\,4\int\limits_{0}^{1}\mbox{d}x\,x\,(1-x)\ln\left[\frac{m^{2}-
k^{2}\,x\,(1-x)}
{\mu^{2}}\right],
\end{equation}
where $m$ is the mass of the quark running in the loop of the penguin diagram
shown in Fig.~\ref{pen-me} and $k$ denotes the four-momentum of the virtual 
gluons and photons appearing in that figure. 

The elements of the matrices $\hat r_{s}$ and $\hat r_{e}$
corresponding to (\ref{e630}) are given as 
follows\cite{buras-nlo}$^-$\cite{rfewp1}:
\begin{eqnarray}
(\hat r_{s})_{23}&=&(\hat r_{s})_{25}=
\,\,\,\,\,\frac{1}{6}\left[\frac{2}{3}\kappa+\frac{10}{9}\right],\,\,\,
(\hat r_{s})_{24}=(\hat r_{s})_{26}=
-\frac{1}{2}\left[\frac{2}{3}\kappa+\frac{10}{9}\right]\nonumber\\
(\hat r_{e})_{17}&=&(\hat r_{e})_{19}=
-\frac{4}{3}\left[\frac{2}{3}\kappa+\frac{10}{9}\right],\,\,\,
(\hat r_{e})_{27}=(\hat r_{e})_{29}=
-\frac{4}{9}\left[\frac{2}{3}\kappa+\frac{10}{9}\right]\label{e634}\,.
\end{eqnarray}
Combining (\ref{e629}) with (\ref{e630}) and (\ref{e634}), we observe
explicitly
that the renormalization scheme dependent terms parametrized by
$\kappa$ cancel each other and obtain the following {\it
renormalization scheme independent} expression:
\begin{eqnarray}
\lefteqn{\left\langle \vec Q^{T}(\mu)\cdot\vec
C(\mu)\right\rangle^{\mbox{{\scriptsize pen}}}=
\sum\limits_{k=3}^{10}\langle Q_{k}\rangle_{0}\overline{C}_{k}(\mu)+
\frac{\alpha_{s}(\mu)}{8\pi}\left[\frac{10}{9}-G(m,k,\mu)
\right]}\nonumber\\
&&\times\Biggr\{\left(-\frac{1}{3}\langle Q_{3}\rangle_{0}+
\langle Q_{4}\rangle_{0}-\frac{1}{3}\langle Q_{5}\rangle_{0}+
\langle Q_{6}\rangle_{0}\right)\overline{C}_{2}(\mu)\nonumber\\
&&+\frac{8}{9}\frac{\alpha}{\alpha_{s}(\mu)}\left(
\langle Q_{7}\rangle_{0}+\langle Q_{9}\rangle_{0}\right)\left(
3\overline{C}_{1}(\mu)+\overline{C}_{2}(\mu)\right)\Biggr\}\,.\label{e635}
\end{eqnarray}
The problems related to renormalization scheme dependences arising at
NLO and their cancellation through certain matrix elements have also been
investigated in Ref.\cite{kps}. There additional subtleties, which are
beyond the scope of this review, have been analyzed.

For later discussions it is useful to consider also the case where 
the proper renormalization group evolution from $\mu={\cal O}(M_W)$ down to 
$\mu={\cal O}(m_b)$ is neglected. The advantage of the corresponding Wilson 
coefficients is the point that they exhibit the top-quark mass dependence 
in a transparent way and allow moreover to investigate the importance of
NLO renormalization group effects. 

\runninghead{Non-leptonic $B$ Decays and Low Energy Effective Hamiltonians} 
{Neglect of the Proper Renormalization Group Evolution}
\subsection{Neglect of the Proper Renormalization Group Evolution}
\noindent
If one does not perform the NLO renormalization group evolution from 
$\mu={\cal O}(M_{W})$ down to $\mu={\cal O}(m_{b})$ but calculates the 
relevant Feynman diagrams directly at a scale of ${\cal O}(M_{W})$ 
with full $W$ and $Z$ propagators and internal top-quark exchanges (see 
Fig.~\ref{feyndiags}), one obtains the following set of coefficient 
functions\cite{rfewp1}:
\begin{eqnarray}
\overline{C}^{(0)}_{1}(\mu)&=&{\cal O}(\alpha_{s}(\mu))+
{\cal O}(\alpha),\quad \overline{C}^{(0)}_{2}(\mu)\,=\,1+{\cal O}
(\alpha_{s}(\mu))+{\cal O}(\alpha)\nonumber\\
\overline{C}^{(0)}_{3}(\mu)&=&-\frac{\alpha_{s}(\mu)}{24\pi}\left[E(x_{t})
-\frac{\alpha}{\alpha_{s}(\mu)}\frac{1}{\sin^{2}
\Theta_{\mbox{{\scriptsize W}}}}\left\{
8B(x_{t})+4C(x_{t})\right\}
+\frac{2}{3}\ln\frac{\mu^{2}}{M_{W}^{2}}-\frac{10}{9}\right]
\nonumber\\
\overline{C}^{(0)}_{4}(\mu)&=&-3\,\overline{C}^{(0)}_{5}(\mu)\,=\,
\overline{C}^{(0)}_{6}(\mu)\,=\,
\frac{\alpha_{s}(\mu)}{8\pi}\left[E(x_{t})
+\frac{2}{3}\ln\frac{\mu^{2}}{M_{W}^{2}}-\frac{10}{9}\right]\nonumber\\
\overline{C}^{(0)}_{7}(\mu)&=&\frac{\alpha}{6\pi}\left[4C(x_{t})+D(x_{t})
+\frac{4}{9}\ln\frac{\mu^{2}}{M_{W}^{2}}-\frac{20}{27}\right]
\label{e636}\\
\overline{C}^{(0)}_{8}(\mu)&=&\overline{C}^{(0)}_{10}(\mu)\,=\,0\nonumber\\
\overline{C}^{(0)}_{9}(\mu)&=&\frac{\alpha}{6\pi}\left[4C(x_{t})+D(x_{t})+
\frac{1}{\sin^{2}
\Theta_{\mbox{{\scriptsize W}}}}
\left\{10B(x_{t})-4C(x_{t})\right\}
+\frac{4}{9}\ln\frac{\mu^{2}}{M_{W}^{2}}-\frac{20}{27}\right]
\nonumber,
\end{eqnarray}
where $x_{t}\equiv m_{t}^{2}/M_{W}^{2}$ and 
\begin{displaymath}
B(x)=\frac{1}{4}\left[\frac{x}{1-x}+\frac{x\ln x}{(x-1)^{2}}\right],\quad
C(x)=\frac{x}{8}\left[\frac{x-6}{x-1}+\frac{3x+2}{(x-1)^{2}}\ln
x\right],
\end{displaymath}
\begin{equation}\label{e637}
D(x)=-\frac{4}{9}\ln x+\frac{-19x^{3}+25x^{2}}{36(x-1)^{3}}+
\frac{x^{2}(5x^{2}-2x-6)}{18(x-1)^{4}}\ln x,
\end{equation}
\begin{displaymath}
E(x)=-\frac{2}{3}\ln
x+\frac{x(18-11x-x^{2})}{12(1-x)^{3}}+\frac{x^{2}(15-16x+4x^{2})}{6(1-x)^{4}}
\ln x.
\end{displaymath}
The Inami-Lim functions\cite{il} $B(x_{t})$, $C(x_{t})$, $D(x_{t})$ 
and $E(x_{t})$ describe contributions of box diagrams (which have not been 
shown in Fig.~\ref{feyndiags}),
$Z$ penguins, photon penguins and gluon penguins, respectively. 
Using the coefficients (\ref{e636}), QCD renormalization group effects 
are included only approximately through the rescaling 
$\alpha_{s}(M_{W})\to\alpha_{s}(\mu)$, where $\mu={\cal O}(m_b)$. Note that
the $\mu$-dependence of the coefficients $\overline{C}^{(0)}_{k}(\mu)$ 
originating from the logarithmic terms of the form $\ln(\mu^{2}/M_{W}^{2})$ 
is cancelled in the matrix element (\ref{e635}) by the one of the function 
$G(m,k,\mu)$. The ${\cal O}(\alpha_{s}(\mu))$ and ${\cal O}(\alpha)$ 
corrections to $\overline{C}^{(0)}_{1}(\mu)$ and 
$\overline{C}^{(0)}_{2}(\mu)$ contribute ${\cal O}(\alpha_{s}(\mu)^{2})$, 
${\cal O}(\alpha\,\alpha_{s}(\mu))$ or ${\cal O} (\alpha^{2})$ effects 
to the penguin amplitude (\ref{e635}) and have to be neglected to the 
order we are working at in this review. For most practical applications, the
differences between using the NLO Wilson coefficients or (\ref{e636}),
i.e.\ the NLO renormalization group effects, are of order $(10-20)\%$
depending on the considered observables\cite{rfewp1}.

This remark concludes the brief introduction to low energy effective
Hamiltonians calculated beyond LO. The subject of the subsequent
section is a review of the current theoretical status of CP violation 
in non-leptonic $B$ decays and of strategies for extracting angles 
of the UT making use of these CP-violating effects.

\runninghead{CP Violation in Non-leptonic $B$-Meson Decays} 
{CP Violation in Non-leptonic $B$-Meson Decays}
\section{CP Violation in Non-leptonic $B$-Meson Decays}
\noindent
Whereas CP-violating asymmetries in charged $B$ decays suffer in general
from large hadronic uncertainties and are hence mainly interesting in
respect of ruling out ``superweak'' models\cite{superweak} of CP violation, 
the neutral 
$B_q$-meson systems $(q\in\{d,s\})$ provide excellent laboratories to 
perform stringent tests of the SM description of CP violation\cite{cp-revs}. 
This feature is mainly due to ``mixing-induced'' CP violation which is 
absent in the charged $B$ system and arises from interference between 
decay- and $B^0_q-\overline{B^0_q}$ mixing-processes. In order to derive 
the formulae for the corresponding CP-violating asymmetries, we 
have to discuss $B^0_q-\overline{B^0_q}$ mixing first. 

\runninghead{CP Violation in Non-leptonic $B$-Meson Decays}  
{$B^0_q-\overline{B^0_q}$ Mixing}
\subsection{The Phenomenon of $B^0_q-\overline{B^0_q}$ Mixing}
\begin{figure}[t]
\centerline{
\epsfxsize=8.5truecm
\epsffile{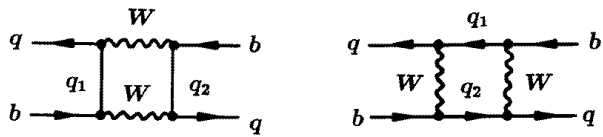}}
\caption{Box diagrams contributing to $B^0_q-\overline{B^0_q}$
mixing $(q_1,q_2\in\{u,c,t\})$.}\label{b-b.bar-mix}
\end{figure}

\noindent
Within the SM, $B^0_q-\overline{B^0_q}$ mixing is induced at lowest order 
through the box diagrams shown in Fig.~\ref{b-b.bar-mix}. Applying a 
matrix notation, the Wigner-Weisskopf formalism\cite{wigwei} yields an 
effective Schr\"odinger equation of the form
\begin{equation}\label{e71}
i\,\frac{\partial}{\partial t}\left(\begin{array}{c} a(t)\\ b(t)
\end{array}
\right)=
\left[\left(\begin{array}{cc}
M_{0}^{(q)} & M_{12}^{(q)}\\ M_{12}^{(q)\ast} & M_{0}^{(q)}
\end{array}\right)-
\frac{i}{2}\left(\begin{array}{cc}
\Gamma_{0}^{(q)} & \Gamma_{12}^{(q)}\\
\Gamma_{12}^{(q)\ast} & \Gamma_{0}^{(q)}
\end{array}\right)\right]
\cdot\left(\begin{array}{c}
a(t)\\ b(t)
\end{array}
\right)
\end{equation}
describing the time evolution of the state vector
\begin{equation}\label{e72}
\left\vert \psi_q(t)\right\rangle=a(t)\left\vert B^{0}_q\right\rangle+
b(t)\left\vert\overline{B^{0}_q}\right\rangle.
\end{equation}
The special form of the mass and decay matrices in (\ref{e71})
follows from invariance under CPT transformations. It is an easy
exercise to evaluate the eigenstates $\left\vert B_{\pm}^{(q)}\right\rangle$ 
with eigenvalues $\lambda_{\pm}^{(q)}$ of that Hamilton operator. They are 
given by
\begin{equation}\label{e73}
\left\vert B_{\pm}^{(q)} \right\rangle  =  
\frac{1}{\sqrt{1+\vert \alpha_q\vert^{2}}}
\left(\left\vert B^{0}_q\right\rangle\pm\alpha_q\left\vert
\overline{B^{0}_q}\right\rangle\right)
\end{equation}
\begin{equation}\label{e74}
\lambda_{\pm}^{(q)}  =
\left(M_{0}^{(q)}-\frac{i}{2}\Gamma_{0}^{(q)}
\right)\pm
\left(M_{12}^{(q)}-\frac{i}{2}\Gamma_{12}^{(q)}\right)\alpha_q,
\end{equation}
where
\begin{equation}\label{e75}
\alpha_q  = \sqrt{\frac{4\vert M_{12}^{(q)}\vert^{2}
e^{-i2\delta\Theta_{M/\Gamma}^{(q)}}+\vert\Gamma_{12}^{(q)}\vert^{2}}
{4\vert M_{12}^{(q)}\vert^{2}+\vert\Gamma_{12}^{(q)}\vert^{2}- 4\vert
M_{12}^{(q)}\vert\vert\Gamma_{12}^{(q)}\vert\sin\delta\Theta_{M
/\Gamma}^{(q)}}}e^{-i\left(\Theta_{\Gamma_{12}}^{(q)}+n'
\pi\right)}.
\end{equation}
Here the notations $M_{12}^{(q)}\equiv e^{i\Theta_{M_{12}}^{(q)}}\vert
M_{12}^{(q)}\vert$, $\Gamma_{12}^{(q)}\equiv
e^{i\Theta_{\Gamma_{12}}^{(q)}}\vert\Gamma_{12}^{(q)}
\vert$ and
$\delta\Theta_{M/\Gamma}^{(q)}\equiv\Theta_{M_{12}}^{(q)}-
\Theta_{\Gamma_{12}}^{(q)}$ have been introduced and $n'\in {\rm Z}$ 
parametrizes the sign of the square root appearing in that expression. 
Calculating the dispersive and absorptive parts of the box diagrams 
depicted in Fig.~\ref{b-b.bar-mix} one obtains\cite{buslst}
\begin{eqnarray}
M_{12}^{(q)}&=&
\lefteqn{\frac{G_{\mbox{{\scriptsize F}}}^{2}M_{W}^{2}M_{B_q}B_{B_q}
f_{B_q}^{2}}{12\pi^{2}}}
\label{e76}\\
&&\times\left[v_{c}^{(q)2}S(x_{c})+v_{t}^{(q)2}S(x_{t})+2v_{c}^{(q)}
v_{t}^{(q)}S(x_{c},x_{t})\right] e^{i(\pi-\phi_{\mbox{{\scriptsize 
CP}}}(B_q))}\nonumber
\end{eqnarray}
and
\begin{eqnarray}
\lefteqn{\Gamma_{12}^{(q)}=\frac{G_{\mbox{{\scriptsize F}}}^{2}m_{b}^{2}
M_{B_q}B_{B_q}f_{B_q}^{2}}{8\pi}
\left[v_{t}^{(q)2}+\frac{8}{3}v_{c}^{(q)}v_{t}^{(q)}
\left(z_{c}+\frac{1}{4}z_{c}^{2}-
\frac{1}{2}z_{c}^{3}\right)\right.}\label{e77}\\
&&+\left.v_{c}^{(q)2}\left\{\sqrt{1-4z_{c}}\left(1-\frac{2}{3}z_{c}\right)+
\frac{8}{3}z_{c}+\frac{2}{3}z_{c}^{2}-\frac{4}{3}z_{c}^{3}-1\right\}\right]
e^{-i\phi_{\mbox{{\scriptsize CP}}}(B_q)},\nonumber
\end{eqnarray}
respectively, where $x_c\equiv m_c^2/M_W^2$, $x_t\equiv m_t^2/M_W^2$,
$z_c\equiv m_c^2/m_b^2$ and $ v_{i}^{(q)}\equiv V_{iq}^{\ast}V_{ib}$. The
non-perturbative ``$B$-parameter'' $B_{B_q}={\cal O}(1)$ is related to
the hadronic matrix element $\bigl\langle\overline{B^0_q}\bigl|
[\bar b\gamma_\mu(1-\gamma_5)q]^2\bigr|B_q^0\bigr\rangle$, $f_{B_q}$ is
the $B_q$-meson decay constant and $M_{B_q}$ denotes the mass of the
$B_q$-meson.  The functions $S(x)$ and $S(x,y)$ are given by
\begin{equation}\label{e78}
S(x)=\left[\frac{1}{4}-\frac{9}{4(x-1)}-\frac{3}{2(x-1)^{2}}\right]x+
\frac{3}{2}\left(\frac{x}{x-1}\right)^{3}\ln{x}
\end{equation}
\begin{equation}\label{e79}
S(x,y)=\frac{xy}{x-y}\left[\frac{1}{4}-\frac{3}{2(x-1)}-
\frac{3}{4(x-1)^{2}}\right]\ln x+(x\leftrightarrow y)-
\frac{3xy}{4(x-1)(y-1)}
\end{equation}
and the phase $\phi_{\mbox{{\scriptsize CP}}}(B_q)$ 
parametrizing the applied CP phase convention is defined through 
\begin{equation}\label{e710}
({\cal CP})\left\vert B^{0}_q\right\rangle=
e^{i\phi_{\mbox{{\scriptsize CP}}}(B_q)}
\left\vert\overline{B^{0}_q}\right\rangle.
\end{equation}
In the presence of a heavy top-quark of mass
$m_t={\cal O}(170\,\mbox{GeV})$ we have
$S(x_{t})={\cal O}(1)$, $S(x_{c})={\cal O}(10^{-4})$ and
$S(x_{c},x_{t})={\cal O}(10^{-3})$. Consequently, since $v_c^{(q)}$ and
$v_t^{(q)}$ are of the same order in $\lambda$ (${\cal O}(\lambda^3)$
and ${\cal O}(\lambda^2)$ for $q=d$ and $s$, respectively),
$M_{12}^{(q)}$ is governed by internal top-quark exchanges and can be
approximated as
\begin{equation}\label{e711}
M_{12}^{(q)}=\frac{G_{\mbox{{\scriptsize F}}}^{2}M_{W}^{2}
M_{B_q}B_{B_q}f_{B_q}^{2}}{12\pi^{2}}\,v_{t}^{(q)2}S(x_{t})\,
e^{i\left(\pi-\phi_{\mbox{{\scriptsize CP}}}(B_q)\right)}.
\end{equation}
On the other hand, since the expression (\ref{e77}) for the off-diagonal
element $\Gamma_{12}^{(q)}$ of the decay matrix is dominated by the 
term proportional to $v_t^{(q)2}$, we have
\begin{equation}\label{e712}
\frac{\Gamma_{12}^{(q)}}{M_{12}^{(q)}}\approx
-\frac{3\pi}{2S(x_{t})}\frac{m_b^2}{M_W^2}.
\end{equation}
Therefore, $|\Gamma_{12}^{(q)}|/|M_{12}^{(q)}|={\cal O}(m_b^2/m_t^2)\ll1$.
Expanding (\ref{e75}) in powers of this small quantity gives
\begin{equation}\label{e713}
\alpha_q=\left[1+\frac{|\Gamma_{12}^{(q)}|}{2|M_{12}^{(q)}|}\sin\delta
\Theta_{M/\Gamma}^{(q)}\right]e^{-i\left(\Theta_{M_{12}}^{(q)}+n'\pi\right)}
+{\cal O}\left(\left(\frac{|\Gamma_{12}^{(q)}|}{|M_{12}^{(q)}|}
\right)^2\right).
\end{equation}
The deviation of $|\alpha_q|$ from 1 describes CP-violating effects in
$B^0_q-\overline{B^0_q}$ oscillations. This type of CP violation is
probed by rate asymmetries in semileptonic decays of neutral
$B_q$-mesons into ``wrong charge'' leptons, i.e.\ by comparing the rate of 
an initially pure $B_q^0$-meson decaying into $l^-\overline{\nu}_l X$ 
with that of an initially pure $\overline{B^0_q}$ decaying into 
$l^+\nu_l X$:
\begin{equation}\label{e714}
{\cal A}^{(q)}_{\mbox{{\scriptsize SL}}}\equiv
\frac{\Gamma(B^0_q(t)\to l^-\overline{\nu}_l X)-\Gamma(\overline{B^0_q}(t)\to
l^+\nu_l X)}{\Gamma(B^0_q(t)\to l^-\overline{\nu}_l X)+
\Gamma(\overline{B^0_q}(t)\to l^+\nu_l X)}=
\frac{|\alpha_q|^4-1}{|\alpha_q|^4+1}\approx\frac{|\Gamma_{12}^{(q)}|}
{|M_{12}^{(q)}|}\sin\delta\Theta^{(q)}_{M/\Gamma}.
\end{equation}
Note that the time dependences cancel in (\ref{e714}). Because of
$|\Gamma_{12}^{(q)}|/|M_{12}^{(q)}|\propto
m_b^2/m_t^2$ and $\sin\delta\Theta^{(q)}_{M/\Gamma}\propto
m_c^2/m_b^2$, the asymmetry (\ref{e714}) is suppressed by 
a factor $m_c^2/m_t^2={\cal O}(10^{-4})$ and is hence expected to 
be very small within the SM. At present there exists an
experimental upper bound $|\mbox{Re}(\varepsilon_{B_d})|\equiv |{\cal
A}^{(d)}_{\mbox{{\scriptsize SL}}}/4|<45\cdot10^{-3}$ (90\% C.L.) from
the CLEO collaboration\cite{cleo} which is about two orders of 
magnitudes above the SM prediction. 

The time-evolution of initially, i.e.\ at $t=0$, pure $\left|B^0_q\right
\rangle$ and $\left|\overline{B^0_q}\right\rangle$ meson states is given by
\begin{eqnarray}
\left|B^0_q(t)\right\rangle&=&f_+^{(q)}(t)\left|B^{0}_q\right\rangle
+\alpha_qf_-^{(q)}(t)\left|\overline{B^{0}_q}\right\rangle\label{e715}\\
\left|\overline{B^0_q}(t)\right\rangle&=&\frac{1}{\alpha_q}f_-^{(q)}(t)
\left|B^{0}_q\right\rangle+f_+^{(q)}(t)\left|\overline{B^{0}_q}\right
\rangle,\label{e716}
\end{eqnarray}
where
\begin{equation}
f_{\pm}^{(q)}(t)=\frac{1}{2}\left(e^{-i\lambda_+^{(q)}t}\pm
e^{-i\lambda_-^{(q)}t}\right).
\end{equation}
Using these time-dependent state vectors and neglecting the very
small CP-violating effects in $B^0_q-\overline{B^0_q}$ mixing that 
are described by $|\alpha_q|\not=1$ (see (\ref{e713})), a straightforward 
calculation yields\cite{time-evol}
\begin{eqnarray}
\Gamma(B^0_q(t)\to f)&=&\left[\left|g_+^{(q)}(t)\right|^2+\left|\xi_f^{(q)}
\right|^2\left|g_-^{(q)}(t)\right|^2-
2\mbox{\,Re}\left\{\xi_f^{(q)}g_-^{(q)}(t)g_+^{(q)}(t)^\ast\right\}
\right]\tilde\Gamma\label{ratebf}\\
\Gamma(\overline{B^0_q}(t)\to f)&=&\left[\left|g_-^{(q)}(t)\right|^2+
\left|\xi_f^{(q)}\right|^2\left|g_+^{(q)}(t)\right|^2-
2\mbox{\,Re}\left\{\xi_f^{(q)}g_+^{(q)}(t)g_-^{(q)}(t)^\ast\right\}
\right]\tilde\Gamma\quad\label{ratebbf}\\
\Gamma(B^0_q(t)\to\overline{f})&=&\left[\left|g_+^{(q)}(t)\right|^2+
\left|\xi_{\overline{f}}^{(q)}\right|^2\left|g_-^{(q)}(t)\right|^2-
2\mbox{\,Re}\left\{\xi_{\overline{f}}^{(q)}g_-^{(q)}(t)g_+^{(q)}(t)^\ast
\right\}\right]\tilde{\overline{\Gamma}}\label{ratebfb}\\
\Gamma(\overline{B^0_q}(t)\to\overline{f})&=&\left[\left|g_-^{(q)}(t)
\right|^2+\left|\xi_{\overline{f}}^{(q)}\right|^2\left|g_+^{(q)}(t)\right|^2-
2\mbox{\,Re}\left\{\xi_{\overline{f}}^{(q)}g_+^{(q)}(t)g_-^{(q)}(t)^\ast
\right\}\right]\tilde{\overline{\Gamma}},\label{ratebbfb}
\end{eqnarray}
where
\begin{equation}
\left|g^{(q)}_{\pm}(t)\right|^2=\frac{1}{4}\left[e^{-\Gamma_L^{(q)}t}+
e^{-\Gamma_H^{(q)}t}\pm2\,e^{-\Gamma_q t}\cos(\Delta M_qt)\right]
\end{equation}
\begin{equation}
g_-^{(q)}(t)\,g_+^{(q)}(t)^\ast=\frac{1}{4}\left[e^{-\Gamma_L^{(q)}t}-
e^{-\Gamma_H^{(q)}t}+2\,i\,e^{-\Gamma_q t}\sin(\Delta M_qt)\right]
\end{equation}
and
\begin{equation}\label{e731}
\xi_f^{(q)}=e^{-i\Theta_{M_{12}}^{(q)}}
\frac{A(\overline{B_q^0}\to f)}{A(B_q^0\to f)},\quad
\xi_{\overline{f}}^{(q)}=e^{-i\Theta_{M_{12}}^{(q)}}
\frac{A(\overline{B_q^0}\to \overline{f})}{A(B_q^0\to \overline{f})}.
\end{equation}
In the time-dependent rates (\ref{ratebf})-(\ref{ratebbfb}), the
time-independent transition rates $\tilde \Gamma$ and 
$\tilde{\overline{\Gamma}}$ correspond to the ``unevolved'' decay 
amplitudes $A(B^0_q\to f)$ and $A(B^0_q\to\overline{f})$, respectively, 
and can be calculated by performing
the usual phase space integrations. The functions $g_{\pm}^{(q)}(t)$
are related to $f_{\pm}^{(q)}(t)$. However, whereas the latter functions
depend through $\alpha_q$ on the quantity $n'$ parametrizing the
sign of the square root appearing in (\ref{e75}), $g_{\pm}^{(q)}(t)$ 
and the rates (\ref{ratebf})-(\ref{ratebbfb}) do not depend on that 
parameter. The $n'$-dependence is cancelled by introducing the 
{\it positive} mass difference
\begin{equation}
\Delta M_q\equiv M_H^{(q)}-M_L^{(q)}=2\left|M_{12}^{(q)}\right|>0
\end{equation}
of the $B_q$ mass eigenstates, where $H$ and $L$ refer to ``heavy''
and ``light'', respectively. The quantities $\Gamma_H^{(q)}$ and 
$\Gamma_L^{(q)}$ denote the corresponding decay widths. Their difference 
can be expressed as
\begin{equation}\label{deltagamma}
\Delta\Gamma_q\equiv\Gamma_H^{(q)}-\Gamma_L^{(q)}=\frac{4\mbox{\,Re}
\left[M_{12}^{(q)}\Gamma_{12}^{(q)\ast}\right]}{\Delta M_q},
\end{equation}
while the average decay width of the $B_q$ mass eigenstates is given by
\begin{equation}
\Gamma_q\equiv\frac{\Gamma^{(q)}_H+\Gamma^{(q)}_L}{2}=\Gamma^{(q)}_0.
\end{equation}
Whereas both the mixing phase $\Theta_{M_{12}}^{(q)}$ and
the amplitude ratios appearing in (\ref{e731}) 
depend on the chosen CP phase convention parametrized through 
$\phi_{\mbox{{\scriptsize CP}}}(B_q)$, the quantities $\xi_f^{(q)}$
and $\xi_{\overline{f}}^{(q)}$ are {\it convention independent observables}.
We shall see the cancellation of $\phi_{\mbox{{\scriptsize CP}}}(B_q)$ 
explicitly in a moment.

The $B^0_q-\overline{B^0_q}$ mixing phase $\Theta_{M_{12}}^{(q)}$
appearing in the equations given above is essential for the later 
discussion of ``mixing-induced'' CP violation. As can be read off 
from the expression (\ref{e711}) for the off-diagonal element
$M_{12}^{(q)}$ of the mass matrix, $\Theta_{M_{12}}^{(q)}$ is related 
to complex phases of CKM matrix elements through
\begin{equation}\label{e717}
\Theta_{M_{12}}^{(q)}=\pi+2\,\mbox{arg}\left(V_{tq}^\ast V_{tb}\right)-
\phi_{\mbox{{\scriptsize CP}}}(B_q).
\end{equation}
In (\ref{e711}), perturbative QCD corrections to $B^0_q-\overline{B^0_q}$ 
mixing have been neglected. Since these corrections, which are presently
known up to NLO\cite{bujw}, show up as a factor
$\eta_{\mbox{{\scriptsize QCD}}}\approx0.55$ multiplying the r.h.s.\ of
(\ref{e711}), they do not affect the mixing phase $\Theta_{M_{12}}^{(q)}$ 
and have therefore no significance for mixing-induced CP violation. 

A measure of the strength of the $B^0_q-\overline{B^0_q}$ oscillations 
is provided by the ``mixing parameter''
\begin{equation}\label{e718}
x_q\equiv\frac{\Delta M_q}{\Gamma_q}.
\end{equation}
The present ranges for $x_d$ and $x_s$ can be summarized as 
\begin{equation}\label{e719}
x_q=\left\{
\begin{array}{ll}
0.72\pm0.03 & \mbox{\quad for $q=d$}\\
\quad{\cal O}(20) & \mbox{\quad for $q=s$,}
\end{array}
\right.
\end{equation}
where $1/\Gamma_d=1.55\,\mbox{ps}$ has been used to evaluate $x_d$ from the
present experimental values for $\Delta M_d$ summarized recently in
Ref.\cite{gibbons}. So far the mixing parameter $x_s$ has not been 
measured directly and only an experimental lower bound 
$x_s\cdot\Gamma_s=\Delta M_s>9.2/\mbox{ps}$, which is based in 
particular on recent ALEPH and DELPHI results\cite{gibbons}, is available. 
Within the SM one expects\cite{al} $x_s$ to be of ${\cal O}(20)$. That
information can be obtained with the help of the relation
\begin{equation}\label{e720}
x_s=x_d\,\frac{1}{|V_{us}|^2R_t^2}\,\frac{1}{R_{ds}}\quad\mbox{with}\quad
R_{ds}=\frac{\Gamma_s}{\Gamma_d}\cdot\frac{M_{B_d}}{M_{B_s}}
\left[\frac{f_{B_d}\sqrt{B_{B_d}}}{f_{B_s}\sqrt{B_{B_s}}}\right]^2,
\end{equation}
where $R_{ds}$ describes $SU(3)$ flavor-breaking effects. 
Note that $R_{ds}=1$ in the strict $SU(3)$ limit.

The mixing parameters listed in (\ref{e719}) have interesting 
phenomenological consequences for the width differences 
$\Delta\Gamma_{d,s}$ defined by (\ref{deltagamma}). Using this 
expression we obtain
\begin{equation}\label{dgam}
\frac{\Delta\Gamma_q}{\Gamma_q}\approx-\frac{3\pi}{2S(x_t)}\frac{m_b^2}
{M_W^2}\,x_q.
\end{equation}
Consequently $\Delta\Gamma_q$ is negative so that the decay width
$\Gamma_H^{(q)}$ of the ``heavy'' mixing eigenstate is smaller than that 
of the ``light'' eigenstate. Since the numerical factor in 
(\ref{dgam}) multiplying the mixing parameter $x_q$ is ${\cal O}(10^{-2})$, 
the width difference $\Delta\Gamma_d$ is very small within the SM. 
On the other hand, the expected large value of $x_s$ implies a sizable 
$\Delta\Gamma_s$ which may be as large as ${\cal O}(20\%)$. The dynamical
origin of this width difference is related to CKM favored $\bar b\to\bar 
cc\bar s$ quark-level transitions into final states that are common both 
to $B^0_s$ and $\overline{B^0_s}$ mesons. Theoretical analyses of 
$\Delta\Gamma_s/\Gamma_s$ indicate that it may indeed be as large as
${\cal O}(20\%)$. These studies are based on box diagram 
calculations\cite{boxes}, on a complementary approach\cite{excl} 
where one sums over many exclusive $\bar b\to\bar cc\bar s$ modes, and
on the Heavy Quark Expansion yielding the most recent result\cite{HQE}
$\Delta\Gamma_s/\Gamma_s=0.16^{+0.11}_{-0.09}$. This width difference 
can be determined experimentally e.g.\ from angular correlations in $B_s\to 
J/\psi\,\phi$ decays\cite{ddlr}. One expects $10^3-10^4$ reconstructed 
$B_s\to J/\psi\,\phi$ events both at Tevatron Run II and at HERA-B which
may allow a precise measurement of $\Delta\Gamma_s$. As was pointed out
by Dunietz\cite{dunietz}, $\Delta\Gamma_s$ may lead to interesting 
CP-violating effects in {\it untagged} data samples of time-evolved 
$B_s$ decays where one does not distinguish between initially 
present $B^0_s$ and $\overline{B^0_s}$ mesons. Before we
shall turn to detailed discussions of CP-violating asymmetries in the
$B_d$ system and of the $B_s$ system in light of $\Delta\Gamma_s$, let us 
focus on $B_q$ decays ($q\in\{d,s\}$) into final CP eigenstates first. For
an analysis of transitions into non CP eigenstates the reader is referred to
Ref.\cite{non-CP}.

\runninghead{CP Violation in Non-leptonic $B$-Meson Decays}  
{$B_q$ Decays into CP Eigenstates}
\subsection{$B_q$ Decays into CP Eigenstates}
\noindent
A very promising special case in respect of extracting CKM phases from
CP-violating effects in neutral $B_q$ decays are transitions into final
states $|f\rangle$ that are eigenstates of the CP operator and hence
satisfy 
\begin{equation}\label{e733a}
({\cal CP})|f\rangle=\pm|f\rangle. 
\end{equation}
Consequently we have $\xi_f^{(q)}=\xi_{\overline{f}}^{(q)}$ in that case
(see (\ref{e731})) and have to deal only with a single observable 
$\xi_f^{(q)}$ containing essentially all the information that is needed
to evaluate the time-dependent decay rates (\ref{ratebf})-(\ref{ratebbfb}).
Decays into final states that are not eigenstates of the CP operator
play an important role in the case of the $B_s$ system to extract the UT
angle $\gamma$ and are discussed in 3.4.5. 

\subsubsection{Calculation of $\xi_f^{(q)}$}
\noindent
Whereas the $B^0_q-\overline{B^0_q}$ mixing phase $\Theta_{M_{12}}^{(q)}$
entering the expression (\ref{e731}) for $\xi_f^{(q)}$ is simply given 
as a function of complex phases of certain CKM matrix elements 
(see (\ref{e717})), 
the amplitude ratio $A(\overline{B^0_q}\to f)/A(B^0_q\to f)$ requires the 
calculation of hadronic matrix elements which are poorly known at present. 
In order to investigate this amplitude ratio, we shall employ the low
energy effective Hamiltonian for $|\Delta B|=1$, $\Delta C=\Delta U=0$
transitions discussed in Section~2. Using (\ref{LEham}) we get
\begin{eqnarray}
\lefteqn{A\left(\overline{B^0_q}\to f\right)=\Bigl\langle f\Bigl\vert
{\cal H}_{\mbox{{\scriptsize eff}}}(\Delta B=-1)\Bigr\vert\overline{B^0_q}
\Bigr\rangle}\label{e735a}\\
&&=\Biggl\langle f\left|
\frac{G_{\mbox{{\scriptsize F}}}}{\sqrt{2}}\left[
\sum\limits_{j=u,c}V_{jr}^\ast V_{jb}\left\{\sum\limits_{k=1}^2
Q_{k}^{jr}(\mu)C_{k}(\mu)
+\sum\limits_{k=3}^{10}Q_{k}^r(\mu)C_{k}(\mu)\right\}\right]\right|
\overline{B^0_q}\Biggr\rangle,\nonumber
\end{eqnarray}
where the flavor label $r\in\{d,s\}$ distinguishes -- as in the whole
subsection -- between $b\to d$ and $b\to s$ transitions. 
On the other hand, the transition amplitude $A\left(B^0_q\to f\right)$ 
is given by
\begin{eqnarray}
\lefteqn{A\left(B^0_q\to f\right)=\left\langle f\left|
{\cal H}_{\mbox{{\scriptsize 
eff}}}(\Delta B=-1)^\dagger\right|B^0_q\right\rangle}\label{e736a}\\
&&=\Biggl\langle f\left|\frac{G_{\mbox{{\scriptsize F}}}}{\sqrt{2}}
\left[\sum\limits_{j=u,c}V_{jr}V_{jb}^\ast \left\{\sum\limits_{k=1}^2
Q_{k}^{jr\dagger}(\mu)C_{k}(\mu)+\sum\limits_{k=3}^{10}
Q_k^{r\dagger}(\mu)C_{k}(\mu)\right\}\right]\right|B^0_q
\Biggr\rangle.\nonumber
\end{eqnarray}
Performing appropriate CP transformations in this equation, i.e.\
inserting the operator $({\cal CP})^\dagger({\cal CP})=\hat 1$ both
after the bra $\langle f|$ and in front of the ket $|B^0_q\rangle$,
yields
\begin{eqnarray}
\lefteqn{A\left(B^0_q\to f\right)=\pm e^{i\phi_{\mbox{{\scriptsize CP}}}
(B_q)}}\label{e737}\\
&&\times\Biggl\langle f\left|
\frac{G_{\mbox{{\scriptsize F}}}}{\sqrt{2}}\left[\sum\limits_{j=u,c}
V_{jr}V_{jb}^\ast\left\{\sum\limits_{k=1}^2
Q_{k}^{jr}(\mu)C_{k}(\mu)+\sum\limits_{k=3}^{10}
Q_{k}^r(\mu)C_{k}(\mu)\right\}\right]\right|\overline{B^0_q}
\Biggr\rangle,\nonumber
\end{eqnarray}
where we have applied the relation
\begin{equation}\label{e738}
({\cal CP})Q_k^{jr\dagger}({\cal CP})^\dagger=Q_k^{jr}
\end{equation}
and have furthermore taken into account (\ref{e710}) and
(\ref{e733a}). Consequently we obtain
\begin{equation}\label{e739}
\frac{A(\overline{B^0_q}\to f)}{A(B^0_q\to f)}=\pm\,
e^{-i\phi_{\mbox{{\scriptsize CP}}}(B_q)}\,\frac{\sum\limits_{j=u,c} 
v_j^{(r)}\Bigl\langle f\Bigl|{\cal Q}^{jr}\Bigr|\overline{B^0_q}
\Bigr\rangle}{\sum
\limits_{j=u,c}v_j^{(r)\ast}\Bigl\langle f\Bigl|{\cal Q}^{jr}\Bigr|
\overline{B^0_q}\Bigr\rangle},
\end{equation}
where $v_j^{(r)}\equiv V_{jr}^\ast V_{jb}$ and the operators 
${\cal Q}^{jr}$ are defined by 
\begin{equation}\label{e740}
{\cal Q}^{jr}\equiv
\sum\limits_{k=1}^2Q_k^{jr}C_k(\mu)+\sum\limits_{k=3}^{10}Q_k^rC_k(\mu).
\end{equation}
Inserting (\ref{e717}) and (\ref{e739}) into the expression
(\ref{e731}) for $\xi_f^{(q)}$, we observe explicitly that the convention
dependent phases $\phi_{\mbox{{\scriptsize CP}}}(B_q)$ appearing in the 
former two equations cancel each other and arrive at 
the {\it convention independent} result
\begin{equation}\label{e741}
\xi_f^{(q)}=\mp\,e^{-i\phi_{\mbox{{\scriptsize M}}}^{(q)}}
\frac{\sum\limits_{j=u,c} v_j^{(r)}
\Bigl\langle f\Bigl|{\cal Q}^{jr}\Bigr|\overline{B^0_q}\Bigr\rangle}{\sum
\limits_{j=u,c}v_j^{(r)\ast}\Bigl\langle f\Bigl|{\cal Q}^{jr}\Bigr|
\overline{B^0_q}\Bigr\rangle}.
\end{equation}
Here the phase $\phi_{\mbox{{\scriptsize M}}}^{(q)}\equiv2\,
\mbox{arg}(V_{tq}^\ast V_{tb})$ arises from the 
$B^0_q-\overline{B^0_q}$ mixing phase $\Theta_{M_{12}}^{(q)}$. 
Applying the modified Wolfenstein 
parametrization (\ref{wolf2}), $\phi_{\mbox{{\scriptsize M}}}^{(q)}$ can be 
related to angles of the UT as follows:
\begin{equation}\label{e742}
\phi_{\mbox{{\scriptsize M}}}^{(q)}=\left\{\begin{array}{cl}
2\beta & \mbox{for $q=d$}\\
0 & \mbox{for $q=s$.}
\end{array}\right.
\end{equation}
Consequently a non-trivial mixing phase arises only in the $B_d$ system.

In general the observable $\xi_f^{(q)}$ suffers from large hadronic 
uncertainties that are introduced through the hadronic matrix elements 
appearing in (\ref{e741}). However, there is a very important special case 
where these uncertainties cancel and theoretical clean predictions of 
$\xi_f^{(q)}$ are possible.

\subsubsection{Dominance of a Single CKM Amplitude}
\noindent
If the transition matrix elements appearing in (\ref{e741}) are
dominated by a single CKM amplitude, the observable $\xi_f^{(q)}$ 
takes the very simple form
\begin{equation}\label{e743}
\xi_f^{(q)}=\mp\exp\left[-i\left\{\phi_{\mbox{{\scriptsize M}}}^{(q)}-
\phi_{\mbox{{\scriptsize D}}}^{(f)}\right\}\right],
\end{equation}
where the characteristic ``decay'' phase 
$\phi_{\mbox{{\scriptsize D}}}^{(f)}$ can be expressed in terms of angles 
of the UT as follows:
\begin{equation}\label{e746}
\phi_{\mbox{{\scriptsize D}}}^{(f)}=\left\{\begin{array}{cl}
-2\gamma & \mbox{for dominant $\bar b\to\bar uu\bar r$ CKM amplitudes 
in $B_q^0\to f$}\\
0 & \mbox{for dominant $\bar b\to\bar cc\bar r$\,\, CKM amplitudes 
in $B_q^0\to f$.}
\end{array}\right.
\end{equation}
The validity of dominance of a single CKM amplitude 
and important phenomenological applications of (\ref{e743})
will be discussed in the following subsections. 

\runninghead{CP Violation in Non-leptonic $B$-Meson Decays}  
{The $B_d$ System}
\subsection{The $B_d$ System}
\noindent
In contrast to the $B_s$ system, the width difference is negligibly small
in the $B_d$ system. Consequently the expressions for the decay rates 
(\ref{ratebf})-(\ref{ratebbfb}) simplify considerably in that case. 

\subsubsection{CP Asymmetries in $B_d$ Decays}
\noindent
Restricting ourselves, as in the previous subsection, to decays into 
final CP eigenstates $|f\rangle$ satisfying (\ref{e733a}), we obtain 
the following expressions for the time-dependent and time-integrated 
CP asymmetries:
\begin{eqnarray}
\lefteqn{a_{\mbox{{\scriptsize CP}}}(B_d\to f;t)\equiv\frac{\Gamma(B_d^0(t)
\to f)-\Gamma(\overline{B_d^0}(t)\to f)}{\Gamma(B_d^0(t)\to f)+
\Gamma(\overline{B_d^0}(t)\to f)}}\nonumber\\
&&={\cal A}_{\mbox{{\scriptsize CP}}}^{\mbox{{\scriptsize dir}}}(B_d\to f)
\cos(\Delta M_d t)+{\cal A}_{\mbox{{\scriptsize CP}}}^{\mbox{{\scriptsize 
mix-ind}}}(B_d\to f)\sin(\Delta M_d t)\label{acptimedep}
\end{eqnarray}
\begin{eqnarray}
\lefteqn{a_{\mbox{{\scriptsize CP}}}(B_d\to f)\equiv
\frac{\int\limits_0^\infty
\mbox{d}t\left[\Gamma(B_d^0(t)\to f)-
\Gamma(\overline{B_d^0}(t)\to f)\right]}
{\int\limits_0^\infty \mbox{d}t \left[\Gamma(B_d^0(t)\to f)+
\Gamma(\overline{B_d^0}(t)\to f)\right]}}\nonumber\\
&&=\frac{1}{1+x_d^2}\left[{\cal A}_{\mbox{{\scriptsize 
CP}}}^{\mbox{{\scriptsize dir}}}(B_d\to f)+x_d\,
{\cal A}_{\mbox{{\scriptsize CP}}}^{\mbox{{\scriptsize mix-ind}}}
(B_d\to f)\right],\label{acptimeint}
\end{eqnarray}
where the {\it direct} CP-violating contributions 
\begin{equation}\label{acpdir}
{\cal A}_{\mbox{{\scriptsize CP}}}^{\mbox{{\scriptsize dir}}}
(B_d\to f)\equiv
\frac{1-\left|\xi_f^{(d)}\right|^2}{1+\left|\xi_f^{(d)}\right|^2}
\end{equation}
have been separated from the {\it mixing-induced} CP-violating contributions
\begin{equation}\label{acpmi}
{\cal A}_{\mbox{{\scriptsize CP}}}^{\mbox{{\scriptsize mix-ind}}}(B_d\to 
f)\equiv\frac{2\,\mbox{Im}\,\xi_f^{(d)}}{1+\left|\xi_f^{(d)}\right|^2}.
\end{equation}
Whereas the former observables describe CP violation arising directly in
the corresponding decay amplitudes, the latter ones are due to interference
between $B^0_d-\overline{B^0_d}$ mixing- and decay-processes. Needless to
say, the expressions (\ref{acptimedep}) and (\ref{acptimeint}) have to be 
modified appropriately for the $B_s$ system because of $\Delta\Gamma_s/
\Gamma_s={\cal O}(20\%)$. In the case of the time-dependent CP asymmetry 
(\ref{acptimedep}) these effects start to become important for 
$t\,\mbox{{\scriptsize $\stackrel{>}{\sim}$}}\,2/\Delta\Gamma_s$. 

\subsubsection{CP Violation in $B_d\to J/\psi\, K_{\mbox{{\scriptsize S}}}$:
the ``Gold-plated'' Way to Extract $\beta$}
\noindent
The channel $B_d\to J/\psi\, K_{\mbox{{\scriptsize S}}}$ is a transition into
a CP eigenstate with eigenvalue $-1$ and originates from a $\bar b\to\bar 
cc\bar s$ quark-level decay\cite{csbs}. Consequently the corresponding 
observable $\xi^{(d)}_{\psi K_{\mbox{{\scriptsize S}}}}$ can be 
expressed as
\begin{equation}\label{xipsiks}
\xi^{(d)}_{\psi K_{\mbox{{\scriptsize S}}}}=+e^{-2i\beta}\left[
\frac{v_u^{(s)}A^{ut'}_{\mbox{{\scriptsize pen}}}+v_c^{(s)}\left(
A_{\mbox{{\scriptsize cc}}}^{c'}+A^{ct'}_{\mbox{{\scriptsize pen}}}\right)}
{v_u^{(s)\ast}A^{ut'}_{\mbox{{\scriptsize pen}}}+v_c^{(s)\ast}\left(
A_{\mbox{{\scriptsize cc}}}^{c'}+A^{ct'}_{\mbox{{\scriptsize pen}}}\right)}
\right],
\end{equation}
where $A_{\mbox{{\scriptsize cc}}}^{c'}$ denotes the $Q_{1/2}^{cs}$ 
current-current operator amplitude and $A^{ut'}_{\mbox{{\scriptsize pen}}}$ 
$(A^{ct'}_{\mbox{{\scriptsize pen}}})$ corresponds to contributions of
the penguin-type with up- and top-quarks (charm- and top-quarks) running
as virtual particles in the loops. Note that within this notation 
penguin-like matrix elements of the $Q_{1/2}^{cs}$ operators like those
depicted in Fig.~\ref{pen-me} are included by definition in the 
$A^{ct'}_{\mbox{{\scriptsize pen}}}$ amplitude, whereas those of 
$Q_{1/2}^{us}$ show up in $A^{ut'}_{\mbox{{\scriptsize pen}}}$.
The primes in (\ref{xipsiks}) 
have been introduced to remind us that we are dealing with a 
$\bar b\to\bar s$ mode. Using the modified Wolfenstein parametrization 
(\ref{wolf2}), the relevant CKM factors take the form 
\begin{equation}
v_u^{(s)}=A\lambda^4R_b\,e^{-i\gamma},\quad
v_c^{(s)}=A\lambda^2\left(1-\lambda^2/2\right)
\end{equation}
and imply that the $A^{ut'}_{\mbox{{\scriptsize pen}}}$ contribution is 
highly CKM suppressed with respect to the part containing the current-current 
amplitude. The suppression factor is given by  
\begin{equation}
\left|v_u^{(s)}/v_c^{(s)}\right|=\lambda^2R_b\approx0.02.
\end{equation}
An additional suppression arises from the fact that 
$A^{ut'}_{\mbox{{\scriptsize pen}}}$ is related to 
loop processes that are governed by Wilson coefficients\cite{bbl-rev} 
of ${\cal O}(10^{-2})$. Moreover the color-structure of 
$B_d\to J/\psi\, K_{\mbox{{\scriptsize S}}}$ leads to further suppression!
The point is that the $\bar c$- and $c$-quarks emerging from the gluons of 
the usual QCD penguin diagrams form a color-octet state and consequently
cannot build up the $J/\psi$ which is a $\bar cc$ color-singlet state. 
Therefore additional gluons are needed -- the corresponding 
contributions are very hard to estimate -- and EW penguins, where
the former color-argument does not hold, may be the most important penguin
contributions to $B_d\to J/\psi\, K_{\mbox{{\scriptsize S}}}$. 
The suppression of $v_u^{(s)}A_{\mbox{{\scriptsize pen}}}^{ut'}$
relative to $v_c^{(s)}(A_{\mbox{{\scriptsize cc}}}^{c'}+
A_{\mbox{{\scriptsize pen}}}^{ct'})$ is compensated slightly since 
the dominant $Q_{1/2}^{cs}$ current-current amplitude 
$A_{\mbox{{\scriptsize cc}}}^{c'}$ 
is color-suppressed by a phenomenological color-suppression 
factor\cite{BSW}$^-$\cite{a1a2-exp} $a_2\approx0.2$. However, since 
$v_u^{(s)}A_{\mbox{{\scriptsize pen}}}^{ut'}$ is 
suppressed by {\it three} sources (CKM-structure, 
loop effects, color-structure), we conclude that $\xi^{(d)}_{\psi 
K_{\mbox{{\scriptsize S}}}}$ is nevertheless given to an excellent 
approximation by
\begin{equation}
\xi^{(d)}_{\psi K_{\mbox{{\scriptsize S}}}}=e^{-2i\beta}
\left[\frac{v_c^{(s)}\left(A_{\mbox{{\scriptsize cc}}}^{c'}+
A^{ct'}_{\mbox{{\scriptsize pen}}}\right)}{v_c^{(s)\ast}\left(
A_{\mbox{{\scriptsize cc}}}^{c'}+A^{ct'}_{\mbox{{\scriptsize pen}}}
\right)}\right]=e^{-2i\beta}
\end{equation}
yielding 
\begin{equation}
{\cal A}_{\mbox{{\scriptsize CP}}}^{\mbox{{\scriptsize dir}}}
(B_d\to J/\psi\, K_{\mbox{{\scriptsize S}}})=0,\quad 
{\cal A}_{\mbox{{\scriptsize CP}}}^{\mbox{{\scriptsize mix-ind}}}
(B_d\to J/\psi\, K_{\mbox{{\scriptsize S}}})=-\sin(2\beta)\,.
\end{equation}
Consequently mixing-induced CP violation in $B_d\to J/\psi\, 
K_{\mbox{{\scriptsize S}}}$ measures $\sin(2\beta)$ to excellent accuracy.
Therefore that decay is usually referred to as the ``gold-plated'' mode
to determine the UT angle $\beta$. For other methods see e.g.\ 
Refs.\cite{non-CP,beta-refs}.

\subsubsection{CP Violation in $B_d\to \pi^+\pi^-$ and Extractions of
$\alpha$}
\noindent
In the case of $B_d\to\pi^+\pi^-$ we have to deal with the decay of a 
$B_d$-meson into a final CP eigenstate with eigenvalue $+1$ that is caused 
by the quark-level process $\bar b\to\bar uu\bar d$. Therefore we may write
\begin{equation}\label{xipipi}
\xi^{(d)}_{\pi^+\pi^-}=-e^{-2i\beta}\left[
\frac{v_u^{(d)}\left(A_{\mbox{{\scriptsize cc}}}^{u}+
A^{ut}_{\mbox{{\scriptsize pen}}}\right)+v_c^{(d)}
A^{ct}_{\mbox{{\scriptsize pen}}}}
{v_u^{(d)\ast}\left(A_{\mbox{{\scriptsize cc}}}^{u}+
A^{ut}_{\mbox{{\scriptsize pen}}}\right)+v_c^{(d)\ast}
A^{ct}_{\mbox{{\scriptsize pen}}}}
\right],
\end{equation}
where the notation of decay amplitudes is as in the previous discussion 
of $B_d\to J/\psi\, K_{\mbox{{\scriptsize S}}}$. Using again (\ref{wolf2}),
the CKM factors are given by 
\begin{equation}\label{CKMbd}
v_u^{(d)}=A\lambda^3R_b\,e^{-i\gamma},\quad
v_c^{(d)}=-A\lambda^3.
\end{equation}
The CKM structure of (\ref{xipipi}) is very different form 
$\xi^{(d)}_{\psi K_{\mbox{{\scriptsize S}}}}$. In particular the pieces 
containing the dominant $Q_{1/2}^{ud}$ current-current contributions 
$A_{\mbox{{\scriptsize cc}}}^{u}$ are CKM suppressed with respect to 
the penguin contributions $A^{ct}_{\mbox{{\scriptsize pen}}}$ by 
\begin{equation}\label{bdckm}
\left|v_u^{(d)}/v_c^{(d)}\right| = R_b\approx0.36\,.
\end{equation}
In contrast to $B_d\to J/\psi\, K_{\mbox{{\scriptsize S}}}$, in the 
$B_d\to\pi^+\pi^-$ case the penguin amplitudes are only suppressed by the 
corresponding Wilson coefficients ${\cal O}(10^{-2})$ and not additionally 
by the color-structure of that decay. Taking into account that the 
current-current amplitude $A_{\mbox{{\scriptsize cc}}}^{u}$ is color-allowed 
and using both (\ref{bdckm}) and characteristic values of the Wilson 
coefficient functions, one obtains
\begin{equation}
\left|\frac{v_c^{(d)}A^{ct}_{\mbox{{\scriptsize pen}}}}
{v_u^{(d)}\left(A_{\mbox{{\scriptsize cc}}}^{u}+
A^{ut}_{\mbox{{\scriptsize pen}}}\right)}\right|={\cal O}(0.15)
\end{equation}
and concludes that
\begin{equation}
\xi^{(d)}_{\pi^+\pi^-}\approx-e^{-2i\beta}\left[\frac{v_u^{(d)}\left(
A_{\mbox{{\scriptsize cc}}}^{u}+A^{ut}_{\mbox{{\scriptsize pen}}}\right)}
{v_u^{(d)\ast}\left(A_{\mbox{{\scriptsize cc}}}^{u}+
A^{ut}_{\mbox{{\scriptsize pen}}}\right)}\right]=-e^{2i\alpha}
\end{equation}
may be a reasonable approximation to obtain an estimate for the UT angle
$\alpha$ from the mixing-induced CP-violating observable 
\begin{equation}\label{BdpipiCP}
{\cal A}_{\mbox{{\scriptsize CP}}}^{\mbox{{\scriptsize mix-ind}}}
(B_d\to\pi^+\pi^-)\approx-\sin(2\alpha).
\end{equation}
Note that a measurement of ${\cal A}_{\mbox{{\scriptsize 
CP}}}^{\mbox{{\scriptsize dir}}}(B_d\to\pi^+\pi^-)\not=0$ would signal the
presence of penguins. We shall come back to this feature later.

The hadronic uncertainties affecting the extraction of $\alpha$ from CP 
violation in $B_d\to\pi^+\pi^-$ were analyzed by many authors in 
the previous literature. A selection of papers is given in 
Refs.\cite{gro-pen,uncert-alpha}. As was pointed out by Gronau and 
London\cite{gl}, the uncertainties related to QCD penguins\cite{pens} 
can be eliminated with the help of isospin relations
involving in addition to $B_d\to\pi^+\pi^-$ also the modes 
$B_d\to\pi^0\pi^0$ 
and $B^{\pm}\to\pi^{\pm}\pi^0$. The isospin relations among the 
corresponding decay amplitudes are given by
\begin{eqnarray}
A(B^0_d\to\pi^+\pi^-)+\sqrt{2}A(B^0_d\to\pi^0\pi^0)&=&\sqrt{2}A(B^+\to
\pi^+\pi^0)\label{isospin1}\\
A(\overline{B^0_d}\to\pi^+\pi^-)+\sqrt{2}A(\overline{B^0_d}\to\pi^0
\pi^0)&=&\sqrt{2}A(B^-\to\pi^-\pi^0)\label{isospin2}
\end{eqnarray}
and can be represented as two triangles in the complex plane that allow
the extraction of a value of $\alpha$ that does not suffer from QCD
penguin uncertainties. It is, however, not possible to control also the
EW penguin uncertainties using that isospin approach. The point is 
up- and down-quarks are coupled differently in EW penguin diagrams
because of their different electrical charges (see (\ref{ew-def})).
Hence one has also to think about the role of these contributions. 
We shall come back to that issue in Section~5, where a more detailed 
discussion of the GL method\cite{gl} in light of EW penguin effects
will be given. 

An experimental problem of the GL method is related to the fact that it
requires a measurement of BR$(B_d\to\pi^0\pi^0)$ which may be smaller
than ${\cal O}(10^{-6})$ because of color-suppression effects\cite{kp}.
Therefore, despite of its attractiveness, that approach may be quite
difficult from an experimental point of view and it is important to have 
alternatives available to determine $\alpha$. Needless to say, that is 
also required in order to over-constrain the UT angle $\alpha$ as much 
as possible at future $B$-physics experiments. Fortunately such methods 
are already on the market. For example, Snyder and Quinn\cite{sq} suggested 
to use $B\to\rho\,\pi$ modes to extract $\alpha$. Another method was proposed 
by Buras and myself\,\cite{PAPII}.
It requires a simultaneous measurement of ${\cal A}_{\mbox{{\scriptsize 
CP}}}^{\mbox{{\scriptsize mix-ind}}}(B_d\to\pi^+\pi^-)$ and ${\cal 
A}_{\mbox{{\scriptsize CP}}}^{\mbox{{\scriptsize mix-ind}}}(B_d\to K^0
\overline{K^0})$ and determines $\alpha$ with the help of a geometrical
triangle construction using the $SU(3)$ flavor symmetry of strong 
interactions. The accuracy of that approach is limited by $SU(3)$-breaking 
corrections which cannot be estimated reliably at present. 
Interestingly the penguin-induced decay $B_d\to K^0
\overline{K^0}$ may exhibit CP asymmetries as large as ${\cal O}(30\%)$ 
within the SM\cite{rf-k0k0.bar}. This feature is due to interference 
between QCD penguins with internal up- and charm-quark exchanges\cite{bf1}. 
In the absence of these contributions, the CP-violating asymmetries of 
$B_d\to K^0\overline{K^0}$ would vanish and ``New Physics'' would be 
required (see e.g.\ Ref.\cite{mw}) to induce CP violation in that decay. 

Before discussing other methods to deal with the penguin uncertainties 
affecting the extraction of $\alpha$ from the CP-violating observables of 
$B_d\to\pi^+\pi^-$, let us next have a closer look at the above
mentioned QCD penguins with up- and charm-quarks running as virtual 
particles in the loops.

\subsubsection{Penguin Zoology}
\noindent
The general structure of a generic $\bar b\to\bar q$ ($q\in\{d,s\}$) penguin
amplitude is given by
\begin{equation}\label{pen-amp}
P^{(q)}=V_{uq}V_{ub}^\ast \,P_u^{(q)}+V_{cq}V_{cb}^\ast \,P_c^{(q)}+
V_{tq}V_{tb}^\ast \,P_t^{(q)},
\end{equation} 
where $P_u^{(q)}$, $P_c^{(q)}$ and $P_t^{(q)}$ are the amplitudes of penguin 
processes with internal up-, charm- and top-quark exchanges, respectively, 
omitting CKM factors. The penguin amplitudes introduced in (\ref{xipsiks})
and (\ref{xipipi}) are related to these quantities through
\begin{equation}
\begin{array}{rclcrcl}
A^{ut}_{\mbox{{\scriptsize pen}}}&=&P_u^{(d)}-P_t^{(d)},&\quad&
A^{ct}_{\mbox{{\scriptsize pen}}}&=&P_c^{(d)}-P_t^{(d)}\\
A^{ut'}_{\mbox{{\scriptsize pen}}}&=&P_u^{(s)}-P_t^{(s)},&\quad&
A^{ct'}_{\mbox{{\scriptsize pen}}}&=&P_c^{(s)}-P_t^{(s)}.
\end{array}
\end{equation}
Using unitarity of the CKM matrix yields
\begin{equation}\label{pen-unit}
P^{(q)}=V_{cq} V_{cb}^\ast\left[P_c^{(q)}-P_u^{(q)}\right]+
V_{tq} V_{tb}^\ast\left[P_t^{(q)}-P_u^{(q)}\right],
\end{equation}
where the relevant CKM factors can be expressed with the help of 
the Wolfenstein parametrization as follows:
\begin{equation}\label{CKMd}
V_{cd}V_{cb}^\ast=-\lambda|V_{cb}|\left(1+{\cal O}\left(\lambda^4\right)
\right),\quad V_{td}V_{tb}^\ast=|V_{td}|e^{-i\beta},
\end{equation}
\begin{equation}\label{CKMs}
V_{cs}V_{cb}^\ast=|V_{cb}|\left(1+{\cal O}\left(\lambda^2\right)\right),\quad
V_{ts}V_{tb}^\ast=-|V_{cb}|\left(1+{\cal O}\left(\lambda^2\right)\right).\quad
\end{equation}
The estimate of the non-leading terms in $\lambda$ follows Ref.\cite{blo}.
Omitting these terms and combining (\ref{pen-unit}) with (\ref{CKMd}) 
and (\ref{CKMs}), the $\bar b\to\bar d$ and $\bar b\to\bar s$ penguin 
amplitudes take the form
\begin{equation}\label{bdpenamp}
P^{(d)}=\left[e^{-i\beta}-\frac{1}{R_t}\Delta P^{(d)}\right]
\left|V_{td}\right|\left|P_{tu}^{(d)}\right|
e^{i\delta_{tu}^{(d)}}
\end{equation}
\begin{equation}\label{bspenamp}
P^{(s)}=\left[1-\Delta P^{(s)}\right]e^{-i\pi}\left|V_{cb}\right|
\left|P_{tu}^{(s)}\right|e^{i\delta_{tu}^{(s)}},
\end{equation}
where the notation
\begin{equation}
P_{q_1q_2}^{(q)}\equiv P_{q_1}^{(q)}-P_{q_2}^{(q)}
\end{equation}
has been introduced and
\begin{equation}\label{DelP}
\Delta P^{(q)}\equiv\frac{P^{(q)}_{cu}}{P^{(q)}_{tu}}
\end{equation}
describes the contributions of ``subdominant'' penguins with up- and 
charm-quarks running as virtual particles in the loops. In the limit 
of degenerate up- and charm-quark masses, $\Delta P^{(q)}$ would vanish 
because of the GIM mechanism\cite{gim}. However, since $m_u\approx4.5$~MeV, 
whereas $m_c\approx1.4$~GeV, this GIM cancellation is incomplete and in 
principle sizable effects arising from $\Delta P^{(q)}$ could be expected.

\begin{figure}[t]
\centerline{
\rotate[r]{
\epsfysize=7truecm
\epsffile{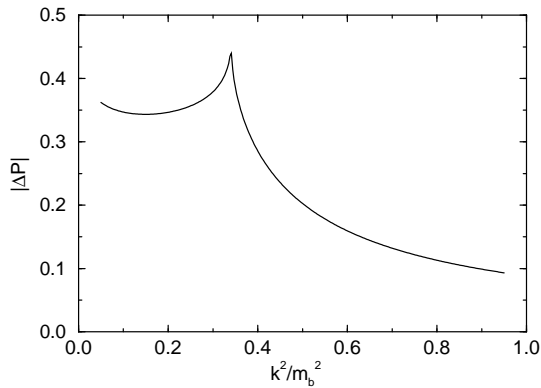}
}}
\caption{The dependence of $\left|\Delta P^{(q)}\right|$ on 
$k^2/m_b^2$.}\label{delp-fig}
\end{figure}

Usually it is assumed that the penguin amplitudes (\ref{bdpenamp}) 
and (\ref{bspenamp}) are dominated by internal top-quark exchanges, 
i.e.\ $\Delta P^{(q)}\approx0$. That is an excellent approximation for
EW penguin contributions which play an important role in certain
$B$ decays only because of the large top-quark mass (see Section~4). 
However, QCD penguins with internal 
up- and charm-quarks may become important as is indicated by 
model calculations at the perturbative quark-level\cite{bf1}. Neglecting 
the -- in that case -- tiny EW penguin contributions and using (\ref{e635}) 
with the Wilson coefficient functions (\ref{e636}) to simplify the 
following discussion, a straightforward calculation yields\cite{bf1}
\begin{equation}\label{DelP-approx}
\Delta P^{(q)}\approx\frac{G(m_c,k,\mu)-G(m_u,k,\mu)}{E(x_t)+
\frac{2}{3}\ln\left(\frac{\mu^2}{M_W^2}\right)-G(m_u,k,\mu)}\,.
\end{equation}
Note that the $1^{\mbox{{\scriptsize st}}}$ sum in the penguin amplitude 
(\ref{e635}) corresponds to penguins with internal top-quarks, whereas the 
piece containing $[10/9-G(m_j,k,\mu)]$ describes the penguin contributions 
with internal $j$-quarks ($j\in\{u,c\}$). 

Within the approximation (\ref{DelP-approx}), which does not depend 
on the flavor label $q$, the strong phase of $\Delta P^{(q)}$ is 
generated exclusively through absorptive parts of time-like 
penguin diagrams with internal up- and charm-quarks 
following the pioneering approach of Bander, Silverman and Soni\cite{bss}. 
Whereas the $\mu$-dependence cancels exactly in (\ref{DelP-approx}), 
this estimate of $\Delta P^{(q)}$ depends strongly on the value of 
$k^2$ denoting the four-momentum of the gluon appearing in the QCD 
penguin diagram depicted in Fig.~\ref{pen-me}. This feature 
can be seen in Fig.~\ref{delp-fig}, where that
dependence is shown. Simple kinematical considerations at the 
quark-level imply that $k^2$ should lie within 
the ``physical'' range\cite{rf1,detr,gh}
\begin{equation}\label{phys-range}
\frac{1}{4}\,\mbox{{\scriptsize $\stackrel{<}{\sim}$}}\,\frac{k^2}{m_b^2}
\,\mbox{{\scriptsize $\stackrel{<}{\sim}$}}\,\frac{1}{2}.
\end{equation}
A detailed discussion of the $k^2$-dependence can be found in Ref.\cite{gh}.

Looking at Fig.~\ref{delp-fig}, we observe that $\Delta P^{(q)}$ may lead 
to sizable effects for such values of $k^2$. Moreover QCD penguin topologies 
with internal up- and charm-quarks contain -- as can be seen easily by 
drawing the corresponding Feynman diagrams -- also long-distance 
contributions, such as the rescattering process 
$B^0_d\to\{D^+D^-\}\to\pi^+\pi^-$ (see e.g.\ Ref.\cite{kamal}), 
which are very hard to estimate. Such long-distance contributions were 
discussed in the context of extracting $V_{td}$ from radiative $B$ decays
in Ref.\cite{abs} and are potentially very serious. 
Consequently it may not be justified to neglect
the $\Delta P^{(q)}$ terms in (\ref{bdpenamp}) and (\ref{bspenamp}).
An important difference arises, however, between these two amplitudes. While
the UT angle $\beta$ shows up in the $\bar b\to\bar d$ case, there is 
only a trivial CP-violating weak phase present in the $\bar b\to\bar s$ 
case. Consequently $\Delta P^{(s)}$ cannot change the general phase 
structure of the $\bar b\to\bar s$ penguin amplitude 
$P^{(s)}$. However, if one takes into account also QCD penguins with 
internal up- and charm-quarks, the $\bar b\to \bar d$ penguin amplitude 
$P^{(d)}$ is no longer related in a simple and ``clean'' way through
\begin{equation}\label{bdpenapprox}
P^{(d)}=e^{-i\beta}e^{i\delta_P^{(d)}}\left|P^{(d)}\right|
\end{equation}
to $\beta$, where $\delta_P^{(d)}$ is a CP-conserving strong phase. This 
feature may affect\cite{bf1} some of the strategies to extract CKM phases 
with the help of $SU(3)$ amplitude relations that will be discussed later 
in this review. 

An interesting consequence of (\ref{bspenamp}) is the relation 
$P^{(s)}=\overline{P^{(s)}}$ between the $\bar b\to\bar s$ QCD penguin
amplitude and its charge-conjugate implying that penguin-induced modes of 
this type, e.g.\ the decay $B_d\to\phi\, K_{\mbox{{\scriptsize S}}}$, 
should exhibit no direct CP violation. Applying the formalism developed in
Subsection~3.2, one finds that 
\begin{equation}\label{acpphiks}
{\cal A}_{\mbox{{\scriptsize CP}}}^{\mbox{{\scriptsize mix-ind}}}(B_d\to\phi\, 
K_{\mbox{{\scriptsize S}}})=-\sin(2\beta)
\end{equation}
measures the UT angle $\beta$. Within the SM, small direct CP violation --
model calculations (see e.g.\ Refs.\cite{rfewp1,kps,gh}) indicate 
asymmetries at the ${\cal O}(1\%)$ level -- may arise from the neglected 
${\cal O}(\lambda^2)$ terms in (\ref{CKMs}) which also
limit the theoretical accuracy of (\ref{acpphiks}). An experimental
comparison between the mixing-induced CP asymmetries of $B_d\to J/\psi\, 
K_{\mbox{{\scriptsize S}}}$ and $B_d\to\phi\,K_{\mbox{{\scriptsize S}}}$,
which should be equal to very good accuracy within the SM, would be 
extremely interesting since the latter decay is a ``rare'' FCNC process 
and may hence be very sensitive to physics beyond the SM.

\subsubsection{Another Look at $B_d\to\pi^+\pi^-$ and the Extraction of
$\alpha$}
\noindent
The ``penguin zoology'' discussed above led Mannel and myself to reanalyze 
the decay $B_d\to\pi^+\pi^-$ without assuming dominance of QCD penguins with 
internal top-quark exchanges\cite{fm1}. To this end it is useful to 
introduce 
\begin{equation}
T\equiv V_{ud}V_{ub}^\ast\, A_{\mbox{{\scriptsize cc}}}^{u}
\end{equation}
and to expand the CP-violating observables (\ref{acpdir}) and (\ref{acpmi}) 
corresponding to $B_d\to\pi^+\pi^-$ in powers of $\overline{P^{(d)}}/T$ and 
$P^{(d)}/T$, which should satisfy the estimate\cite{fm1}
\begin{equation}
\left|\frac{\overline{P^{(d)}}}{T}\right|
\approx\left|\frac{P^{(d)}}{T}\right|
\approx0.07-0.23,
\end{equation}
and to keep only the leading terms in that expansion:
\begin{eqnarray}
&& {\cal A}_{\mbox{{\scriptsize CP}}}^{\mbox{{\scriptsize dir}}}
(B_d\to\pi^+\pi^-) = 2 \lambda R_t\frac{|\tilde P|}{|T|}  
\sin \delta \sin \alpha + {\cal O}\left((P^{(d)}/T)^2\right)\label{cpasym}\\
&& {\cal A}_{\mbox{{\scriptsize CP}}}^{\mbox{{\scriptsize mix-ind}}}
(B_d\to\pi^+\pi^-) \nonumber\\
&&\qquad=  - \sin 2 \alpha - 2 \lambda R_t 
\frac{|\tilde P|}{|T|}\cos \delta\, \cos 2 
\alpha\,\sin \alpha+{\cal O}\left((P^{(d)}/T)^2
\right).\label{cpasym1}
\end{eqnarray}
Similar expressions were also derived by Gronau in Ref.\cite{gro-pen}.
However, it has {\it not} been assumed in (\ref{cpasym}) and (\ref{cpasym1}) 
that QCD penguins are dominated by internal top-quark exchanges and the
physical interpretation of the amplitude $\tilde P$ is quite different from
Ref.\cite{gro-pen}. This quantity is given by
\begin{equation}\label{Pprime}
\tilde P\equiv\left[1-\Delta P^{(d)}\right]|V_{cb}|\left|P_{tu}^{(d)}\right|
e^{i\delta_{tu}^{(d)}}\,,
\end{equation}
and $\delta$ appearing in (\ref{cpasym}) and (\ref{cpasym1}) is simply the 
CP-conserving strong phase of $\tilde P/T$.
If we compare (\ref{Pprime}) with (\ref{bdpenamp}) and (\ref{bspenamp}),
we observe that it is not equal to the amplitude $P^{(d)}$ -- as one would
expect na\"\i vely -- but that its phase structure corresponds exactly
to the $\bar b\to\bar s$ QCD penguin amplitude $e^{i\pi}P^{(s)}$. 

The two CP-violating observables (\ref{cpasym}) and (\ref{cpasym1}) 
depend on the three ``unknowns'' $\alpha$, $\delta$ and $|\tilde 
P|/|T|$ (strategies to extract the CKM factor $R_t$ are discussed in 
Ref.\cite{buras-ichep96} and $\lambda$ is the usual Wolfenstein parameter). 
Consequently an additional input is needed to determine $\alpha$ from 
(\ref{cpasym}) and (\ref{cpasym1}). Taking into account the discussion given 
in the previous paragraph, it is very natural to use the $SU(3)$ flavor 
symmetry of strong interactions to accomplish this task. In the strict 
$SU(3)$ limit one does not distinguish between down- and strange-quarks 
and $|\tilde P|$ corresponds simply to the magnitude of the decay amplitude 
of a penguin-induced $\bar b\to\bar s$ transition such as $B^+\to\pi^+ K^0$ 
with an expected branching ratio\cite{kp} of ${\cal O}(10^{-5})$. On the 
other hand, $|T|$ can be estimated from the rate of $B^+\to\pi^+\pi^0$ by 
neglecting color-suppressed current-current operator contributions. 
Following these lines one obtains
\begin{equation}\label{BR-rat}
\frac{|\tilde P|}{|T|}\approx\frac{f_\pi}{f_K}\sqrt{\frac{1}{2}
\frac{\mbox{BR}(B^+\to\pi^+K^0)}{\mbox{BR}(B^+\to\pi^+\pi^0)}},
\end{equation}
where $f_\pi$ and $f_K$ are the $\pi$- and $K$-meson decay constants,
respectively, taking into account factorizable $SU(3)$-breaking.
That relation allows the extraction both of $\alpha$ and $\delta$ from 
the measured CP-violating observables (\ref{cpasym}) and (\ref{cpasym1}). 
Problems of this approach arise if $\alpha$ is close to $45^\circ$ or 
$135^\circ$, where the expansion (\ref{cpasym1}) for 
${\cal A}_{\mbox{{\scriptsize CP}}}^{\mbox{{\scriptsize mix-ind}}}
(B_d\to\pi^+\pi^-)$ breaks down.
Assuming a total theoretical uncertainty of 30\% in the quantity 
$r\equiv2\lambda R_t|\tilde P|/|T|$ governing (\ref{cpasym}) and 
(\ref{cpasym1}), an uncertainty of $\pm\,3^\circ$ in the extracted 
value of $\alpha$ is expected if $\alpha$ is not too close to these 
singular points\cite{fm1}. For values of $\alpha$ far away from $45^\circ$ 
and $135^\circ$, one may even have an uncertainty of $\pm\,1^\circ$
as is indicated by the following example: Let us assume that the 
CP asymmetries are measured to be ${\cal A}_{\mbox{{\scriptsize 
CP}}}^{\mbox{{\scriptsize dir}}}(B_d\to\pi^+\pi^-)=+\,0.1$ and  
${\cal A}_{\mbox{{\scriptsize CP}}}^{\mbox{{\scriptsize mix-ind}}}
(B_d\to\pi^+\pi^-)=-\,0.25$ and that (\ref{BR-rat}) gives $r=0.26$. 
Assuming a theoretical uncertainty of 30\% in $r$, i.e.\ $\Delta r=0.04$, 
and inserting these numbers into (\ref{cpasym}) and (\ref{cpasym1}) gives
$\alpha=(76\pm1)^\circ$ and $\delta=(24\pm4)^\circ$. On the other hand, 
a na\"\i ve analysis 
using (\ref{BdpipiCP}) where the penguin contributions are neglected would 
yield $\alpha=83^\circ$. Consequently the theoretical uncertainty of the 
extracted value of $\alpha$ is expected to be significantly smaller than
the shift through the penguin contributions. Since this method of 
extracting $\alpha$ requires neither difficult measurements of very small 
branching ratios nor complicated geometrical constructions it may turn 
out to be very useful for the early days of the $B$-factory era beginning at
the end of this millennium.

\subsubsection{A Simultaneous Extraction of $\alpha$ and $\gamma$}
\noindent
Recently it has been pointed out by Dighe, Gronau and Rosner\cite{dgr}
that a time-dependent measurement of $B_d\to\pi^+\pi^-$ in combination with 
the branching ratios for $B^0_d\to\pi^- K^+$, $B^+\to\pi^+ K^0$ and their
charge-conjugates may allow a simultaneous determination of the UT angles
$\alpha$ and $\gamma$. These decays provide the following six observables
$A_1,\ldots,A_6$:
\begin{eqnarray}
\Gamma(B^0_d(t)\to\pi^+\pi^-)+\Gamma(\overline{B^0_d}(t)\to\pi^+\pi^-)&=&
e^{-\Gamma_dt}\,A_1\\
\Gamma(B^0_d(t)\to\pi^+\pi^-)-\Gamma(\overline{B^0_d}(t)\to\pi^+\pi^-)
\nonumber&=&e^{-\Gamma_dt}\left[A_2\cos(\Delta M_dt)\right.\nonumber\\
&&\quad+\left.A_3\sin(\Delta M_dt)\right]\\
\Gamma(B^0_d\to\pi^-K^+)+\Gamma(\overline{B^0_d}\to\pi^+K^-)&=&
A_4\\
\Gamma(B^0_d\to\pi^-K^+)-\Gamma(\overline{B^0_d}\to\pi^+K^-)&=&
A_5\\
\Gamma(B^+\to\pi^+K^0)+\Gamma(B^-\to\pi^-\overline{K^0})&=&
A_6.
\end{eqnarray}
Using $SU(3)$ flavor symmetry of strong interactions, neglecting
annihilation amplitudes, which should be suppressed by ${\cal O}(f_{B_d}/
M_{B_d})$ with $f_{B_d}\approx180\,\mbox{MeV}$, and assuming moreover
that the $\bar b\to\bar d$ QCD penguin amplitude is related in a simple
way to $\beta$ through (\ref{bdpenapprox}), i.e.\ assuming top-quark
dominance, the observables $A_1,\ldots,A_6$ can be expressed in terms
of six ``unknowns'' including $\alpha$ and $\gamma$. However, as we have 
outlined above, it is questionable whether the last assumption is justified 
since (\ref{bdpenapprox}) may be affected by QCD penguins 
with internal up- and charm-quark exchanges\cite{bf1}. Consequently the 
method proposed in Ref.\cite{dgr} suffers from theoretical limitations. 
Nevertheless it is an interesting approach, probably mainly in view of 
constraining $\gamma$ which is the most difficult to measure angle of the 
UT. In order to extract that angle, $B_s$ decays play an important role as
we will see in the following subsection. 

\newpage

\runninghead{CP Violation in Non-leptonic $B$-Meson Decays}  
{The $B_s$ System}
\subsection{The $B_s$ System}
\noindent
The major phenomenological differences between the $B_d$ and $B_s$
systems arise from their mixing parameters (\ref{e719}) and from the 
fact that at leading order in the Wolfenstein expansion only a trivial 
weak mixing phase (\ref{e742}) is present in the $B_s$ case. 

\subsubsection{CP Violation in $B_s\to\rho^0K_{\mbox{{\scriptsize S}}}$:
the ``Wrong'' Way to Extract $\gamma$}
\noindent
Let us begin our discussion of the $B_s$ system by having a closer look
at the transition $B_s\to\rho^0K_{\mbox{{\scriptsize S}}}$ which appears
frequently in the literature as a tool to extract $\gamma$. It is a $B_s$
decay into a final CP eigenstate with eigenvalue $-1$ that is (similarly 
as the $B_d\to\pi^+\pi^-$ mode) caused by the quark-level process 
$\bar b\to\bar uu\bar d$. Hence the corresponding observable 
$\xi^{(s)}_{\rho^0K_{\mbox{{\scriptsize S}}}}$ can be expressed as
\begin{equation}\label{xirhoks}
\xi^{(s)}_{\rho^0K_{\mbox{{\scriptsize S}}}}=+e^{-i0}\left[
\frac{v_u^{(d)}\left({\cal A}_{\mbox{{\scriptsize cc}}}^{u}+
{\cal A}^{ut}_{\mbox{{\scriptsize pen}}}\right)+v_c^{(d)}
{\cal A}^{ct}_{\mbox{{\scriptsize pen}}}}
{v_u^{(d)\ast}\left({\cal A}_{\mbox{{\scriptsize cc}}}^{u}+
{\cal A}^{ut}_{\mbox{{\scriptsize pen}}}\right)+v_c^{(d)\ast}
{\cal A}^{ct}_{\mbox{{\scriptsize pen}}}}
\right],
\end{equation}
where the notation is as in 3.3.3. The structure of (\ref{xirhoks}) is 
very similar to that of the observable $\xi^{(d)}_{\pi^+\pi^-}$ given in 
(\ref{xipipi}). However, an important difference arises between 
$B_d\to\pi^+\pi^-$ and $B_s\to \rho^0K_{\mbox{{\scriptsize S}}}$:
although the penguin contributions are expected to be of equal order 
of magnitude in (\ref{xipipi}) and (\ref{xirhoks}), their importance 
is enhanced in the latter case since the current-current amplitude 
${\cal A}_{\mbox{{\scriptsize cc}}}^{u}$ is color-suppressed
by a phenomenological color-suppression factor\cite{BSW}$^-$\cite{a1a2-exp} 
$a_2\approx0.2$. Consequently, \mbox{using} in addition to that value of 
$a_2$ characteristic Wilson coefficient functions for the penguin operators 
and (\ref{bdckm}) for the ratio of CKM factors, one obtains
\begin{equation}
\left|\frac{v_c^{(d)}{\cal A}^{ct}_{\mbox{{\scriptsize pen}}}}
{v_u^{(d)}\left({\cal A}_{\mbox{{\scriptsize cc}}}^{u}+
{\cal A}^{ut}_{\mbox{{\scriptsize pen}}}\right)}\right|={\cal O}(0.5).
\end{equation}
This estimate implies that
\begin{equation}
\xi^{(s)}_{\rho^0K_{\mbox{{\scriptsize S}}}}\approx+e^{-i0}
\left[\frac{v_u^{(d)}\left({\cal A}_{\mbox{{\scriptsize cc}}}^{u}+
{\cal A}^{ut}_{\mbox{{\scriptsize pen}}}\right)}
{v_u^{(d)\ast}\left({\cal A}_{\mbox{{\scriptsize cc}}}^{u}+
{\cal A}^{ut}_{\mbox{{\scriptsize pen}}}\right)}\right]=e^{-2i\gamma}
\end{equation}
is a {\it very bad} approximation which should {\it not} allow a meaningful 
determination of $\gamma$ from the mixing-induced CP-violating asymmetry
arising in $B_s\to\rho^0K_{\mbox{{\scriptsize S}}}$. Needless to note, 
the branching ratio of that decay is expected to be of ${\cal O}(10^{-7})$
which makes its experimental investigation very difficult. Interestingly
there are other $B_s$ decays -- some of them receive also penguin 
contributions -- which {\it do} allow extractions of $\gamma$. Some of 
these strategies are even theoretically clean and suffer from no hadronic 
uncertainties. Before focussing on these modes, let us discuss an 
experimental problem of $B_s$ decays that is related to time-dependent 
measurements. 

\subsubsection{The $B_s$ System in Light of $\Delta\Gamma_s$}
\noindent
The large mixing parameter $x_s={\cal O}(20)$ that is expected\cite{al}
within the SM implies very rapid $B^0_s-\overline{B^0_s}$ 
oscillations requiring an excellent vertex resolution system to 
keep track of the $\Delta M_s t$ terms. That is obviously a formidable 
experimental task. It may, however, not be necessary to trace the rapid 
$\Delta M_s t$ oscillations in order to shed light on the mechanism of 
CP violation\cite{dunietz}. This remarkable feature is due 
to the expected sizable width difference $\Delta\Gamma_s$ which has
been discussed at the end of Subsection~3.1. Because of that width 
difference already {\it untagged} $B_s$ rates, which are defined by 
\begin{equation}\label{untagged}
\Gamma[f(t)]\equiv\Gamma(B_s^0(t)\to f)+\Gamma(\overline{B^0_s}(t)\to f),
\end{equation}
may provide valuable information about the phase structure of the observable
$\xi_f^{(s)}$. This can be seen nicely by rewriting (\ref{untagged}) 
with the help of (\ref{ratebf}) and (\ref{ratebbf}) in a more explicit way 
as follows:
\begin{equation}\label{EE3}
\Gamma[f(t)]\propto\left[\left(1+\left|\xi_f^{(s)}
\right|^2\right)\left(e^{-\Gamma_L^{(s)} t}+e^{-\Gamma_H^{(s)} t}\right)
-2\mbox{\,Re\,}\xi_f^{(s)}\left(e^{-\Gamma_L^{(s)} t}-
e^{-\Gamma_H^{(s)} t}\right)\right].
\end{equation}
In this expression the rapid oscillatory $\Delta M_s t$ terms, which
show up in the {\it tagged} rates (\ref{ratebf}) and (\ref{ratebbf}), 
cancel\cite{dunietz}. Therefore it depends only on the two exponents 
$e^{-\Gamma_L^{(s)} t}$ and $e^{-\Gamma_H^{(s)} t}$. From an experimental 
point of view, such untagged analyses are clearly much more promising than 
tagged ones in respect of efficiency, acceptance and purity. 

In order to illustrate these untagged rates in more detail, let us 
consider an estimate of $\gamma$ using untagged $B_s\to K^+K^-$ and 
$B_s\to K^0\overline{K^0}$ decays that has been proposed 
recently by Dunietz and myself\,\cite{fd1}. Using the $SU(2)$ isospin
symmetry of strong interactions to relate the QCD penguin contributions to
these decays (EW penguins are color-suppressed in these modes
and should therefore play a minor role as we will see in Sections~4 and 5), 
we obtain
\begin{equation}\label{EE4}
\Gamma[K^+K^-(t)]\propto |P'|^2\Bigl[\bigl(1-2\,|r|\cos\varrho\,
\cos\gamma+|r|^2\cos^2\gamma\bigr)e^{-\Gamma_L^{(s)} t}+|r|^2\sin^2\gamma\, 
e^{-\Gamma_H^{(s)} t}\Bigr]
\end{equation}
and
\begin{equation}\label{EE5}
\Gamma[K^0\overline{K^0}(t)]\propto |P'|^2\,e^{-\Gamma_L^{(s)} t},
\end{equation}
where 
\begin{equation}\label{EE6}
r\equiv|r|e^{i\varrho}=\frac{|T'|}{|P'|}e^{i(\delta_{T'}-\delta_{P'})}.
\end{equation}
Here we have used the same notation as Gronau et al.\ in Ref.\cite{grl-gam}
which will turn out to be very useful for later discussions: $P'$ 
denotes the $\bar b\to\bar s$ QCD penguin amplitude corresponding to
(\ref{bspenamp}), $T'$ is the color-allowed $\bar b\to\bar uu\bar s$ 
current-current amplitude, and $\delta_{P'}$ and $\delta_{T'}$ denote the 
corresponding CP-conserving strong phases. The primes remind us that we are 
dealing with $\bar b\to\bar s$ amplitudes. In order to determine $\gamma$ 
from the untagged rates (\ref{EE4}) and (\ref{EE5}), we need an additional 
input that is provided by the $SU(3)$ flavor symmetry of strong
interactions. Using that symmetry and neglecting as in (\ref{BR-rat})
the color-suppressed current-current contributions to $B^+\to\pi^+\pi^0$, 
one finds\cite{grl-gam}
\begin{equation}\label{EE7}
|T'|\approx\lambda\,\frac{f_K}{f_\pi}\,\sqrt{2}\,|A(B^+\to\pi^+\pi^0)|,
\end{equation}
where $\lambda$ is the usual Wolfenstein parameter, $f_K/f_\pi$ 
takes into account factorizable $SU(3)$-breaking, and $A(B^+\to\pi^+\pi^0)$ 
denotes the appropriately normalized decay amplitude of $B^+\to\pi^+\pi^0$. 
Since $|P'|$ is known from the untagged $B_s\to K^0\,\overline{K^0}$ rate
(\ref{EE5}), the quantity $|r|=|T'|/|P'|$ can be estimated 
with the help of (\ref{EE7}) and allows the extraction of $\gamma$ from 
the part of (\ref{EE4}) evolving with exponent $e^{-\Gamma_H^{(s)} t}$. 
As we will see in a moment, one can even do better, i.e.\ without using an
$SU(3)$-based estimate like (\ref{EE7}), by considering the decays 
corresponding to $B_s\to K \overline{K}$ where two vector mesons 
or appropriate higher resonances are present in the final states\cite{fd1}.  

\subsubsection{An Extraction of $\gamma$ using $B_s\to K^{\ast+}
K^{\ast-}$ and $B_s\to K^{\ast0}\overline{K^{\ast0}}$}\label{kkbar}
\noindent
The untagged angular distributions of these decays, which take the general 
form
\begin{equation}\label{EE8}
[f(\theta,\phi,\psi;t)]=\sum_k\left[\overline{b^{(k)}}(t)+b^{(k)}(t)\right]
g^{(k)}(\theta,\phi,\psi),
\end{equation}
provide many more observables than the untagged modes
$B_s\to K^+K^-$ and $B_s\to K^0\overline{K^0}$ discussed in 3.4.2. 
Here $\theta$, $\phi$ and $\psi$ are generic decay angles describing 
the kinematics of the decay products arising in the
decay chain $B_s\to K^\ast(\to\pi K)\,\overline{K^\ast}(\to \pi\overline{K})$.
The observables $\left[\overline{b^{(k)}}(t)+b^{(k)}(t)\right]$ governing the
time-evolution of the untagged angular distribution (\ref{EE8}) are given
by real or imaginary parts of bilinear combinations of decay amplitudes
that are of the following structure:
\begin{eqnarray}
\lefteqn{\left[A_{\tilde f}^\ast(t)\,A_f(t)\right]\equiv\left\langle
\left(K^\ast\overline{K^\ast}\right)_{\tilde f}\left|{\cal 
H}_{\mbox{{\scriptsize eff}}}\right|\overline{B_s^0}(t)\right\rangle^\ast
\left\langle\left(K^\ast\overline{K^\ast}\right)_{f}\left|{\cal 
H}_{\mbox{{\scriptsize eff}}}\right|\overline{B_s^0}(t)\right\rangle}
\nonumber\\
&&+\left\langle\left(K^\ast\overline{K^\ast}\right)_{\tilde f}\left|{\cal 
H}_{\mbox{{\scriptsize eff}}}\right|B_s^0(t)\right\rangle^\ast
\left\langle\left(K^\ast\overline{K^\ast}\right)_{f}\left|{\cal 
H}_{\mbox{{\scriptsize eff}}}\right|B_s^0(t)\right\rangle.\label{EE9}
\end{eqnarray}
In this expression, $f$ and $\tilde f$ are labels that define the relative
polarizations of $K^\ast$ and $\overline{K^\ast}$ in final state 
configurations $\left(K^\ast\overline{K^\ast}\right)_f$ (e.g.\ linear 
polarization states\cite{rosner} $\{0,\parallel,\perp\}$) with CP 
eigenvalues $\eta_{\mbox{{\scriptsize CP}}}^f$:
\begin{equation}\label{EE10}
({\cal CP})\left|\left(K^\ast\overline{K^\ast}\right)_f\right\rangle
=\eta_{\mbox{{\scriptsize CP}}}^f\left|
\left(K^\ast\overline{K^\ast}\right)_f\right\rangle.
\end{equation}
An analogous relation holds for $\tilde f$. The observables of the 
angular distributions for $B_s\to K^{\ast+}
K^{\ast-}$ and $B_s\to K^{\ast0}\overline{K^{\ast0}}$
are given explicitly in Ref.\cite{fd1}. In the case of the latter decay
the formulae simplify considerably since it is a penguin-induced 
$\bar b\to\bar sd\bar d$ mode and receives therefore no tree contributions.
Using, as in (\ref{EE4}) and (\ref{EE5}), the $SU(2)$ isospin symmetry of 
strong interactions, the QCD penguin contributions to these decays can be 
related to each other. If one takes into account these relations and goes
very carefully through the observables of the corresponding untagged angular 
distributions, one finds that they allow the extraction of 
$\gamma$ without any additional theoretical input\cite{fd1}. 
In particular no $SU(3)$ symmetry arguments are needed and the $SU(2)$ 
isospin symmetry suffices to accomplish this task. The angular distributions 
provide moreover information about the 
hadronization dynamics of the corresponding decays, and the 
formalism\cite{fd1} developed for $B_s\to K^{\ast+}K^{\ast-}$ applies 
also to $B_s\to\rho^0\phi$ if one performs a suitable replacement of 
variables. Since that channel is expected to be dominated 
by EW penguins as discussed in Subsection~4.3, it may allow 
interesting insights into the physics of these operators.

\subsubsection{$B_s\to D_s^{\ast+}D_s^{\ast-}$ and $B_s\to J/\psi\,
\phi$: the ``Gold-plated'' Transitions to Extract $\eta$}\label{gold}
\noindent
The following discussion is devoted to an analysis\cite{fd1} of 
the decays $B_s\to D_s^{\ast+}(\to D_s^+\gamma)\,D_s^{\ast-}
(\to D_s^-\gamma)$ and $B_s\to J/\psi(\to l^+l^-)\,\phi(\to K^+K^-)$, 
which is the counterpart of the ``gold-plated'' mode $B_d\to J/\psi\,
K_{\mbox{{\scriptsize S}}}$ to measure $\beta$. Since these decays are 
dominated by a single CKM amplitude, the hadronic uncertainties cancel in 
$\xi_f^{(s)}$ (see 3.2.2) taking in that particular case the following 
form:
\begin{equation}\label{EE11}
\xi_f^{(s)}=-\eta_{\mbox{{\scriptsize CP}}}^f
\,e^{i\,\phi_{\mbox{{\scriptsize CKM}}}}\,.
\end{equation}
Consequently the observables of the untagged angular distributions, which 
have the same general structure as (\ref{EE8}), simplify 
considerably\cite{fd1}. In (\ref{EE11}), $f$ is -- as in (\ref{EE9}) and 
(\ref{EE10}) -- a label defining the 
relative polarizations of $X_1$ and $X_2$ in final state configurations 
$\left(X_1\,X_2\right)_f$ with CP eigenvalue 
$\eta_{\mbox{{\scriptsize CP}}}^f$, where $(X_1,X_2)\in
\left\{(D_s^{\ast+},D_s^{\ast-}),(J/\psi,\phi)\right\}$. Applying 
(\ref{e743}) in combination with (\ref{e742}) and (\ref{e746}), the 
CP-violating weak phase $\phi_{\mbox{{\scriptsize CKM}}}$ would vanish. 
In order to obtain a non-vanishing result for that phase, its exact 
definition is
\begin{equation}\label{phickm}
\phi_{\mbox{{\scriptsize 
CKM}}}\equiv-2\left[\mbox{arg}(V_{ts}^\ast V_{tb})-\mbox{arg}(V_{cs}^\ast 
V_{cb})\right],
\end{equation} 
we have to take into account higher order terms in the Wolfenstein 
expansion of the CKM matrix yielding $\phi_{\mbox{{\scriptsize CKM}}}=
2\lambda^2\eta={\cal O}(0.03)$. Consequently the small weak phase 
$\phi_{\mbox{{\scriptsize CKM}}}$ measures simply $\eta$ which fixes the 
height of the UT. Another interesting interpretation of (\ref{phickm}) is 
the fact that it is related to an angle in a rather squashed and 
therefore ``unpopular'' unitarity triangle\cite{akl}. Other useful 
expressions for (\ref{phickm}) can be found in Ref.\cite{snowmass}.

A characteristic feature of the angular distributions for 
$B_s\to D_s^{\ast+}D_s^{\ast-}$ and $B_s\to J/\psi\,\phi$ is 
interference between CP-even and CP-odd final state configurations leading 
to untagged observables that are proportional to  
\begin{equation}\label{EE12}
\left(e^{-\Gamma_L^{(s)}t}-e^{-\Gamma_H^{(s)}t}
\right)\sin\phi_{\mbox{{\scriptsize CKM}}}.
\end{equation}
As was shown in Ref.\cite{fd1}, the angular distributions for both the 
color-allowed channel $B_s\to D_s^{\ast+} D_s^{\ast-}$ and the 
color-suppressed transition $B_s\to J/\psi\,\phi$ each provide separately
sufficient information to determine $\phi_{\mbox{{\scriptsize CKM}}}$ from 
their untagged data samples. The extraction of 
$\phi_{\mbox{{\scriptsize CKM}}}$ is, however, not as clean 
as that of $\beta$ from $B_d\to J/\psi\,K_{\mbox{{\scriptsize S}}}$. This 
feature is due to the smallness of $\phi_{\mbox{{\scriptsize CKM}}}$ with 
respect to $\beta$, enhancing the importance of the unmixed amplitudes 
proportional to the CKM factor $V_{us}^\ast V_{ub}$ which are similarly 
suppressed in both cases. 

Within the SM one expects a very small value of $\phi_{\mbox{{\scriptsize 
CKM}}}$ and $\Gamma_H^{(s)}<\Gamma_L^{(s)}$. However, that need not to
be the case in many scenarios for ``New Physics'' (see e.g.\ Ref.\cite{yg}).
An experimental study of the decays $B_s\to D_s^{\ast+}D_s^{\ast-}$ 
and $B_s\to J/\psi\,\phi$ may shed light on this issue\cite{fd1}, and an
extracted value of $\phi_{\mbox{{\scriptsize CKM}}}$ that is much larger than
${\cal O}(0.03)$ would most probably signal physics beyond the SM.

\subsubsection{$B_s$ Decays caused by $\bar b\to\bar uc\bar s$ ($b\to c\bar 
us$) and Clean Extractions of $\gamma$}\label{nonCP}
\noindent
Exclusive $B_s$ decays caused by $\bar b\to\bar uc\bar s$ ($b\to c\bar u 
s$) quark-level transitions belong to decay class iii) introduced in 
Subsection~2.1, i.e.\ are pure tree decays receiving {\it no} penguin
contributions, and probe\cite{gam} the UT angle $\gamma$. 
Their transition amplitudes can be expressed as hadronic 
matrix elements of low energy effective Hamiltonians having the 
following structures\cite{fd2}:
\begin{eqnarray}
{\cal H}_{\mbox{{\scriptsize eff}}}(\overline{B^0_s}\to f)&=&
\frac{G_{\mbox{{\scriptsize F}}}}{\sqrt{2}}\,\overline{v}
\left[\overline{O}_1\,{\cal C}_1(\mu)+
\overline{O}_2\,{\cal C}_2(\mu)\right]\\
{\cal H}_{\mbox{{\scriptsize eff}}}(B^0_s\to f)&=&
\frac{G_{\mbox{{\scriptsize F}}}}{\sqrt{2}}\,v^\ast
\left[O_1^\dagger\,{\cal C}_1(\mu)+O_2^\dagger\,{\cal C}_2(\mu)\right].
\end{eqnarray}
Here $f$ denotes a final state with valence-quark content 
$s\bar u\, c\bar s$, the relevant CKM factors take the form
\begin{equation}
\overline{v}\equiv V_{us}^\ast V_{cb}=A\lambda^3,\quad
v\equiv V_{cs}^\ast V_{ub}=A\lambda^3R_b\,e^{-i\gamma},
\end{equation} 
where the modified Wolfenstein parametrization (\ref{wolf2}) has been used, 
and $\overline{O}_k$ and $O_k$ denote current-current operators (see 
(\ref{cc-def})) that are given by
\begin{equation}
\begin{array}{rclrcl}
\overline{O}_1&=&\left(\bar s_\alpha u_\beta\right)_{\mbox{{\scriptsize 
V--A}}}\left(\bar c_\beta b_\alpha\right)_{\mbox{{\scriptsize V--A}}},
\quad&\overline{O}_2&=&\left(\bar s_\alpha u_\alpha\right)_{\mbox{{\scriptsize 
V--A}}}\left(\bar c_\beta b_\beta\right)_{\mbox{{\scriptsize V--A}}},
\label{O-cc}\\
O_1&=&\left(\bar s_\alpha c_\beta\right)_{\mbox{{\scriptsize V--A}}}
\left(\bar u_\beta b_\alpha\right)_{\mbox{{\scriptsize V--A}}},\quad&
O_2&=&\left(\bar s_\alpha c_\alpha\right)_{\mbox{{\scriptsize V--A}}}
\left(\bar u_\beta b_\beta\right)_{\mbox{{\scriptsize V--A}}}\label{Obar-cc}.
\end{array}
\end{equation}
Nowadays the Wilson coefficient functions ${\cal C}_1(\mu)$ and 
${\cal C}_2(\mu)$ are available at NLO and the corresponding results can 
be found in Refs.\cite{bbl-rev,acmp,bw}.

Performing appropriate CP transformations in the matrix element
\begin{eqnarray}
\lefteqn{\left\langle f\left|O_1^\dagger(\mu){\cal C}_1(\mu)+
O_2^\dagger(\mu){\cal C}_2(\mu)\right|B_s^0\right\rangle}
\nonumber\\
&&=\left\langle f\left|({\cal CP})^\dagger({\cal CP})\left[O_1^\dagger(\mu)
{\cal C}_1(\mu)+O_2^\dagger(\mu){\cal C}_2(\mu)\right]({\cal CP})^\dagger
({\cal CP})\right|B_s^0\right\rangle\\
&&=e^{i\phi_{\mbox{{\scriptsize CP}}}(B_s)}\,\Bigl\langle\overline{f}\Bigl|
O_1(\mu){\cal C}_1(\mu)+O_2(\mu){\cal C}_2(\mu)\Bigr|
\overline{B^0_s}\Bigr\rangle,
\nonumber
\end{eqnarray}
where (\ref{e710}) and the analogue of (\ref{e738}) have been taken into
account, gives
\begin{eqnarray}
A(\overline{B^0_s}\to f)&=&\left\langle f\left|{\cal 
H}_{\mbox{{\scriptsize eff}}}(\overline{B^0_s}\to f)\right|
\overline{B^0_s}\right\rangle=\frac{G_{\mbox{{\scriptsize F}}}}{\sqrt{2}}
\,\overline{v}\,\,\overline{M}_f\\
A(B^0_s\to f)&=&\left\langle f\left|{\cal H}_{\mbox{{\scriptsize eff}}}
(B^0_s\to f)\right|B^0_s\right\rangle\,=\,e^{i\phi_{\mbox{{\scriptsize 
CP}}}(B_s)}\frac{G_{\mbox{{\scriptsize F}}}}{\sqrt{2}}\,v^\ast
M_{\overline{f}}
\end{eqnarray}
with the strong hadronic matrix elements
\begin{eqnarray}
\overline{M}_f&\equiv&\Bigl\langle f\Bigl|\overline{O}_1(\mu){\cal C}_1(\mu)+
\overline{O}_2(\mu){\cal C}_2(\mu)\Bigr|\overline{B^0_s}\Bigr\rangle\\
M_{\overline{f}}&\equiv&\Bigl\langle\overline{f}\Bigl|O_1(\mu){\cal C}_1(\mu)+
O_2(\mu){\cal C}_2(\mu)\Bigr|\overline{B^0_s}\Bigr\rangle.
\end{eqnarray}
Consequently, using in addition (\ref{e717}) and (\ref{e742}), the 
observable $\xi_f^{(s)}$ defined in (\ref{e731}) is given by
\begin{equation}\label{EXIF}
\xi_f^{(s)}=-e^{-i\phi_{\mbox{{\scriptsize M}}}^{(s)}}
\frac{\overline{v}}{v^\ast}\frac{\overline{M}_f}{M_{\overline f}}
=-e^{-i\gamma}\frac{1}{R_b}\frac{\overline{M}_f}{M_{\overline f}}\,.
\end{equation}
Note that $\phi_{\mbox{{\scriptsize CP}}}(B_s)$ cancels in (\ref{EXIF})
which is a nice check. An analogous calculation yields
\begin{equation}
\xi_{\overline{f}}^{(s)}=-e^{-i\phi_{\mbox{{\scriptsize M}}}^{(s)}}
\frac{v}{\overline{v}^\ast}\frac{M_{\overline{f}}}{\overline{M}_f}
=-e^{-i\gamma}R_b\,\frac{M_{\overline f}}{\overline{M}_f}.
\end{equation}
If one measures the tagged time-dependent decay rates 
(\ref{ratebf})-(\ref{ratebbfb}), both $\xi_f^{(s)}$ and 
$\xi_{\overline{f}}^{(s)}$ can be determined and allow a {\it theoretically 
clean} determination of $\gamma$ since
\begin{equation}
\xi_f^{(s)}\cdot\xi_{\overline{f}}^{(s)}=e^{-2i\gamma}.
\end{equation}

There are by now well-known strategies on the market using time-evolutions 
of $B_s$ modes originating from $\bar b\to\bar uc\bar s$ ($b\to c\bar us$) 
quark-level transitions, e.g.\ $\stackrel{{\mbox{\tiny (---)}}}{B_s}
\to\stackrel{{\mbox{\tiny (---)}}}{D^0}\phi$\,\cite{gam,glgam} and 
$\stackrel{{\mbox{\tiny (---)}}}{B_s}\to D_s^\pm K^\mp$\,\cite{adk}, 
to extract $\gamma$. However, as we have noted already, in these methods 
tagging is essential and the rapid $\Delta M_s t$ oscillations have to be 
resolved which is an experimental challenge. The question what can be 
learned from {\it untagged} data samples of these decays, where the 
$\Delta M_s t$ terms cancel, has been investigated by Dunietz\cite{dunietz}. 
In the untagged case the determination of $\gamma$ requires additional inputs: 
\begin{itemize}
\item Color-suppressed modes $\stackrel{{\mbox{\tiny (---)}}}{B_s}\to
\stackrel{{\mbox{\tiny (---)}}}{D^0}\phi$: a measurement of the 
untagged $B_s\to D^0_{\pm} \phi$ rate is needed, where 
$D^0_{\pm}$ is a CP eigenstate of the neutral $D$ system.
\item Color-allowed modes $\stackrel{{\mbox{\tiny (---)}}}{B_s}\to 
D_s^\pm K^\mp$: a theoretical input corresponding to the ratio of the 
unmixed rates $\Gamma(B^0_s\to D_s^-K^+)/\Gamma(B^0_s\to D_s^-\pi^+)$ is
needed. This ratio can be estimated with the help of the ``factorization'' 
hypothesis\cite{fact1,fact2} which may work reasonably 
well for these color-allowed channels\cite{bjorken}.
\end{itemize}
Interestingly the untagged data samples may exhibit CP-violating 
effects that are described by observables of the form
\begin{equation}\label{EE13}
\Gamma[f(t)]-\Gamma[\overline{f}(t)]\propto\left(e^{-\Gamma_L^{(s)}t}-
e^{-\Gamma_H^{(s)}t}\right)\sin\varrho_f\,\sin\gamma.
\end{equation}
Here $\varrho_f$ is a CP-conserving strong phase. Because of the 
$\sin\varrho_f$ factor, a non-trivial strong phase is essential in that 
case. Consequently the CP-violating observables (\ref{EE13}) vanish within 
the factorization approximation predicting $\varrho_f\in\{0,\pi\}$. 
Since factorization may be a reasonable working assumption 
for the color-allowed modes $\stackrel{{\mbox{\tiny 
(---)}}}{B_s}\to D_s^\pm K^\mp$, the CP-violating effects in their
untagged data samples are expected to be tiny. On the other hand,
the factorization hypothesis is very questionable for 
the color-suppressed decays $\stackrel{{\mbox{\tiny (---)}}}{B_s}\to
\stackrel{{\mbox{\tiny (---)}}}{D^0}\phi$ and sizable CP violation may 
show up in the corresponding untagged rates\cite{dunietz}. 

Concerning such CP-violating effects and the extraction of $\gamma$ from
untagged rates, the decays $\stackrel{{\mbox{\tiny (---)}}}{B_s}\to 
D_s^{\ast\pm} K^{\ast\mp}$ and $\stackrel{{\mbox{\tiny (---)}}}{B_s}\to
\stackrel{{\mbox{\tiny(---)}}}{D^{\ast0}}\phi$ are expected to be 
more promising than the transitions discussed above. As was
shown in Ref.\cite{fd2}, the time-dependences of their untagged angular
distributions allow a clean extraction of $\gamma$ without any additional 
input. The final state configurations of these decays are not admixtures of 
CP eigenstates as in the case of the decays discussed in 3.4.3 and 3.4.4. 
They can, however, be classified by their parity eigenvalues. 
A characteristic feature of the corresponding angular distributions 
is interference between parity-even and parity-odd configurations that may
lead to potentially large CP-violating effects in the untagged data samples
even when all strong phase shifts vanish. An example of such an
untagged CP-violating observable is the following quantity\cite{fd2}:
\begin{eqnarray}
\lefteqn{\mbox{Im}\left\{\left[A_f^\ast(t)\,A_\perp(t)\right]\right\}+
\mbox{Im}\left\{\left[A_f^{\mbox{{\scriptsize C}}\ast}(t)\,
A_\perp^{\mbox{{\scriptsize C}}}(t)\right]\right\}}\nonumber\\
&&\propto\left(e^{-\Gamma_L^{(s)}t}-e^{-\Gamma_H^{(s)}t}\right)
\bigl\{|R_f|\cos(\delta_f-\vartheta_\perp)+|R_\perp|
\cos(\delta_\perp-\vartheta_f)\bigr\}\,\sin\gamma.\label{EE14}
\end{eqnarray}
In that expression bilinear combinations of certain decay amplitudes 
(see (\ref{EE9})) show up, $f\in\{0,\parallel\}$ denotes a 
linear polarization state\cite{rosner} and $\delta_f$, $\vartheta_f$
are CP-conserving phases that are induced through strong final 
state interaction effects. For the details concerning the 
observable (\ref{EE14}) -- in particular the definition of the 
relevant charge-conjugate amplitudes $A_f^{\mbox{{\scriptsize C}}}$ and the 
quantities $|R_f|$ -- the reader is referred to Ref.\cite{fd2}. 
Here I would like to emphasize only that the strong phases
enter in the form of {\it cosine} terms. Therefore non-trivial 
strong phases are -- in contrast to (\ref{EE13}) -- not essential for 
CP violation in the corresponding untagged data samples and one
expects, even within the factorization approximation, which may apply to
the color-allowed modes $\stackrel{{\mbox{\tiny (---)}}}{B_s}\to 
D_s^{\ast\pm} K^{\ast\mp}$, potentially large effects. 
 
Since the soft photons in the decays $D_s^\ast\to D_s\gamma$, $D^{\ast0}
\to D^0\gamma$ are difficult to detect, certain higher resonances exhibiting
significant all-charged final states, e.g.\ $D_{s1}(2536)^+\to
D^{\ast+}K^0$, $D_1(2420)^0\to D^{\ast+}\pi^-$ with $D^{\ast+}\to 
D^0\pi^+$, may be more promising for certain detector configurations. 
A similar comment applies also to the mode $B_s\to D_s^{\ast+}D_s^{\ast-}$
discussed in 3.4.4.

To finish the presentation of the $B_s$ system, let me stress once
again that the untagged measurements discussed in this Subsection are 
much more promising in view of efficiency, acceptance and purity than 
tagged analyses. Moreover the oscillatory $\Delta M_st$ terms, which
may be too rapid to be resolved with present vertex technology, cancel
in untagged $B_s$ data samples. However, a lot of statistics is required 
and the natural place for these experiments seems to be a hadron collider
(note that the formulae given above have to be modified appropriately for
$e^+-e^-$ machines to take into account coherence of the 
$B^0_s-\overline{B^0_s}$ pair at $\Upsilon(5\mbox{S})$). 
Obviously the feasibility of untagged strategies to extract CKM phases
depends crucially on a sizable width difference $\Delta\Gamma_s$. Even 
if it should turn out to be too small for such untagged analyses, 
once $\Delta\Gamma_s\not=0$ has been established experimentally, 
the formulae developed in Refs.\cite{fd1,fd2} have also 
to be used to determine CKM phases correctly from tagged measurements. 
Clearly time will tell and experimentalists will certainly find out which 
method is most promising from an experimental point of view. 

Let me conclude the review of CP violation in the neutral $B_q$ systems
with the following remark. We have considered 
only {\it exclusive} neutral $B_q$-meson decays. However, also {\it 
inclusive} decay processes with specific quark-flavors, e.g.\ 
$\bar b\to\bar uu\bar d$ or $\bar b\to\bar cc\bar s$, may exhibit 
mixing-induced CP-violating asymmetries\cite{ds}. Recently the determination 
of $\sin(2\alpha)$ from the CP asymmetry arising in inclusive $B_d$ decays 
into charmless final states has been analyzed by assuming local quark-hadron
duality\cite{bbd-inclCP}. Compared to exclusive 
transitions, inclusive decay processes have of course rates that are 
larger by orders of magnitudes. However, due to the summation over processes 
with asymmetries of alternating signs, the inclusive CP asymmetries are 
unfortunately diluted with respect to the exclusive case. The calculation
of the dilution factor suffers in general from large hadronic uncertainties.
Progress has been made in Ref.\cite{bbd-inclCP}, where local quark-hadron 
duality has been used to evaluate this quantity. From an experimental point 
of view, inclusive measurements, e.g.\ of inclusive $B_d^0$ decays caused by 
$\bar b\to\bar uu\bar d$, are very difficult (see also M. Gronau's talk
in Ref.\cite{cp-revs}) and their practical usefulness is unclear at present.

\runninghead{CP Violation in Non-leptonic $B$-Meson Decays}  
{The Charged $B$ System}
\subsection{The Charged $B$ System}
\noindent
Since mixing-effects are absent in the charged $B$-meson system, 
non-vanishing CP-violating asymmetries of charged $B$ decays would 
give unambiguous evidence for direct CP violation. Due to the unitarity of 
the CKM matrix, the transition amplitude of a charged $B$ decay can be 
written in the following general form:
\begin{equation}\label{ED81}
A(B^-\to f)=v_{1}\,A_{1}\,e^{i\alpha_{1}}+v_{2}\,A_{2}\,e^{i\alpha_{2}},
\end{equation}
where $v_{1}$, $v_{2}$ are CKM factors, $A_{1}$, $A_{2}$ are ``reduced'',
i.e.\ real, hadronic matrix elements of weak transition operators and
$\alpha_{1}$, $\alpha_{2}$ denote CP-conserving phases generated 
through strong final state interaction effects. On the other hand, the 
transition amplitude of the CP-conjugate decay $B^+\to\overline{f}$ is 
given by
\begin{equation}\label{ED82}
A(B^+\to\overline{f})=v_{1}^{\ast}A_{1}\,e^{i\alpha_{1}}+v_{2}^{\ast}\,
A_{2}\,e^{i\alpha_{2}}.
\end{equation}
If the CP-violating asymmetry of the decay $B\to f$ is defined through
\begin{equation}\label{ED83}
{\cal A}_{\mbox{{\scriptsize CP}}}\equiv\frac{\Gamma(B^+\to\overline{f})-
\Gamma(B^-\to f)}{\Gamma(B^+\to\overline{f})+\Gamma(B^-\to f)},
\end{equation}
the transition amplitudes (\ref{ED81}) and (\ref{ED82}) yield
\begin{equation}\label{ED84}
{\cal A}_{\mbox{{\scriptsize CP}}}=\frac{2\,\mbox{Im}(v_{1}v_{2}^\ast)
\sin(\alpha_{1}-\alpha_{2})A_{1}A_{2}}{|v_{1}|^{2}A_{1}^{2}
+|v_{2}|^{2}A_{2}^{2}+2\,\mbox{Re}(v_{1}
v_{2}^\ast)\cos(\alpha_{1}-\alpha_{2})A_{1}A_{2}}.
\end{equation}
Consequently there are two conditions that have to be met
simultaneously in order to get a non-zero CP asymmetry 
${\cal A}_{\mbox{{\scriptsize CP}}}$:
\begin{itemize}
\item[i)] There has to be a relative {\it CP-violating} weak phase, i.e.\
$\mbox{Im}(v_1 v_2^\ast)\not=0$, between the two amplitudes contributing 
to $B\to f$. This phase difference can be expressed in terms of
complex phases of CKM matrix elements and is thus calculable.
\item[ii)] There has to be a relative {\it CP-conserving} strong phase, 
i.e.\ $\sin(\alpha_{1}-\alpha_{2})\not=0$, generated by strong final 
state interaction effects. In contrast to the CP-violating weak phase 
difference, the calculation of $\alpha_{1}-\alpha_{2}$ is very involved 
and suffers in general from large theoretical uncertainties.
\end{itemize}
These general requirements for the appearance of direct CP violation apply 
of course also to neutral $B_q$ decays, where direct CP
violation shows up as ${\cal A}_{\mbox{{\scriptsize 
CP}}}^{\mbox{{\scriptsize dir}}}\not=0$ (see (\ref{acpdir})).

Semileptonic decays of charged $B$-mesons obviously do not fulfil 
point i) and exhibit therefore no CP violation within the SM. However, 
there are non-leptonic modes of charged $B$-mesons 
corresponding to decay classes i) and ii) introduced in Subsection~2.1 
that are very promising in respect of direct CP violation. In decays 
belonging to class i), e.g.\ in $B^+\to\pi^0K^+$,
non-zero CP asymmetries (\ref{ED83}) may arise 
from interference between current-current and penguin operator contributions, 
while non-vanishing CP-violating effects may be generated in the pure 
penguin-induced decays of class ii), e.g.\ in $B^+\to K^+\overline{K^0}$, 
through interference between penguins with internal up- and charm-quark 
exchanges (see 3.3.4).

In the case of $\bar b\to \bar cc\bar s$ modes, e.g.\ $B^+\to J/\psi\,
K^+$, {\it vanishing} CP violation can be predicted to excellent 
accuracy within the SM because of the arguments given in 3.3.2, where 
the ``gold-plated'' mode $B_d\to J/\psi\,K_{\mbox{{\scriptsize S}}}$ has 
been discussed exhibiting the same decay structure. In general, however, 
the CP-violating asymmetries (\ref{ED84}) suffer from large theoretical 
uncertainties arising in particular from the strong final state 
interaction phases $\alpha_1$ and $\alpha_2$. Therefore CP violation
in charged $B$ decays does in general not allow a clean determination
of CKM phases. The theoretical situation is a bit similar to
Re$(\varepsilon'/\varepsilon)$ discussed in Subsection 1.1, and the major 
goal of a possible future measurement of non-zero CP asymmetries in 
charged $B$ decays is related to the fact that these effects would
immediately rule out ``superweak'' models of CP violation\cite{superweak}.
A detailed discussion of the corresponding calculations, which are rather 
technical, is beyond the scope of this review and the interested reader
is referred to Refs.\cite{rf1}$^-$\cite{kps,gh,siwy}$^-$\cite{remt} where 
further references can be found. 

Concerning theoretical cleanliness, there is, however, an important 
exception. In respect of extracting $\gamma$, charged $B$ decays belonging 
to decay class iii), i.e.\ pure tree decays, play an outstanding role. 
Using certain triangle relations among their decay amplitudes, a 
theoretical clean determination of this angle is possible. 

\runninghead{CP Violation in Non-leptonic $B$-Meson Decays}  
{Relations among Non-leptonic $B$ Decay Amplitudes}
\subsection{Relations among Non-leptonic $B$ Decay Amplitudes}
\noindent
During recent years, relations among amplitudes of non-leptonic $B$ decays
have been very popular to develop strategies for extracting UT angles,
in particular for the ``hard'' angle $\gamma$. There are both {\it exact} 
relations and {\it approximate} relations which are based on the $SU(3)$ 
flavor symmetry of strong interactions and certain plausible dynamical 
assumptions. Let us turn to the ``prototype'' of this approach first. 

\subsubsection{$B\to DK$ Triangles}
\noindent
Applying an appropriate CP phase convention to simplify the following
discussion, the CP eigenstates $|D^0_\pm\rangle$ of the neutral $D$-meson 
system with CP eigenvalues $\pm1$ are given by
\begin{equation}\label{ED85}
\left|D^0_\pm\right\rangle=\frac{1}{\sqrt{2}}\left(\left|D^0\right
\rangle\pm\left|\overline{D^0}\right\rangle\right),
\end{equation}
so that the $B^\pm\to D^0_+K^\pm$ transition amplitudes can be expressed
as\cite{gw}
\begin{eqnarray}
\sqrt{2}A(B^+\to D^0_+K^+)&=&A(B^+\to D^0 K^+)+A(B^+\to\overline{D^0}K^+)
\label{ED86}\\
\sqrt{2}A(B^-\to D^0_+K^-)&=&A(B^-\to\overline{D^0} K^-)+A(B^-\to D^0K^-).
\label{ED87}
\end{eqnarray}
These relations, which are valid {\it exactly}, can be represented as two 
triangles in the complex plane. Taking into account that the $B^+\to DK^+$ 
decays originate from $\bar b\to\bar uc\bar s,\, \bar c u\bar s$ quark-level 
transitions yields
\begin{eqnarray}
A(B^+\to D^0K^+)&=&e^{i\gamma}\lambda\,|V_{cb}|R_b\,|a|\,e^{i\Delta_a}\,=\,
e^{2i\gamma}\,A(B^-\to\overline{D^0}K^-)\label{ED88}\\
A(B^+\to\overline{D^0}K^+)&=&\lambda\,|V_{cb}||A|\,e^{i\Delta_A}\,=\,A(B^-\to 
D^0K^-)\label{ED89},
\end{eqnarray}
where $|a|$, $|A|$ are magnitudes of hadronic matrix elements of the 
current-current operators (\ref{O-cc}) and $\Delta_a$, $\Delta_A$ 
denote the corresponding CP-conserving strong phases. Consequently 
the modes $B^+\to D^0 K^+$ and $B^+\to\overline{D^0}K^+$ exhibit no 
CP-violating effects. However, since the 
requirements for direct CP violation discussed in the previous 
subsection are fulfilled in the $B^\pm\to D^0_+K^\pm$ case because 
of (\ref{ED86}), (\ref{ED87}) and (\ref{ED88}), (\ref{ED89}), we expect
\begin{equation}\label{ED810}
|A(B^+\to D^0_+K^+)|\not=|A(B^-\to D^0_+K^-)|,
\end{equation}
i.e.\ non-vanishing CP violation in that charged $B$ decay. 

\begin{figure}[t]
\centerline{
\epsfxsize=10.8truecm
\epsffile{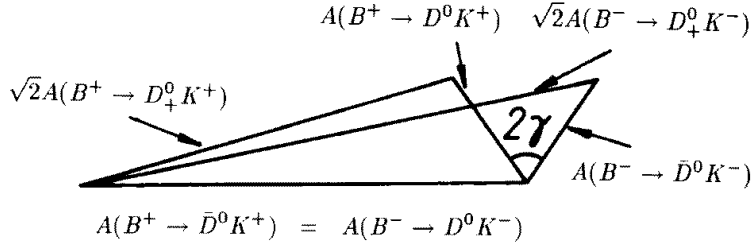}}
\caption{Triangle relations among $B^\pm\to DK^\pm$ decay 
amplitudes.}\label{BDK-triangles}
\end{figure}

Combining all these considerations, we conclude that the triangle
relations (\ref{ED86}) and (\ref{ED87}), which are depicted in 
Fig.~\ref{BDK-triangles}, can be used to extract $\gamma$ by measuring 
only the rates of the corresponding six processes. This approach was
proposed by Gronau and Wyler in Ref.\cite{gw}. It is theoretically clean 
and suffers from no hadronic uncertainties. Unfortunately the 
triangles are expected to be very squashed ones since $B^+\to D^0 K^+$
is both color- and CKM-suppressed with respect to $B^+\to\overline{D^0}K^+$:
\begin{equation}
\frac{|A(B^+\to D^0K^+)|}{|A(B^+\to\overline{D^0}K^+)|}=
R_b\,\frac{|a|}{|A|}\approx0.36\,\frac{a_2}{a_1}\approx0.08.
\end{equation}
Here $a_1$, $a_2$ are the usual phenomenological 
color-factors\cite{BSW,nrsx} satisfying\cite{a1a2-exp} 
$a_2/a_1=0.26\pm0.05\pm0.09$. Using the $SU(3)$ flavor symmetry, 
the corresponding branching ratios can be estimated from the measured 
value\cite{PDG} $(5.3\pm0.5)\cdot10^{-3}$ of 
BR$(B^+\to\overline{D^0}\pi^+)$ to be
BR$(B^+\to\overline{D^0}K^+)\approx4\cdot10^{-4}$ and 
BR$(B^+\to D^0 K^+)\approx2\cdot10^{-6}$.
Another problem is related to the CP eigenstate of the neutral $D$ system.
It is detected through $D^0_+\to\pi^+\pi^-,K^+K^-,\ldots$ and is 
experimentally challenging since the corresponding BR$\times$(detection 
efficiency) is expected to be at most of ${\cal O}(1\%)$. Therefore the 
Gronau-Wyler method\cite{gw} will unfortunately be very difficult from 
an experimental point of view. A feasibility study can be found e.g.\ 
in Ref.\cite{stone}.
  
A variant of the clean determination of $\gamma$ discussed above was 
proposed by Dunietz\cite{isi} and uses the decays $B^0_d\to D^0_+ 
K^{\ast0}$, $B^0_d\to\overline{D^0} K^{\ast0}$, $B^0_d\to D^0 
K^{\ast0}$ and their charge-conjugates. Since these modes 
are ``self-tagging'' through $K^{\ast0}\to K^+\pi^-$, no time-dependent 
measurements are needed in this method although neutral $B_d$ decays
are involved. Compared to the Gronau-Wyler approach\cite{gw}, both 
$B^0_d\to\overline{D^0} K^{\ast0}$ and $B^0_d\to D^0 K^{\ast0}$ are 
color-suppressed, i.e.\
\begin{equation}
\frac{|A(B^0_d\to D^0K^{\ast0})|}{|A(B^0_d\to\overline{D^0}K^{\ast0})|}
\approx R_b\,\frac{a_2}{a_2}\approx0.36.
\end{equation}
Consequently the amplitude triangles are probably not as as squashed as 
in the $B^\pm\to D K^\pm$ case. The corresponding branching ratios are 
expected to be of ${\cal O}(10^{-5})$. Unfortunately one has also to 
deal with the difficulties of detecting the neutral $D$-meson 
CP eigenstate $D^0_+$.

\subsubsection{$SU(3)$ Amplitude Relations}
\noindent
In a series of interesting papers\cite{grl-gam,ghlr}, Gronau, Hern\'andez, 
London and Rosner (GHLR) pointed out that the $SU(3)$ flavor symmetry of 
strong interactions\cite{su3-sym} -- which appeared already several times
in this review -- can be combined with certain plausible dynamical 
assumptions, e.g.\ neglect of annihilation topologies, to derive 
amplitude relations among $B$ decays into $\pi\pi$, $\pi K$ and 
$K\overline{K}$ final states. These relations may allow determinations
both of weak phases of the CKM matrix and of strong final state interaction
phases by measuring {\it only} the corresponding branching ratios. 

In order to illustrate this approach, let me describe briefly 
the ``state of the art'' one had about 3 years ago. At that time it 
was assumed that EW penguins should play a very minor role in non-leptonic 
$B$ decays and consequently their contributions were not taken into account.  
Within that approximation, which will be analyzed very carefully in 
Sections 4 and 5, the decay 
amplitudes for $B\to\{\pi\pi,\pi K,K\overline{K}\}$ transitions can be 
represented in the limit of an exact $SU(3)$ flavor symmetry in terms 
of five reduced matrix elements. This decomposition can also be performed 
in terms of diagrams. At the quark-level one finds six different topologies 
of Feynman diagrams contributing to $B\to\{\pi\pi,\pi K,K\overline{K}\}$ 
that show up in the corresponding decay amplitudes only as five independent 
linear combinations\cite{grl-gam,ghlr}. In contrast to the 
classification of non-leptonic $B$ decays performed in Subsection~2.1, these 
six topologies of Feynman diagrams include also three non-spectator
diagrams, i.e.\ annihilation processes, where the decaying $b$-quark
interacts with its partner anti-quark in the $B$-meson. However, due
to dynamical reasons, these three contributions are expected to be
suppressed relative to the others and hence should play a very minor role. 
Consequently, neglecting these diagrams, \mbox{$6-3=3$} topologies of 
Feynman diagrams suffice to represent the transition amplitudes of $B$ 
decays into $\pi\pi$, $\pi K$ and $K\overline{K}$ final states. To be 
specific, these diagrams describe ``color-allowed'' and ``color-suppressed'' 
current-current processes $T$ $(T')$ and $C$ $(C')$, respectively, and QCD 
penguins $P$ $(P')$. As in Refs.\cite{grl-gam,ghlr} and in 3.4.2, an unprimed 
amplitude denotes strangeness-preserving decays, whereas a primed amplitude 
stands for strangeness-changing transitions. Note that the color-suppressed
topologies $C$ and $C'$ involve the color-suppression 
factor\cite{BSW}$^-$\cite{a1a2-exp} $a_2\approx0.2$.

\begin{figure}[t]
\centerline{
\epsfxsize=12.4truecm
\epsffile{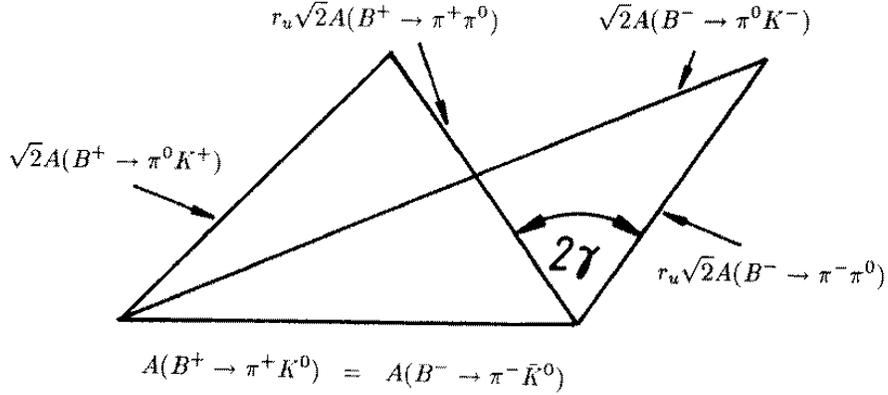}}
\caption{Na\"\i ve $SU(3)$ triangle relations among 
$B^+\to\{\pi^+\pi^0,\pi^+K^0,\pi^0K^+\}$ and charge-conjugate 
decay amplitudes {\it neglecting} EW penguin contributions.}\label{grl-const}
\end{figure}

Let us consider the decays $B^+\to\{\pi^+\pi^0,\pi^+ K^0,\pi^0 K^+\}$, 
i.e.\ the ``original'' GRL method\cite{grl-gam}, as an example.
Neglecting both EW penguins, which will be discussed later, and the 
dynamically suppressed non-spectator contributions mentioned above, the 
decay amplitudes of these modes can be expressed as
\begin{equation}\label{ED1201}
\begin{array}{rcl}
\sqrt{2}\,A(B^+\to\pi^+\pi^0)&=&-\left(T+C\right)\\
A(B^+\to\pi^+ K^0)&=&P'\\
\sqrt{2}\,A(B^+\to\pi^0 K^+)&=&-\left(T'+C'+P'\right)
\end{array}
\end{equation}
with
\begin{equation}\label{ED1202}
T=|T|\,e^{i\gamma}\,e^{i\delta_T},\quad
C=|C|\,e^{i\gamma}\,e^{i\delta_C}.
\end{equation}
Here $\delta_T$ and $\delta_C$ denote CP-conserving strong phases. 
Using the $SU(3)$ flavor symmetry, the strangeness-changing amplitudes 
$T'$ and $C'$ can be obtained easily from the strangeness-preserving ones 
through
\begin{equation}\label{ED1203}
\frac{T'}{T}\approx\frac{C'}{C}\approx\lambda\frac{f_K}{f_\pi}\equiv r_u,
\end{equation}
where $f_K$ and $f_\pi$ take into account factorizable $SU(3)$-breaking
corrections as in (\ref{EE7}). The structures of the $\bar b\to\bar d$
and $\bar b\to\bar s$ QCD penguin amplitudes $P$ and $P'$ corresponding 
to $P^{(d)}$ and $P^{(s)}$ (see (\ref{bdpenamp}) and (\ref{bspenamp})),
respectively, have been discussed in 3.3.4. It is an easy exercise 
to combine the decay amplitudes given in (\ref{ED1201}) appropriately to 
derive the relations
\begin{eqnarray}
\sqrt{2}\,A(B^+\to\pi^0 K^+)+A(B^+\to\pi^+
K^0)&=&r_u\sqrt{2}\,A(B^+\to\pi^+\pi^0)\label{ED1204a}\\
\sqrt{2}\,A(B^-\to\pi^0 K^-)+A(B^-\to\pi^-
\overline{K^0})&=&r_u\sqrt{2}\,A(B^-\to\pi^-\pi^0),\label{ED1204b}
\end{eqnarray}
which can be represented as two triangles in the complex plane. If one
measures the rates of the corresponding six decays, these triangles can
easily be constructed. Their relative orientation is fixed through
$A(B^+\to\pi^+ K^0)=A(B^-\to\pi^-\overline{K^0})$, which is due to the fact 
that there is no non-trivial CP-violating weak phase present in the 
$\bar b\to\bar s$ QCD penguin amplitude governing $B^+\to\pi^+ K^0$
as we have seen in 3.3.4. Taking into account moreover (\ref{ED1202}), 
we conclude that these triangles should allow a determination of $\gamma$ 
as can be seen in Fig.~\ref{grl-const}. From the geometrical point of view, 
that GRL approach\cite{grl-gam} is very similar to the $B^\pm\to DK^\pm$ 
construction\cite{gw} shown in Fig.~\ref{BDK-triangles}.
Furthermore it involves also only charged $B$ decays and therefore neither 
time-dependent measurements nor tagging are required. In comparison 
with the Gronau-Wyler method\cite{gw}, at first sight the major 
advantage of the GRL strategy seems to be that all branching 
ratios are expected to be of the same order of magnitude ${\cal O}(10^{-5})$, 
i.e.\ the corresponding triangles are not squashed ones, and that the 
difficult to measure CP eigenstate $D^0_+$ is not required.

However, things are unfortunately not that simple and -- despite of its 
attractiveness -- the general GHLR approach\cite{grl-gam,ghlr} to extract 
CKM phases from $SU(3)$ amplitude relations suffers from theoretical 
limitations. The most obvious limitation is of course related to the fact 
that the relations are not, as e.g.\ (\ref{ED86}) or (\ref{ED87}), valid 
exactly but suffer from $SU(3)$-breaking corrections\cite{ghlrsu3}. 
While factorizable $SU(3)$-breaking can be included straightforwardly
through certain meson decay constants or form factors, non-factorizable
$SU(3)$-breaking corrections cannot be described in a reliable quantitative
way at present. Another limitation is related to $\bar b\to\bar d$ QCD 
penguin topologies with internal up- and charm-quark exchanges which may 
affect the simple relation (\ref{bdpenapprox}) between $\beta$ and the 
$\bar b\to\bar d$ QCD penguin amplitude $P$ significantly as we have seen 
in 3.3.4. Consequently these contributions may preclude reliable extractions 
of $\beta$ using $SU(3)$ 
amplitude relations and the assumption that $\bar b\to\bar d$ QCD penguin
amplitudes are dominated by internal top-quark exchanges\cite{bf1} (see 
also 3.3.6). Remarkably also EW penguins\cite{rfewp1,rfewp2,rfewp3}, which 
we have neglected in our discussion of $SU(3)$ amplitude relations so far, 
have a very important impact on some $SU(3)$ constructions, in particular 
on the GRL method\cite{grl-gam} of determining $\gamma$. As we will see in 
Section~5, this approach is even {\it spoiled} by these 
contributions\cite{dh,ghlrewp}. However, there are other -- generally more 
involved -- $SU(3)$ methods\cite{ghlrewp}$^-$\cite{PAPIII} that are 
not affected by EW penguins. Interestingly it is in principle also possible 
to shed light on the physics of these operators by using $SU(3)$ amplitude 
relations\cite{PAPIII,PAPI}. This issue has been one of the ``hot topics'' 
in $B$ physics over the last few years and will be the subject of the 
remainder of this review. Before we shall investigate the role of EW penguins 
in methods for extracting angles of the UT in Section~5, let us in the 
following section have a closer look at a few non-leptonic $B$ decays that 
are affected significantly by EW penguin operators.

\begin{figure}[t]
\centerline{
\rotate[r]{
\epsfysize=8.5truecm
\epsffile{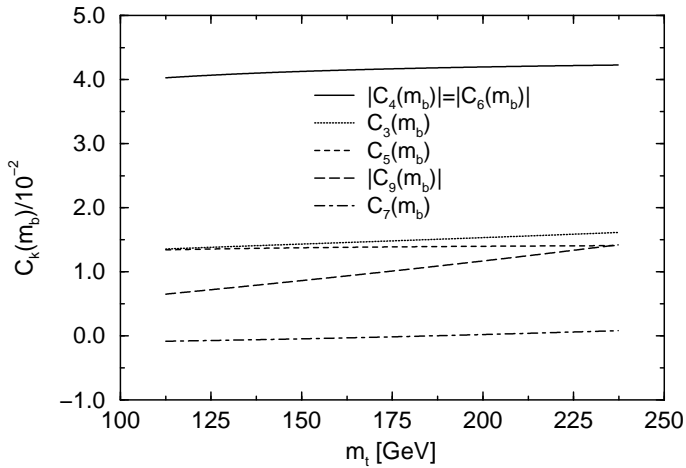}}}
\caption{The dependence of the Wilson coefficients 
(\ref{e636}) on the top-quark mass $m_t$ for $\mu=m_b$ and  
$\Lambda_{\overline{\mbox{{\scriptsize MS}}}}=0.3\,
\mbox{GeV}$ corresponding to four active quark flavors.}\label{fig-clomt}
\end{figure}

\runninghead{Electroweak Penguin Effects in Non-leptonic $B$-Meson Decays}
{Electroweak Penguin Effects in Non-leptonic $B$-Meson Decays}
\section{Electroweak Penguin Effects in Non-leptonic $B$-Meson Decays}
\noindent
Since the ratio $\alpha/\alpha_s={\cal O}(10^{-2})$ of the QED and QCD
couplings is very small, one would expect that EW penguins should only
play a minor role in comparison with QCD penguins. That would indeed be
the case if the top-quark was not ``heavy''. However, the Wilson coefficient 
of one EW penguin operator -- the operator $Q_9$ specified in 
(\ref{ew-def}) -- increases strongly with the top-quark mass as can be 
seen nicely in Fig.~\ref{fig-clomt}. There the $m_t$-dependence of the 
coefficients (\ref{e636}), which correspond to the case where the
proper renormalization group evolution from $\mu={\cal O}(M_W)$ down to
$\mu={\cal O}(m_b)$ has been neglected, is shown. A very similar behavior 
is also exhibited by the NLO Wilson coefficients\cite{bbl-rev}. 
Consequently interesting EW penguin effects may arise from this feature 
in certain non-leptonic $B$ decays because of the large top-quark 
mass that has been measured\cite{CDF,D0} recently with impressive 
accuracy\cite{mtop} by the CDF and D0 collaborations to be 
$m_t^{\mbox{{\scriptsize Pole}}}=(175\pm6)\,\mbox{GeV}$. The parameter 
$m_t$ used in analyses of non-leptonic weak decays is, however, not equal 
to that measured ``pole'' mass\cite{buras-ichep96}. In NLO calculations, 
$m_t$ refers to the running top-quark current-mass normalized at the 
scale $\mu=m_t$, i.e.\ $\overline{m}_t(m_t)$, which is typically by 
$8\,\mbox{GeV}$ smaller than $m_t^{\mbox{{\scriptsize Pole}}}$ for 
$m_t={\cal O}(170\,\mbox{GeV})$. The EW penguin effects discussed in the 
following subsections were pointed out first in 
Refs.\cite{rfewp1,rfewp2,rfewp3}. Meanwhile they were confirmed
by several other authors\cite{kp,ghlrewp,dh-ewp}$^-$\cite{dy}.

\begin{figure}[t]
\centerline{
\rotate[r]{
\epsfysize=8.8truecm
\epsffile{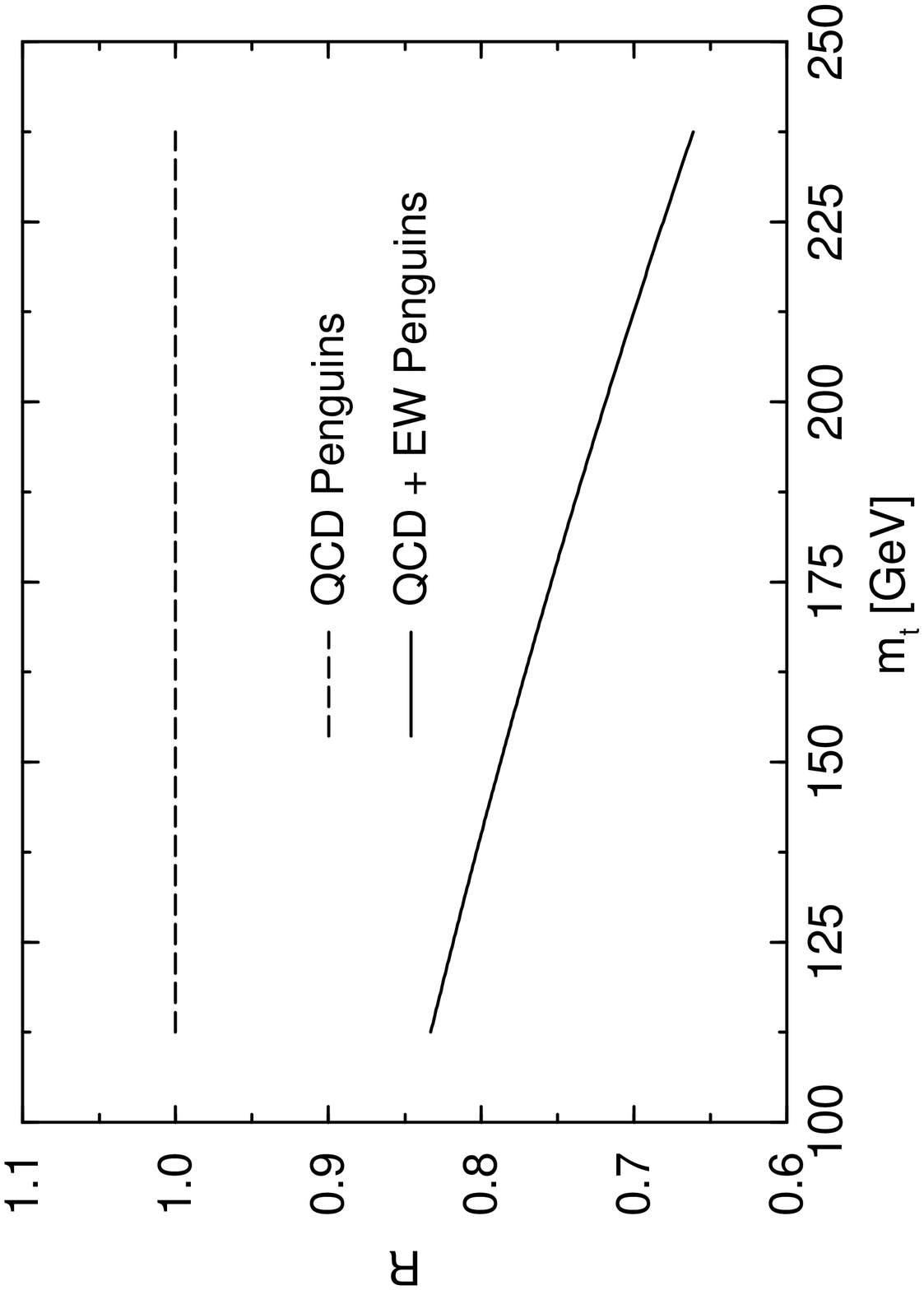}}}
\caption{The dependence of the ratio 
$R\propto\mbox{BR}(B^+\to K^+\phi)/\mbox{BR}(B^+\to\pi^+K^{\ast0})$ 
on $m_t$ using the Wilson coefficients (\ref{e636}).}\label{fig-Rmt}
\end{figure}

\runninghead{Electroweak Penguin Effects in Non-leptonic $B$-Meson Decays}
{EW Penguin Effects in $B^+\to K^+\phi$ and $B^+\to\pi^+
K^{\ast0}$}
\subsection{EW Penguin Effects in $B^+\to K^+\phi$ and $B^+\to\pi^+
K^{\ast0}$}
\noindent
The channels $B^+\to K^+\phi$ and $B^+\to\pi^+K^{\ast0}$ originating from 
the penguin-induced $\bar b$-quark decays $\bar b\to \bar ss\bar s$ 
and $\bar b\to\bar s d\bar d$, respectively, are very similar from a 
QCD point of view, i.e.\ as far as their QCD penguin contributions are 
concerned. This feature is obvious if one draws the corresponding 
Feynman diagrams which is an easy exercise. However, an important 
difference arises in respect of EW penguin contributions. We have to deal 
both with small color-suppressed and sizable color-allowed EW penguin 
diagrams. Whereas the former contributions are again very similar for 
$B^+\to K^+\phi$ and $B^+\to\pi^+K^{\ast0}$, the color-allowed EW penguin 
contributions are absent in the $B^+\to\pi^+K^{\ast0}$ case and contribute 
only to $B^+\to K^+\phi$. Consequently significant EW penguin 
effects\cite{rfewp1} are expected in the mode $B^+\to K^+\phi$, 
while these effects should be negligible in the decay 
$B^+\to\pi^+K^{\ast0}$. 

This rather qualitative kind of reasoning is in agreement with the results
of certain model calculations\cite{rfewp1,dh-ewp}, where the formalism 
scetched in Section~2 was applied in combination with the ``factorization'' 
hypothesis\cite{fact1,fact2}. By factorization one means that the 
hadronic matrix elements of the four-quark operators appearing in the
low energy effective Hamiltonian (\ref{LEham}) are factorized into the
product of hadronic matrix elements of two quark-currents that are 
described by a set of form factors. Usually the model proposed by 
Bauer, Stech and Wirbel\cite{BSW} (BSW) is used for these form factors and
was also applied in Refs.\cite{rfewp1,dh-ewp}. In contrast to color-allowed
current-current processes, where ``factorization'' may work reasonably 
well\cite{bjorken}, this assumption is questionable for penguin processes 
which are classical examples of non-factorizable diagrams. Nevertheless this 
approach may give us a feeling for the expected orders of magnitudes. 
Unfortunately a more reliable analytical way of dealing with non-leptonic 
$B$ decays is not available at present. 

The corresponding calculations are quite complicated and a discussion of
their technicalities would not be useful in the context of this review. 
Let me therefore just briefly discuss the main results. The model 
calculations indicate that EW penguins lead to a reduction of 
BR$(B^+\to K^+\phi)$ by ${\cal O}(30\%)$ for $m_t={\cal O}(170\,
\mbox{GeV})$, while these effects are below $2\%$ in the case of 
BR$(B^+\to\pi^+K^{\ast0})$. As in Fig.~\ref{delp-fig}, the 
branching ratios, which are both of ${\cal O}(10^{-5})$, depend strongly 
on $k^2$, the four-momentum of the gluons and photons appearing in the 
penguin diagram depicted in Fig.~\ref{pen-me}. This ``unphysical'' 
$k^2$-dependence\cite{gh} is due to the use of the above mentioned model.
In order to reduce this dependence as well as other hadronic uncertainties, 
the ratio\cite{rfewp1}
\begin{eqnarray}
\lefteqn{R\equiv\left[\frac{f_{K^{\ast}}F_{B\pi}(M_{K^\ast}^{2};1^{-})}
{f_{\phi}F_{BK}(M_{\phi}^{2};1^{-})}\right]^{2}
\left[\frac{\Phi(M_{\pi}/M_{B},M_{K^{\ast}}/M_{B})}
{\Phi(M_{K}/M_{B},M_{\phi}/M_{B})}
\right]^{3}}\nonumber\\
&&\times\left[\frac{\mbox{BR}(B^{+}\to K^{+}\phi)}{\mbox{BR}(B^{+}\to 
\pi^{+}K^{\ast0})}\right]\approx0.5\times
\left[\frac{\mbox{BR}(B^{+}\to K^{+}\phi)}{\mbox{BR}(B^{+}\to \pi^{+}
K^{\ast0})}\right],\label{e919}
\end{eqnarray}
where $f_V$ are meson decay constants, $F_{PP'}$ are quark-current form
factors and $\Phi(x,y)$ is the usual two-body phase space function,
turns out to be very useful. Although $R$ is affected in almost the same 
way by EW penguins as the branching ratio BR$(B^{+}\to K^{+}\phi)$, it 
suffers much less from hadronic uncertainties, is very stable against 
variations both of the momentum transfer $k^{2}/m_{b}^{2}$ and of the QCD 
scale parameter $\Lambda_{\overline{\mbox{{\scriptsize MS}}}}$, and
does not depend on CKM factors if the ${\cal O}(\lambda^2)$ terms in 
(\ref{CKMs}) are neglected. These terms play a minor role and may lead to 
tiny direct CP-violating asymmetries of ${\cal O}(1\%)$. One should keep
in mind, however, that 
BR$(B^{+}\to K^{+}\phi)$ and BR$(B^{+}\to\pi^{+} K^{\ast0})$ could 
receive quite different contributions in principle if ``factorization'' does 
not hold. Therefore $R$ could be affected by such unknown corrections.

The effect of the NLO renormalization group evolution from 
$\mu={\cal O}(M_W)$ down to $\mu={\cal O}(m_b)$, i.e.\ the difference 
between using exact NLO Wilson coefficients\cite{bbl-rev} or (\ref{e636}),  
is rather small and gives an enhancement of the branching ratios by
about ${\cal O}(10\%)$ and an even smaller enhancement in the case of $R$. 
Hence the use of the approximate
Wilson coefficients (\ref{e636}) is justified. The effects of the EW penguins
can be seen nicely in Fig.~\ref{fig-Rmt}, where the top-quark mass
dependence of $R$ calculated with the help of these coefficients is  
shown. Whereas the dashed line corresponds to the case where only QCD
penguins are included, the solid line describes the calculation taking
into account both QCD and EW penguin operators. 

There are not only some non-leptonic $B$ decays that are affected 
significantly by EW penguins. There are even a few channels where the 
corresponding operators may play the {\it dominant} role as we will 
see in the following two subsections. 

\runninghead{Electroweak Penguin Effects in Non-leptonic $B$-Meson Decays}
{EW Penguin Effects in $B^+\to\pi^+\phi$}
\subsection{EW Penguin Effects in $B^+\to\pi^+\phi$}
\noindent
In respect of EW penguin effects, the mode $B^+\to\pi^+\phi$ is also quite
interesting\cite{rfewp2}. Within the spectator model, it originates from the 
penguin-induced $\bar b$-quark decay $\bar b\to\bar d s\bar s$, where the 
$s\bar s$ pair hadronizes into the $\phi$-meson which is present in a 
color-singlet state. The $s$- and $\bar s$-quarks emerging from the gluons 
of the usual QCD penguin diagrams form, however, a color-octet state and 
consequently cannot build up that $\phi$-meson (see also 3.3.2). Thus, 
using both an appropriate NLO low energy effective Hamiltonian and the BSW 
model in combination with the factorization assumption to estimate the 
relevant hadronic matrix elements of the QCD penguin operators, 
one finds a very small branching ratio 
BR$\left.(B^+\to\pi^+\phi)\right|_{\mbox{{\scriptsize QCD}}}={\cal 
O}(10^{-10})$. The non-vanishing result is due to the renormalization group 
evolution from $\mu={\cal O}(M_W)$ down to $\mu={\cal O}(m_b)$. Neglecting 
this evolution, i.e.\ applying the approximate Wilson coefficients 
(\ref{e636}), would give a vanishing branching ratio because of the 
color-arguments given above. Since these arguments do not apply to EW 
penguins, their contributions are expected to become important\cite{rfewp2}. 
In fact, taking into account also these operators gives a branching ratio 
BR$\left.(B^+\to\pi^+\phi)\right|_{\mbox{{\scriptsize QCD+EW}}}={\cal 
O}(10^{-8})$ for $m_t={\cal O}(170\,\mbox{GeV})$ that increases strongly 
with the top-quark mass. Unfortunately the enhancement by a factor of 
${\cal O}(10^2)$ through EW penguins is not strong enough to make the 
decay $B^+\to\pi^+\phi$ measurable in the foreseeable future.
 
The color-arguments for the QCD penguins may be affected by additional 
soft gluon exchanges which are not under quantitative 
control at present. These contributions would show up as non-factorizable 
contributions to the hadronic matrix elements of the penguin operators 
which were neglected in Ref.\cite{rfewp2}. Nevertheless there is no 
doubt that EW penguins play a very important -- probably even dominant -- 
role in the decay $B^+\to\pi^+\phi$ and related modes like $B^+\to\rho^+\phi$.

\runninghead{Electroweak Penguin Effects in Non-leptonic $B$-Meson Decays}
{EW Penguin Effects in $B_s\to\pi^0\phi$}
\subsection{EW Penguin Effects in $B_s\to\pi^0\phi$}
\noindent
The theoretical situation arising in the decay $B_s^0\to\pi^0\phi$ caused by
$\bar b\to\bar s\,(u\bar u,d\bar d)$ quark-level transitions is much more 
favorable than in the previous two subsections because of the $SU(2)$ 
isospin symmetry of strong interactions. Let me therefore be more detailed 
in the presentation of that transition which is expected to be dominated 
by EW penguins\cite{rfewp3}. In contrast to the decays discussed in 4.1 
and 4.2, it receives not only penguin but also current-current operator 
contributions at the tree level. The final state is an eigenstate of the CP 
operator with eigenvalue +1 and has strong isospin quantum numbers 
$(I,I_3)=(1,0)$, whereas the initial state is an isospin singlet. Thus we 
have to deal with a $\Delta I=1$ transition. 

Looking at the operator basis given in (\ref{cc-def})-(\ref{ew-def}),
we observe that the current-current operators $Q^{us}_{1/2}$ and the
EW penguin operators can lead to final states both with isospin $I=0$ 
and $I=1$, whereas the QCD penguin operators give only final states 
with $I=0$. Therefore the $\Delta I=1$ transition $B_s\to\pi^0\phi$ 
receives no QCD penguin contributions and arises purely from the 
current-current operators $Q^{us}_{1/2}$  and the EW penguin operators. 
For the same reason, QCD penguin matrix elements of the current-current 
operators $Q_{2}^{us}$ and $Q_{2}^{cs}$ (see (\ref{e630})) with up- and 
charm-quarks running as virtual particles in the loops, respectively, 
do not contribute to that decay.
Consequently, using in addition the unitarity of the CKM matrix and
applying the modified Wolfenstein parametrization (\ref{wolf2}) yielding
\begin{equation}\label{ED1105}
V_{us}^{\ast}V_{ub}=\lambda|V_{ub}|\,e^{-i\gamma},\quad
V_{ts}^{\ast}V_{tb}=-|V_{ts}|=-|V_{cb}|(1+{\cal O}(\lambda^2)),
\end{equation}
the hadronic matrix element of the Hamiltonian (\ref{LEham}) can be
expressed as
\begin{eqnarray}
\lefteqn{\Bigl\langle\pi^0\phi\Bigl|{\cal H}_{\mbox{{\scriptsize eff}}}
(\Delta B=-1)\Bigr|\overline{B^0_s}\Bigr\rangle=
\frac{G_{\mbox{{\scriptsize F}}}}{\sqrt{2}}\,|V_{ts}|}\label{ED1106}\\
&&\times\left[\lambda^2R_b\,e^{-i\gamma}
\sum\limits_{k=1}^2\Bigl\langle\pi^{0}\phi\Bigl|
Q_k^{us}(\mu)\Bigr|\overline{B^0_s}\Bigr\rangle C_k(\mu)
+\sum\limits_{k=7}^{10}\Bigl\langle\pi^0\phi\Bigl|
Q_k^s(\mu)\Bigr|\overline{B^0_s}\Bigr\rangle C_k(\mu)\right],\nonumber
\end{eqnarray}
where the correction of  ${\cal O}(\lambda^{2})$ in (\ref{ED1105}) has
been omitted.

Neglecting EW penguin operators for a moment and applying the 
formalism developed in 3.2.2, we would find ${\cal A}_{\mbox{{\scriptsize 
CP}}}^{\mbox{{\scriptsize mix-ind}}}(B_s\to\pi^0\phi)=\sin(2\gamma)$.
The approximation of neglecting EW penguin operator contributions to 
$B_s\to\pi^0\phi$ is, however, very bad since the current-current 
amplitude $A_{\mbox{{\scriptsize CC}}}$ is suppressed relative to the
EW penguin part $A_{\mbox{{\scriptsize EW}}}$ by the CKM factor
$\lambda^2R_b\approx0.02$. Moreover the current-current operator
contribution is color-suppressed by $a_2\approx0.2$. On the other hand, 
in the presence of a heavy top-quark, the Wilson coefficient of the dominant
EW penguin operator $Q_9^s$ contributing to $B_s\to\pi^0\phi$
in color-allowed form is of ${\cal O}(10^{-2})$ (see Fig.~\ref{fig-clomt}). 
Therefore we expect 
$|A_{\mbox{{\scriptsize EW}}}|/|A_{\mbox{{\scriptsize CC}}}|
={\cal O}(10^{-2}/(0.02\cdot0.2))={\cal O}(2.5)$ and conclude that EW penguins
have not only to be taken into account in an analysis of $B_s\to\pi^0\phi$
but should even give the {\it dominant} contribution to that channel.

In order to simplify the following discussion, let us neglect the 
influence of QCD corrections to EW penguins for a moment. Then the Wilson 
coefficients of the corresponding operators are given by 
$\overline{C}^{(0)}_k(\mu)$ $(k\in\{7,\ldots,10\})$ specified in (\ref{e636}).
Since $\overline{C}^{(0)}_8(\mu)$ and $\overline{C}^{(0)}_{10}(\mu)$
vanish, we have to consider only hadronic matrix elements of the operators
$Q_{1/2}^{us}$, $Q_7^s$ and of the dominant EW penguin operator $Q_9^s$. 
For the evaluation of the hadronic matrix elements, it is convenient to 
perform a Fierz transformation of the current-current operators and to 
consider
\begin{eqnarray}
Q_{1}^{us}&=&(\bar uu)_{\mbox{{\scriptsize V--A}}}(\bar 
sb)_{\mbox{{\scriptsize V--A}}}
\,=\, {\cal Q}^{I=0}_{\mbox{{\scriptsize V--A}}}+{\cal 
Q}^{I=1}_{\mbox{{\scriptsize V--A}}}\nonumber\\
Q_{2}^{us}&=&(\bar u_{\alpha}u_{\beta})_{\mbox{{\scriptsize V--A}}}
(\bar s_{\beta}b_{\alpha})_{\mbox{{\scriptsize V--A}}}\,=\,
\tilde{\cal Q}^{I=0}_{\mbox{{\scriptsize V--A}}} +\tilde{\cal 
Q}^{I=1}_{\mbox{{\scriptsize V--A}}}\nonumber\\
Q_7^s\,\,&=&\left[(\bar uu)_{\mbox{{\scriptsize V+A}}}-
\frac{1}{2}(\bar dd)_{\mbox{{\scriptsize 
V+A}}}\right](\bar sb)_{\mbox{{\scriptsize V--A}}}\,=\, 
\frac{1}{2}{\cal Q}^{I=0}_{\mbox{{\scriptsize V+A}}}+
\frac{3}{2}{\cal Q}^{I=1}_{\mbox{{\scriptsize V+A}}}\label{ED1109}\\
Q_9^s\,\,&=&\left[\,(\bar uu)_{\mbox{{\scriptsize V--A}}}\,-\,
\frac{1}{2}(\bar dd)_{\mbox{{\scriptsize V--A}}}\right](\bar
sb)_{\mbox{{\scriptsize V--A}}}\,=\, \frac{1}{2}{\cal 
Q}^{I=0}_{\mbox{{\scriptsize V--A}}}+
\frac{3}{2}{\cal Q}^{I=1}_{\mbox{{\scriptsize V--A}}}.\nonumber
\end{eqnarray}
Here the parts of $Q_7^s$ and $Q_9^s$ with quark flavors $(\bar ss)
(\bar sb)$ and $(\bar cc)(\bar sb)$, which do not contribute
to $\overline{B_s^0}\to\pi^0\phi$, have been neglected and the following 
isospin operators have been introduced:
\begin{eqnarray}
{\cal Q}^{I=0}_{\mbox{{\scriptsize V$\pm$A}}}&=&\frac{1}{2}\left[(\bar 
uu)_{\mbox{{\scriptsize V$\pm$A}}}+(\bar 
dd)_{\mbox{{\scriptsize V$\pm$A}}}\right](\bar sb)_{\mbox{{\scriptsize 
V--A}}}\nonumber\\
{\cal Q}^{I=1}_{\mbox{{\scriptsize V$\pm$A}}}&=&\frac{1}{2}
\left[(\bar uu)_{\mbox{{\scriptsize V$\pm$A}}}-
(\bar dd)_{\mbox{{\scriptsize V$\pm$A}}}\right]
(\bar sb)_{\mbox{{\scriptsize V--A}}}\nonumber\\
\tilde{\cal Q}^{I=0}_{\mbox{{\scriptsize V$\pm$A}}}&=&\frac{1}{2}\left[(\bar 
u_{\alpha}u_{\beta})_{\mbox{{\scriptsize V$\pm$A}}}+(\bar
d_{\alpha}d_{\beta})_{\mbox{{\scriptsize V$\pm$A}}}\right]
(\bar s_{\beta}b_{\alpha})_{\mbox{{\scriptsize V--A}}}\label{ED1110}\\
\tilde{\cal Q}^{I=1}_{\mbox{{\scriptsize V$\pm$A}}}&=&\frac{1}{2}\left[(\bar 
u_{\alpha}u_{\beta})_{\mbox{{\scriptsize V$\pm$A}}}-(\bar
d_{\alpha}d_{\beta})_{\mbox{{\scriptsize V$\pm$A}}}\right](\bar 
s_{\beta}b_{\alpha})_{\mbox{{\scriptsize V--A}}}.\nonumber
\end{eqnarray}
Taking into account that $B_s\to\pi^0\phi$ is a $\Delta I=1$ transition and
employing non-perturbative ``$B$-parameters'' to parametrize the
hadronic matrix elements yields
\begin{eqnarray}
\Bigl\langle\pi^0\phi\Bigl|Q_1^{us}(\mu)\Bigr|\overline{B_s^0}\Bigr\rangle&=&
\Bigr\langle\pi^0\phi\Bigl|{\cal Q}^{I=1}_{\mbox{{\scriptsize V--A}}}(\mu)
\Bigr|\overline{B_s^0}\Bigr\rangle\,=\,B_{\mbox{{\scriptsize V--A}}}(\mu)\,h
\,=\,\frac{2}{3}\Bigl\langle\pi^0\phi\Bigl|Q_9^s(\mu)\Bigr|\overline{B_s^0}
\Bigr\rangle\nonumber\\
\Bigl\langle\pi^0\phi\Bigl|Q_2^{us}(\mu)\Bigr|\overline{B_s^0}\Bigr\rangle&=&
\Bigl\langle\pi^0\phi\Bigl|\tilde{\cal Q}^{I=1}_{\mbox{{\scriptsize V--A}}}
(\mu)\Bigr|\overline{B_s^0}\Bigr\rangle\,=\,\frac{1}{3}\tilde 
B_{\mbox{{\scriptsize V--A}}}(\mu)\,h\label{ED1111}\\
\Bigl\langle\pi^0\phi\Bigl|Q_7^s(\mu)\Bigr|\overline{B_s^0}\Bigr\rangle&=&
\frac{3}{2}\Bigl\langle\pi^0\phi\Bigl|{\cal Q}^{I=1}_{\mbox{{\scriptsize 
V+A}}}(\mu)\Bigr|\overline{B_s^0}\Bigr\rangle=\frac{3}{2}
B_{\mbox{{\scriptsize V+A}}}(\mu)\,h,\nonumber
\end{eqnarray}
where $h$ corresponds to the ``factorized'' matrix element $\bigl\langle
\pi^0\phi\bigl|Q_1^{us}\bigr|\overline{B_s^0}\bigr\rangle$:
\begin{equation}\label{ED1112}
h=i\,\frac{f_{\pi}}{\sqrt{2}}\,i\,2\,M_{\phi}\,
F_{B_s\phi}(M_{\pi}^2;0^-)\,
(\varepsilon_{\phi}\cdot p_{B_s}).
\end{equation}

Since the $\pi^0$-meson is a pseudoscalar particle and emerges from the 
axial vector parts of the quark-currents 
$\left[(\bar uu)_{\mbox{{\scriptsize V$\pm$A}}}-
(\bar dd)_{\mbox{{\scriptsize V$\pm$A}}}\right]$, it is quite natural to 
assume
\begin{equation}\label{ED1114}
B_{\mbox{{\scriptsize V+A}}}(\mu)\approx-B_{\mbox{{\scriptsize V--A}}}(\mu).
\end{equation}
For a similar reason, the one-loop QED penguin matrix elements of the
current-current operators $Q_{1/2}^{us}$ and $Q_{1/2}^{cs}$, which have to
be taken into account in order to have a consistent calculation
(see Subsection~2.3), vanish, since the virtual photons appearing in 
the QED penguin diagrams generate $(\bar uu)_{\mbox{{\scriptsize V}}}$ and
$(\bar dd)_{\mbox{{\scriptsize V}}}$ vector currents that cannot
create the pseudoscalar $\pi^0$-meson. Consequently we obtain
\begin{eqnarray}
\lefteqn{\Bigl\langle\pi^0\phi\Bigl|{\cal H}_{\mbox{{\scriptsize eff}}}
(\Delta B=-1)\Bigr|\overline{B^0_s}\Bigr\rangle=\frac{G_{\mbox{{\scriptsize 
F}}}}{\sqrt{2}}\,|V_{ts}|\,B_{\mbox{{\scriptsize V--A}}}(\mu)\,h}
\label{ED1115}\\
&&\times\left[\lambda^2R_b\,e^{-i\gamma}\left\{C_1(\mu)+\frac{1}{3}
\frac{\tilde B_{\mbox{{\scriptsize V--A}}}(\mu)}{B_{\mbox{{\scriptsize 
V--A}}}(\mu)}C_2(\mu)\right\}+\frac{3}{2}\left\{\overline{C_9}^{(0)}(\mu)-
\overline{C_7}^{(0)}(\mu)\right\}\right],\nonumber
\end{eqnarray}
so that the CP-violating observables can be expressed as 
\begin{equation}\label{ED1116}
{\cal A}_{\mbox{{\scriptsize CP}}}^{\mbox{{\scriptsize dir}}}
(B_s\to\pi^0\phi)=0,\quad
{\cal A}_{\mbox{{\scriptsize CP}}}^{\mbox{{\scriptsize mix-ind}}}
(B_s\to\pi^0\phi)=\frac{2\,(x+\cos\gamma)\sin\gamma}{x^2+2\,x\cos\gamma+1},
\end{equation}
while the branching ratio BR$(B_s\to\pi^0\phi)$ takes the form 
\begin{figure}[t]
\centerline{
\rotate[r]{
\epsfysize=8truecm
\epsffile{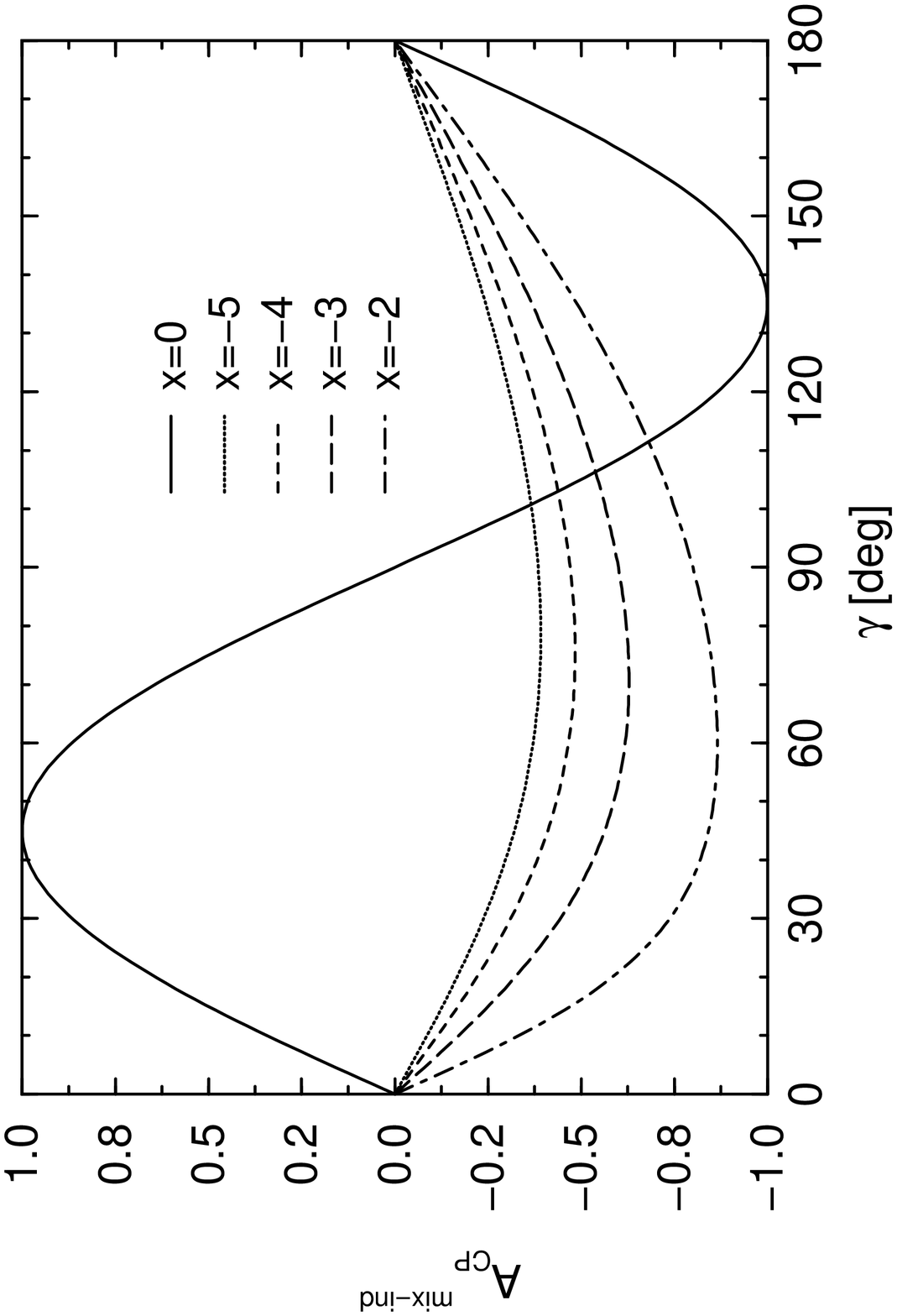}}}
\caption{The dependence of ${\cal A}_{\mbox{{\scriptsize 
CP}}}^{\mbox{{\scriptsize mix-ind}}}(B_s\to\pi^0\phi)$ on 
$\gamma$ for various values of $x$.}\label{fig-Acp}
\end{figure}
\begin{equation}\label{ED1117}
{\cal R}\equiv\frac{\mbox{BR}(B_s\to\pi^0\phi)}{\mbox{BR}_{\mbox{{\scriptsize
CC}}}(B_s\to\pi^0\phi)}=x^2+2\,x\cos\gamma+1,
\end{equation}
where $\mbox{BR}_{\mbox{{\scriptsize CC}}}(B_s\to\pi^0\phi)={\cal O}
(10^{-8})$ denotes the current-current branching ratio. In these equations, 
$x$ describes the ratio of the contribution of the EW penguin operators to 
that of the current-current operators:
\begin{equation}\label{ED1118}
x\equiv\frac{A_{\mbox{{\scriptsize EW}}}}{A_{\mbox{{\scriptsize CC}}}}=
\frac{3\left[\overline{C}_9^{(0)}(\mu)-\overline{C}_7^{(0)}(\mu)
\right]}{2\lambda^2R_b\left[C_1(\mu)+\frac{1}{3}\frac{\tilde
B_{\mbox{{\scriptsize V--A}}}(\mu)}{B_{\mbox{{\scriptsize V--A}}}(\mu)}
C_2(\mu)\right]}\,.
\end{equation}
Note that deviations from the relation (\ref{ED1114}) would only lead to
very small corrections to (\ref{ED1118}) since $\overline{C}_7^{(0)}(\mu)$ 
is suppressed relative to $\overline{C}_9^{(0)}(\mu)$ by a factor of
${\cal O}(10^{-2})$. To eliminate the hadronic uncertainties in 
(\ref{ED1118}), we identify the combination of the Wilson coefficient 
functions $C_{1/2}(\mu)$ and the corresponding $B$-parameters with the 
phenomenological color-suppression factor\cite{BSW}$^-$\cite{a1a2-exp}
\begin{equation}\label{ED1119}
a_2\approx C_1(\mu)+\frac{1}{3}\frac{\tilde B_{\mbox{{\scriptsize V--A}}}
(\mu)}{B_{\mbox{{\scriptsize V--A}}}(\mu)}C_2(\mu).
\end{equation}
Applying the analytical expressions for the Wilson coefficients
$\overline{C}_7^{(0)}(\mu)$ and $\overline{C}_9^{(0)}(\mu)$ given in
(\ref{e636}), their $\mu$-dependences cancel explicitly so that we arrive 
at the $\mu$-independent expression
\begin{equation}\label{ED1121}
x\approx\frac{\alpha}{2\pi\lambda^2R_b\,a_2\sin^2\Theta_{\mbox{{\scriptsize
W}}}}\bigl[5B(x_t)-2C(x_t)\bigr],
\end{equation}
where the Inami-Lim functions\cite{il} $B(x_t)$ and $C(x_t)$ are
given in (\ref{e637}) and describe box diagrams and $Z$ penguins, 
respectively. The photon penguin contributions to 
$\overline{C}^{(0)}_7(\mu)$ and $\overline{C}^{(0)}_9(\mu)$ 
described by the Inami-Lim function $D(x_t)$ cancel in (\ref{ED1121})
because of the remark after (\ref{ED1114}).

\begin{figure}
\centerline{
\rotate[r]{
\epsfysize=8truecm
\epsffile{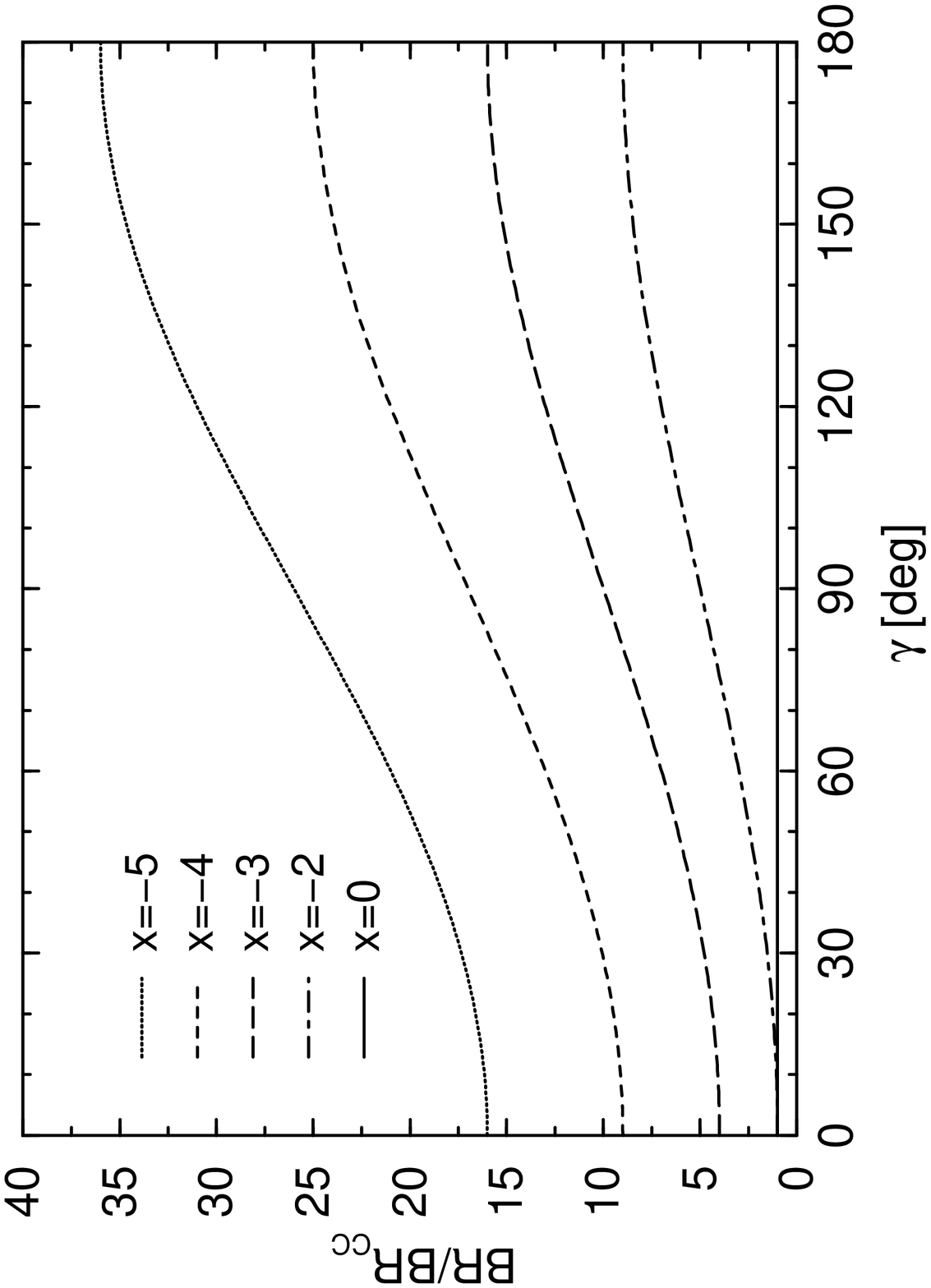}}}
\caption{The dependence of $\mbox{BR}(B_s\to\pi^0\phi)/
\mbox{BR}_{\mbox{{\scriptsize CC}}}(B_s\to\pi^0\phi)$ on $\gamma$ for
various values of $x$.}\label{fig-br}
\end{figure}

Note that (\ref{ED1121}) is rather clean concerning hadronic uncertainties. 
This nice feature is due to the fact that we are in a position to absorb 
all non-perturbative $B$-parameters related to deviations from na\"\i ve 
factorization of the hadronic matrix elements by introducing the 
phenomenological color-suppression factor $a_2$. Concerning short-distance 
QCD corrections, which have been neglected so far, we have to consider only
those affecting the box diagrams and $Z$ penguins contributing to $x$,
since the QCD corrections to the current-current operators are
incorporated effectively in $a_2$. The corresponding short-distance 
QCD corrections are small if we use $\overline{m}_t(m_t)$ as has to
be done in NLO analyses of weak decays\cite{buras-ichep96,buchbur}.
Applying on the other hand the formalism described briefly in Section~2, 
one finds that the QCD corrections to EW penguin operators arising from 
the renormalization group evolution from $\mu={\cal O}(M_W)$ down 
to $\mu={\cal O}(m_b)$ modify $x$ by only a few percent and are hence
also negligibly small.

Using as an example $a_2=0.25$, $R_b=0.36$ and $m_t=170\,
\mbox{GeV}$ yields $x\approx-3$ and confirms nicely our qualitative
expectation that EW penguins should play the dominant role in 
$B_s\to\pi^0\phi$. Varying $a_2$ within $0.2\,\,\mbox{{\scriptsize 
$\stackrel{<}{\sim}$}}\,\,a_2\,\,\mbox{{\scriptsize $\stackrel{<}{\sim}$}}\,\,
0.3$ and $R_b$ and $m_t$ within their presently allowed experimental ranges 
gives $-5\,\,\mbox{{\scriptsize $\stackrel{<}{\sim}$}}\,\,x\,\,
\mbox{{\scriptsize $\stackrel{<}{\sim}$}}-2$. The EW penguin contributions 
lead to dramatic effects in the mixing-induced CP asymmetry as well as in 
the branching ratio as can be seen nicely in Figs.~\ref{fig-Acp} and 
\ref{fig-br}, where the dependences of 
${\cal A}_{\mbox{{\scriptsize CP}}}^{\mbox{{\scriptsize mix-ind}}}
(B_s\to\pi^0\phi)$ and of the ratio ${\cal R}$ on $\gamma$ are shown for
various values of $x$. The solid lines in these figures correspond to 
the case where EW penguins are neglected completely. In the case of 
${\cal A}_{\mbox{{\scriptsize CP}}}^{\mbox{{\scriptsize mix-ind}}}
(B_s\to\pi^0\phi)$ even the sign is changed through the EW penguin 
contributions for $\gamma<90^\circ$, whereas the branching ratio is 
enhanced by a factor of ${\cal O}(10)$ with respect to the pure 
current-current case. The resulting BR$(B_s\to\pi^0\phi)$ is of 
${\cal O}(10^{-7})$, so that an experimental investigation of that 
decay -- which would be interesting to explore EW penguins -- 
will unfortunately be very difficult. Needless to say, the modes 
$B_s\to\rho^0\phi, \pi^0\eta, \rho^0\eta$ exhibiting a very similar 
dynamics should also be dominated by their EW penguin 
contributions\cite{ghlrewp,dht-ewp}. 

\runninghead{The Role of Electroweak Penguins in Strategies for Extracting
CKM Phases}{The Role of Electroweak Penguins in Strategies for Extracting
CKM Phases}
\section{The Role of EW Penguins in Strategies for Extracting CKM Phases}
\noindent
In the strategies for extracting CKM phases reviewed in Section~3, EW 
penguins do not lead to problems wherever it has not been emphasized 
explicitly. That is in fact the case for most of these methods. However,
the GL approach\cite{gl} to eliminate the penguin uncertainties affecting
the determination of $\alpha$ from ${\cal A}_{\mbox{{\scriptsize 
CP}}}^{\mbox{{\scriptsize mix-ind}}}(B_d\to\pi^+\pi^-)$ with the help
of isospin relations among $B\to\pi\pi$ decays (see 3.3.3), as well as
the GRL method\cite{grl-gam} to determine $\gamma$ from $SU(3)$ amplitude 
relations involving $B^+\to\{\pi^+\pi^0,\pi^+K^0,\pi^0K^+\}$ and their
charge-conjugates (see 3.6.2) require a careful 
investigation\cite{dh,ghlrewp,PAPI}.

\begin{figure}[t]
\centerline{
\epsfxsize=9.5truecm
\epsffile{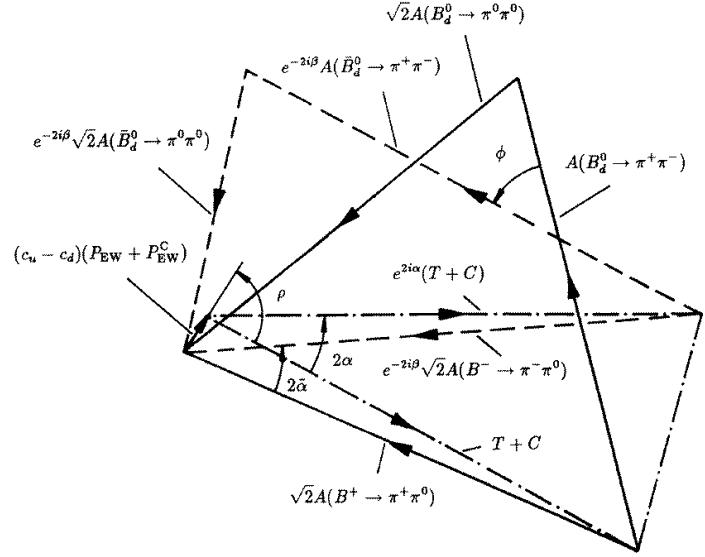}}
\caption{The determination of $\alpha$ from $B\to\pi\pi$ isospin 
triangles in the presence of EW penguins.}\label{fig-Bpipi}
\end{figure}

\runninghead{The Role of Electroweak Penguins in Strategies for Extracting
CKM Phases}{The GL Method of Extracting $\alpha$}
\subsection{The GL Method of Extracting $\alpha$}
\noindent
If one redraws the GL construction\cite{gl} to determine $\alpha$ from 
$B\to\pi\pi$ isospin triangles by taking into account EW penguin 
contributions, one obtains the situation shown in Fig.~\ref{fig-Bpipi}. This 
construction\cite{PAPI} is a bit different from the original one presented 
in Ref.\cite{gl}, since the $A(\overline{B}\to\pi\pi)$ amplitudes have been 
rotated by $e^{-2i\beta}$. The angle $\phi$ fixing the relative orientation 
of the two isospin triangles, which are constructed by measuring only the
corresponding six branching ratios, is determined from the mixing-induced CP 
asymmetry of $B_d\to\pi^+\pi^-$ with the help of the relation
\begin{equation}
{\cal A}_{\mbox{{\scriptsize CP}}}^{\mbox{{\scriptsize mix-ind}}}
(B_d\to\pi^+\pi^-)=-\,\frac{2\,\bigl|A(\overline{B^0_d}\to\pi^+\pi^-)\bigr|
\bigl|A(B^0_d\to\pi^+\pi^-)\bigr|}{\bigl|A(\overline{B^0_d}\to\pi^+
\pi^-)\bigr|^2+\bigl|A(B^0_d\to\pi^+\pi^-)\bigr|^2}\,\sin\phi\,.
\end{equation}
In Fig.~\ref{fig-Bpipi}, the notation of GHLR\cite{ghlrewp} has been used, 
where $P_{\mbox{{\scriptsize EW}}}$ and 
$P_{\mbox{{\scriptsize EW}}}^{\mbox{{\scriptsize C}}}$ denote color-allowed
and color-suppressed $\bar b\to\bar d$ EW penguin amplitudes and $c_u=+2/3$ 
and $c_d=-1/3$ are the electrical up- and down-type quark charges, 
respectively. Because of the presence of EW penguins, the construction shown 
in that figure does {\it not} allow the determination of the {\it exact} 
angle $\alpha$ of the UT. It allows only the extraction of an angle 
$\tilde\alpha$ that is related to $\alpha$ through 
\begin{equation}\label{deltaalpha}
\alpha=\tilde\alpha+\Delta\alpha,
\end{equation}
where $\Delta\alpha$ is given by
\begin{equation}
\Delta\alpha=r\,\sin\alpha\,\cos\,(\rho-\alpha)+{\cal O}(r^2)
\end{equation}
with
\begin{equation}
r\equiv\frac{\left|(c_u-c_d)(P_{\mbox{{\scriptsize EW}}}+
P_{\mbox{{\scriptsize EW}}}^{\mbox{{\scriptsize C}}})\right|}{|T+C|}\approx
\left|\frac{P_{\mbox{{\scriptsize EW}}}}{T}\right|.
\end{equation}
Since $r$ is expected to be of ${\cal O}(0.2^2)$ as can be shown by using
a plausible hierarchy of $\bar b\to\bar d$ decay amplitudes\,\cite{ghlrewp},
EW penguins should not lead to serious problems in the GL method. 
This statement can also be put on more quantitative ground. Unfortunately 
$\rho$ contains strong final state interaction phases and hence cannot be 
calculated at present. However, using $|\cos(\rho-\alpha)|\leq1$, one may 
estimate the following upper bound\cite{PAPIII} for the uncertainty 
$\Delta\alpha$:
\begin{equation}\label{deltaalpha-est}
|\Delta\alpha|\,\,\mbox{{\scriptsize $\stackrel{<}{\sim}$}}\,\,
\frac{\alpha}{2\pi a_1\sin^2\Theta_{\mbox{{\scriptsize
W}}}}\bigl|5B(x_t)-2C(x_t)\bigr|\cdot\left\vert\frac{V_{td}}{V_{ub}}
\right\vert\left\vert\sin\alpha\right\vert\,.
\end{equation}
Taking into account the present status of the CKM matrix yielding
the upper limit\cite{al} $|V_{td}|/|V_{ub}|\leq4.6$ gives 
$|\Delta\alpha|/|\sin\alpha|\,\,
\mbox{{\scriptsize $\stackrel{<}{\sim}$}}\,4^\circ$ for a top-quark 
mass $m_t=170\,\mbox{GeV}$ and a phenomenological color-factor 
$a_1=1$.

\begin{figure}[t]
\centerline{
\epsfxsize=10truecm
\epsffile{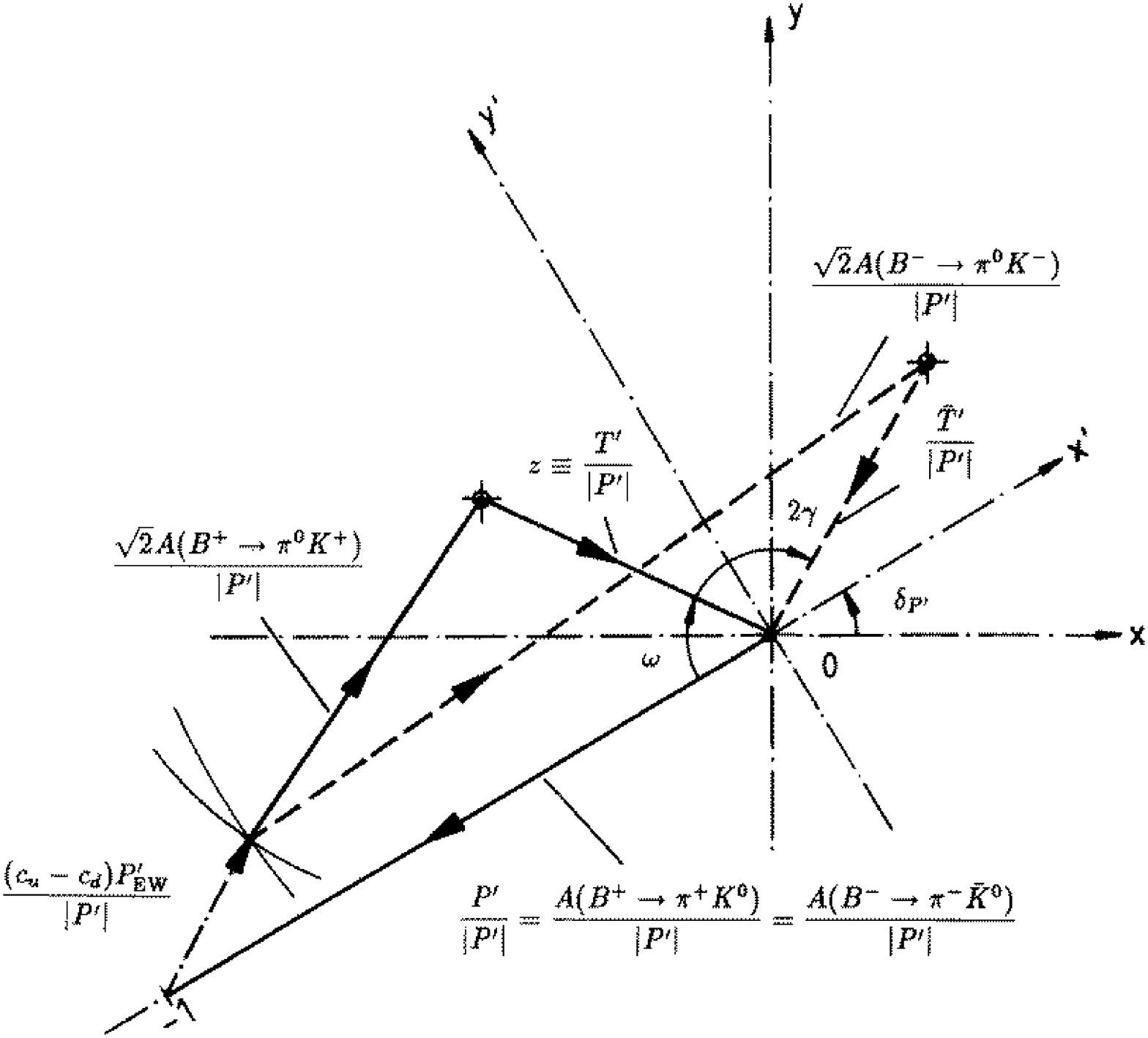}}
\caption{$SU(3)$ relations among $B^+\to\{\pi^+\pi^0,\pi^+K^0,
\pi^0K^+\}$ and charge-conjugate decay amplitudes {\it including} EW penguin
contributions.}\label{fig-GRL-EWP}
\end{figure}

\runninghead{The Role of Electroweak Penguins in Strategies for Extracting
CKM Phases}{The GRL Method of Extracting $\gamma$}
\subsection{The GRL Method of Extracting $\gamma$}
\noindent
In the case of the GRL strategy\cite{grl-gam} for extracting the UT 
angle $\gamma$ from the construction shown in Fig.~\ref{grl-const},
we have to deal with $\bar b\to\bar s$ modes which exhibit an 
interesting hierarchy of decay amplitudes\cite{dh,ghlrewp} that is very
different from the $\bar b\to\bar d$ case discussed in the previous 
subsection. Since the
color-allowed current-current amplitude $T'$ is highly CKM suppressed 
by $\lambda^2R_b\approx0.02$, one expects that the QCD penguin amplitude 
$P'$ plays the dominant role in this decay class and that $T'$ and 
the color-allowed EW penguin amplitude $P'_{\mbox{{\scriptsize EW}}}$
are equally important\cite{ghlrewp}:
\begin{equation}\label{bs-hierarchy}
\left|\frac{T'}{P'}\right|={\cal O}(0.2),\quad
\left|\frac{P'_{\mbox{{\scriptsize EW}}}}{T'}\right|={\cal O}(1)\,.
\end{equation}\label{hier-est}
The last ratio can be estimated more quantitatively\cite{PAPIII} as
\begin{equation}\label{REW-est}
\left|\frac{P'_{\mbox{{\scriptsize EW}}}}{T'}\right|\approx
\frac{\alpha}{2\pi\lambda^2 R_b\, a_1\sin^2\Theta_{\mbox{{\scriptsize
W}}}}\bigl|5B(x_t)-2C(x_t)\bigr|\,r_{SU(3)}.
\end{equation}
Here $r_{SU(3)}$ takes into account $SU(3)$-breaking corrections. 
Factorizable corrections are described by
\begin{equation}
\left.r_{SU(3)}\right\vert_{\mbox{{\scriptsize fact}}}=\frac{f_{\pi}}{f_K}
\frac{F_{BK}(0;0^+)}{F_{B\pi}(0;0^+)},
\end{equation}
where the BSW form factors\cite{BSW} parametrizing the corresponding
quark-current matrix elements yield 
$\left.r_{SU(3)}\right\vert_{\mbox{{\scriptsize fact}}}\approx1$. The
ratio (\ref{REW-est}) increases significantly with the top-quark mass. 
Using $m_t=170\,\mbox{GeV}$, $R_b=0.36$, $a_1=1$ and $r_{SU(3)}=1$ gives 
$|P'_{\mbox{{\scriptsize EW}}}|/|T'|\approx0.8$ and confirms nicely the 
expectation (\ref{bs-hierarchy}). 

Consequently EW penguins are very important in that case and even 
{\it spoil} the GRL approach\cite{grl-gam} to determine $\gamma$ as 
was pointed out by Deshpande and He\cite{dh}. This feature can be seen 
nicely in Fig.~\ref{fig-GRL-EWP}, where color-suppressed EW penguin and 
current-current amplitudes are neglected to simplify the 
presentation\cite{PAPI}. If the EW penguin amplitude $(c_u-c_d)\,
P'_{\mbox{{\scriptsize EW}}}$ were not there, this figure would correspond 
to Fig.~\ref{grl-const} and we would simply have to deal with two triangles 
in the complex plane that could be fixed by measuring only the six branching 
ratios corresponding to 
$B^+\to\{\pi^+\pi^0,\pi^+K^0,\pi^0K^+\}$ and their charge-conjugates.
However, EW penguins do contribute and since the magnitude of their 
``unknown'' amplitude $(c_u-c_d)\,P'_{\mbox{{\scriptsize EW}}}$ is of the 
same size as $|T'|$, it is unfortunately not possible to determine $\gamma$
with the help of this construction.  This feature led to the
development of other methods using $SU(3)$ amplitude relations to extract 
$\gamma$ that require also only measurements of branching ratios, and 
to strategies to control EW penguins in a quantitative way to shed light
on the physics of these FCNC processes. 

\runninghead{The Role of Electroweak Penguins in Strategies for Extracting
CKM Phases}{$SU(3)$ Strategies for Extracting $\gamma$ without EW Penguin 
Problems}
\subsection{$SU(3)$ Strategies for Extracting $\gamma$ without EW Penguin 
Problems}
\noindent
In the recent literature\cite{ghlrewp}$^-$\cite{PAPIII}
some solutions have been proposed to solve the 
problem arising from EW penguins in the GRL approach. Let us have a closer
look at them in this subsection. 

\begin{figure}[t]
\centerline{
\rotate[r]{
\epsfysize=7.5truecm
\epsffile{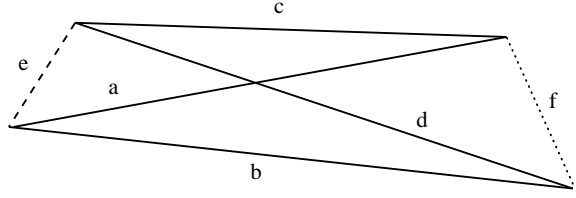}}}
\caption{Amplitude quadrangle for $B\to\pi K$ decays. The labels are 
explained in the text.}\label{fig-quadrangle}
\end{figure}

\subsubsection{Amplitude Quadrangle for $B\to\pi K$ Decays}
\noindent
A quadrangle construction involving $B\to\pi K$ decay amplitudes was 
proposed in Ref.\cite{ghlrewp} that can be used in principle to determine
$\gamma$ irrespectively of the presence of EW penguins. This construction
is shown in Fig.~\ref{fig-quadrangle}, where (a) corresponds to 
$A(B^+\to\pi^+K^0)$, (b) to $\sqrt{2}\,A(B^+\to\pi^0K^+)$,
(c) to $\sqrt{2}\,A(B^0_d\to\pi^0K^0)$, (d) to $A(B^0_d\to\pi^-K^+)$ and 
the dashed line (e) to the decay amplitude $\sqrt{3}\,A(B_s^0\to\pi^0\eta)$. 
The dotted line (f) denotes an $I=3/2$ isospin amplitude $A_{3/2}$ that is 
composed of two parts and can be written as\cite{ghlrewp}
\begin{equation}\label{A32}
A_{3/2}=\bigl|A_{\pi K}^{\mbox{{\scriptsize T}}}\bigr|\,e^{i\tilde\delta_T}
\,e^{i\gamma}-\bigl|A_{\pi K}^{\mbox{{\scriptsize EWP}}}\bigr|\,
e^{i\tilde\delta_{\mbox{{\scriptsize EWP}}}}\,.
\end{equation}
The corresponding charge-conjugate amplitude takes on the other hand the
form
\begin{equation}\label{A32CP}
\overline{A}_{3/2}=\bigl|A_{\pi K}^{\mbox{{\scriptsize T}}}\bigr|\,
e^{i\tilde\delta_T}\,e^{-i\gamma}-\bigl|A_{\pi K}^{\mbox{{\scriptsize 
EWP}}}\bigr|\,e^{i\tilde\delta_{\mbox{{\scriptsize EWP}}}}\,,
\end{equation}
so that the EW penguin contributions cancel in the difference of (\ref{A32})
and (\ref{A32CP}):
\begin{equation}\label{ampl-diff}
A_{3/2}-\overline{A}_{3/2}=2\,i\,e^{i\tilde\delta_T}\,
\bigl|A_{\pi K}^{\mbox{{\scriptsize T}}}\bigr|\,\sin\gamma\,.
\end{equation}
In order to determine this amplitude difference geometrically, both the 
quadrangle depicted in Fig.~\ref{fig-quadrangle} and the one corresponding 
to the charge-conjugate processes have to be constructed by measuring the
branching rations corresponding to (a)--(e). Moreover the relative 
orientation of these two quadrangles in the complex plane has to be fixed. 
This can be done easily through the side (a) as no non-trivial CP-violating 
weak phase is present in the $\bar b\to\bar s$ penguin-induced decay 
$B^+\to\pi^+K^0$, i.e.\  $A(B^-\to\pi^-\overline{K^0})=A(B^+\to\pi^+K^0)$ 
(see 3.3.4). Since the quantity 
$\bigl|A_{\pi K}^{\mbox{{\scriptsize T}}}\bigr|$ 
corresponds to $|T'+C'|$, it can be determined with the help of the 
$SU(3)$ flavor symmetry (note (\ref{ED1201}) and (\ref{ED1203})) by 
measuring the branching ratio for $B^+\to\pi^+\pi^-$, i.e.\ through 
$\bigl|A_{\pi K}^{\mbox{{\scriptsize T}}}\bigr|=r_u\,\sqrt{2}\,
|A(B^+\to\pi^+\pi^0)|$, so that both $\sin\gamma$ and the strong 
phase $\tilde\delta_T$ can be extracted from the amplitude difference 
(\ref{ampl-diff}). Unfortunately the dashed line (e) corresponds to the 
decay $B^0_s\to\pi^0\eta$ that is dominated by EW 
penguins\cite{rfewp3,dht-ewp} (see Subsection~4.3) and is therefore expected 
to exhibit a branching ratio at the ${\cal O}(10^{-7})$ level. Consequently 
the amplitude quadrangles are rather squashed ones and this approach to 
determine $\gamma$ is very difficult from an experimental point of view. 

\begin{figure}[t]
\centerline{
\rotate[r]{
\epsfysize=7.5truecm
\epsffile{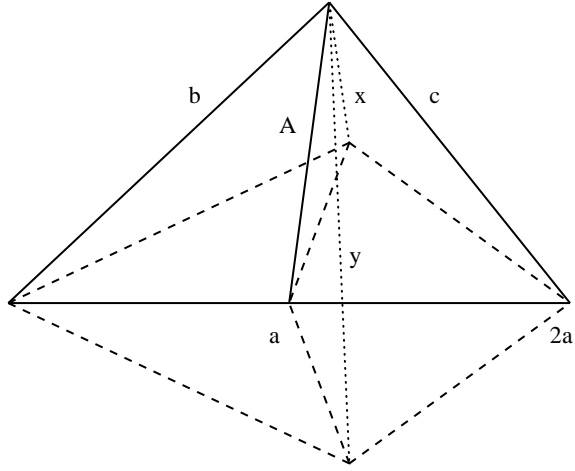}}}
\caption{$SU(3)$ amplitude relations involving $B^+\to\{\pi^+K^0,
\pi^0K^+,\eta_8\,K^+\}$ and charge-conjugates (dashed lines). The labels are 
explained in the text.}\label{fig-dehe}
\end{figure}

\subsubsection{$SU(3)$ Relations among $B^+\to\{\pi^+K^0,
\pi^0K^+,\eta_8\,K^+\}$ Decay Amplitudes}
\noindent
Another approach to extract $\gamma$ involving the decays 
$B^+\to\{\pi^+K^0,\pi^0K^+,\eta_8\,K^+\}$ and their charge-conjugates
was proposed by Deshpande and He in Ref.\cite{dh-gam}. Using $SU(3)$ flavor 
symmetry, it is possible to derive relations among the corresponding decay 
amplitudes that can be represented in the complex plane as shown in 
Fig.~\ref{fig-dehe}. Here the solid lines labelled (a), (b) and (c) 
correspond to the decay amplitudes $A(B^+\to\pi^+K^0)$, $\sqrt{2}\,
A(B^+\to\pi^0 K^+)$ and $\sqrt{6}\,A(B^+\to\eta_8\,K^+)$, respectively, and 
the dashed lines represent the corresponding charge-conjugate amplitudes. 
Note that $A(B^-\to\pi^-\overline{K^0})=A(B^+\to\pi^+K^0)$ has also been 
used in this construction. Similarly as in 5.3.1, the determination 
of $\gamma$ can be accomplished by considering the difference of a 
particularly useful chosen combination $A$ of decay amplitudes and its
charge-conjugate $\overline{A}$, where the penguin contributions cancel:
\begin{equation}
A-\overline{A}=2\,\sqrt{2}\,i\,e^{i\tilde\delta_T}r_u\,|A(B^+\to\pi^+\pi^0)|
\,\sin\gamma\,.
\end{equation}
Here the magnitude of the $B^+\to\pi^+\pi^0$ amplitude is used --  as 
in the $B\to\pi K$ quadrangle approach\cite{ghlrewp} -- to fix $|T'+C'|$.
In Fig.~\ref{fig-dehe}, the dotted lines (x) and (y) represent two possible 
solutions for this amplitude difference. The fact that this construction 
does not give a unique solution for $A-\overline{A}$ is a well-known 
characteristic feature of all geometrical constructions of this kind, i.e.\
one has in general to deal with several discrete ambiguities. 

Compared to the method using $B\to\pi K$ quadrangles discussed in 5.3.1, the
advantage of this strategy is that all branching ratios are expected to
be of the same order of magnitude ${\cal O}(10^{-5})$. In particular one
has not to deal with an EW penguin dominated channel exhibiting a 
branching ratio at the ${\cal O}(10^{-7})$ level. However, the accuracy of
the strategy is limited by $\eta-\eta'$ mixing, i.e.\ the $A(B^\pm\to
\eta_8\,K^\pm)$ amplitudes have to be determined through
\begin{equation}
A(B^\pm\to\eta_8\,K^\pm)=A(B^\pm\to\eta\, K^\pm)\cos\Theta+
A(B^\pm\to\eta'K^{\pm})\sin\Theta
\end{equation}
with a mixing angle $\Theta\approx20^\circ$, and by other $SU(3)$-breaking
effects which cannot be calculated at present. A similar approach to 
determine $\gamma$ was proposed by Gronau and Rosner in Ref.\cite{gr-gam},
where the amplitude construction is expressed in terms of the physical
$\eta$ and $\eta'$ states. A detailed discussion of $SU(3)$ amplitude 
relations for $B$ decays involving $\eta$ and $\eta'$ in light of 
extractions of CKM phases can be found in Ref.\cite{eta-etap}.

\subsubsection{A Simple Strategy for Fixing $\gamma$ and Obtaining Insights
into the World of EW Penguins}
\begin{figure}[t]
\centerline{
\rotate[r]{
\epsfysize=8truecm
\epsffile{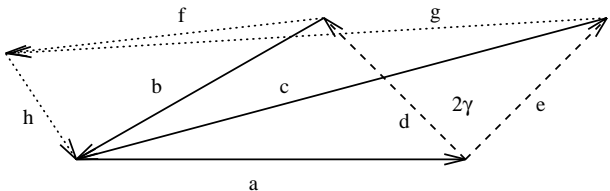}}}
\caption{$SU(2)$ isospin relations among $B^+\to\pi^+K^0$, 
$B^0_d\to\pi^-K^+$ and charge-conjugates. The labels are
explained in the text.}\label{fig-EWdet}
\end{figure}

\noindent
Since the geometrical constructions discussed in 5.3.1 and 5.3.2 are
quite complicated and appear to be very challenging from an experimental
point of view, let us consider a much simpler approach\cite{PAPIII} 
to determine $\gamma$. It uses the decays $B^+\to\pi^+K^0$, 
$B^0_d\to\pi^-K^+$ and their charge-conjugates. In the case of these 
transitions, EW penguins contribute only in color-suppressed form and hence
play a minor role. Neglecting these contributions and using 
the $SU(2)$ isospin symmetry of strong interactions -- not $SU(3)$ -- 
to relate their QCD penguin contributions (note the similarity to the 
example given in 3.4.2), the corresponding decay amplitudes can be
written in the GHLR notation\cite{ghlr} as 
\begin{eqnarray}
A(B^+\to\pi^+K^0)&=&P'\,=\,A(B^-\to\pi^-\overline{K^0})\nonumber\\
A(B^0_d\to\pi^-K^+)&=&-\,(P'+T')\\
A(\overline{B^0_d}\to\pi^+K^-)&=&-\,(P'+e^{-2i\gamma}\,T')\,.\nonumber
\end{eqnarray}
Let me note that these relations are on rather solid ground from a 
theoretical point of view. They can be represented in the complex plane 
as shown in Fig.~\ref{fig-EWdet}. Here (a) corresponds to 
$A(B^+\to\pi^+K^0)=P'=A(B^-\to\pi^-\overline{K^0})$,
(b) to $A(B^0_d\to\pi^-K^+)$, (c) to $A(\overline{B^0_d}\to\pi^+K^-)$ and
the dashed lines (d) and (e) to the color-allowed current-current 
amplitudes $T'$ and $e^{-2i\gamma}\,T'$, respectively. The dotted lines 
(f)--(h) will be discussed in a moment. Note that these $B\to\pi K$ decays
appeared already in 3.3.6. Combining their branching ratios with the 
observables of a time-dependent measurement of $B_d\to\pi^+\pi^-$, a
simultaneous extraction of $\alpha$ and $\gamma$ may be possible\cite{dgr}. 
The information provided by the $B\to\pi K$ modes can, however, also be
used for a quite different approach that may finally allow the determination
of EW penguin amplitudes.

In order to determine $\gamma$ from Fig~\ref{fig-EWdet}, we have to know the 
length $|T'|$ of the dashed lines (d) and (e). In fact, the situation is 
analogous to the 
extraction of $\gamma$ from (\ref{EE4}) and (\ref{EE5}) in 3.4.2. There we 
saw that $B^+\to\pi^+\pi^0$ provides an estimate of that quantity through 
(\ref{EE7}) which is based on two assumptions: $SU(3)$ flavor symmetry and 
neglect of color-suppressed current-current contributions to 
$B^+\to\pi^+\pi^0$. Consequently, following these lines, it is possible to
obtain an estimate of $\gamma$ by measuring only BR$(B^+\to\pi^+K^0)=
\mbox{BR}(B^-\to\pi^-\overline{K^0})$, BR$(B^0_d\to\pi^-K^+)$,
BR$(\overline{B^0_d}\to\pi^+K^-)$ and BR$(B^+\to\pi^+\pi^0)=\mbox{BR}
(B^-\to\pi^-\pi^0)$. Note that the neutral $B_d$ decays are ``self-tagging''
modes so that no time-dependent measurements are needed and that this 
estimate of $\gamma$ is very similar to the ``original'' GRL 
approach\cite{grl-gam} shown in Fig.~\ref{grl-const} that is unfortunately
spoiled by EW penguins. Needless to say, this strategy is very simple from 
a geometrical point of view -- just triangle constructions -- and very 
promising from an experimental point of view since all branching ratios are 
of the same order of magnitude ${\cal O}(10^{-5})$. Moreover no difficult
to measure CP eigenstate of the neutral $D$ system is required as in 3.6.1.

Let me emphasize that the ``weak'' point of this approach -- and of 
the one using untagged $B_s$ decays discussed in 3.4.2 -- is the 
relation (\ref{EE7}) to estimate $|T'|$. Therefore the ``estimate'' of 
$\gamma$ may well turn into a solid ``determination'' if it should become
possible to fix the magnitude of the color-allowed current-current
amplitude contributing to $B^0_d\to\pi^-K^+$ in a more reliable way. 
Another possibility of fixing $|T'|$ is of course the factorization 
hypothesis which may work reasonably well for that color-allowed 
amplitude\cite{bjorken} and could be used as some kind of cross-check 
for (\ref{EE7}). Maybe the ``final'' result for $|T'|$ will come from 
lattice gauge theory one day. 

Interestingly the construction shown in Fig.~\ref{fig-EWdet} provides even
more information if one takes into account the amplitude relations
\begin{eqnarray}
\sqrt{2}\,A(B^+\to\pi^0K^+)&\approx&-\,\left[P'+T'+(c_u-c_d)
P_{\mbox{{\scriptsize EW}}}'\right]\\
\sqrt{2}\,A(B^-\to\pi^0K^-)&\approx&-\,\left[P'+e^{-2i\gamma}\,T'+(c_u-c_d)
P_{\mbox{{\scriptsize EW}}}'\right],
\end{eqnarray}
where color-suppressed current-current and EW penguin amplitudes have
been neglected. Consequently, the dotted lines (f) and (g) corresponding to
$\sqrt{2}\,A(B^+\to\pi^0K^+)$ and $\sqrt{2}\,
A(B^-\to\pi^0K^-)$, respectively, allow a determination 
of the dotted line (h) denoting the color-allowed \mbox{$\bar b\to
\bar s$} EW penguin amplitude $(c_u-c_d)P_{\mbox{{\scriptsize EW}}}'$. 
Since EW penguins are -- in contrast to QCD penguins -- dominated to 
excellent accuracy by internal top-quark exchanges, the $\bar b\to\bar d$
EW penguin amplitude $(c_u-c_d)P_{\mbox{{\scriptsize EW}}}$ is related in 
the limit of an exact $SU(3)$ flavor symmetry to the corresponding 
$\bar b\to\bar s$ amplitude through the simple relation
\begin{equation}
(c_u-c_d)P_{\mbox{{\scriptsize EW}}}=-\lambda\,R_t\,e^{-i\beta}\,(c_u-c_d)
P_{\mbox{{\scriptsize EW}}}'
\end{equation}
and may consequently be determined from the constructed 
$(c_u-c_d)P_{\mbox{{\scriptsize EW}}}'$ amplitude. 

\runninghead{The Role of Electroweak Penguins in Strategies for Extracting
CKM Phases}{Towards Control over Electroweak Penguins}
\subsection{Towards Control over EW Penguins}
\noindent
It would be very useful to determine the EW penguin contributions 
experimentally. That would allow several predictions, consistency checks
and tests of certain SM calculations\cite{PAPIII}. For example, one
may determine the quantity $x$ introduced in (\ref{ED1118}) parametrizing
the EW penguin effects in $B_s\to\pi^0\phi$ experimentally, and may
compare this result with the SM expression (\ref{ED1121}). That way 
one may obtain predictions for ${\cal A}_{\mbox{{\scriptsize 
CP}}}^{\mbox{{\scriptsize mix-ind}}}(B_s\to\pi^0\phi)$ and 
BR$(B_s\to\pi^0\phi)$ long before it might be possible (if it is possible 
at all!) to measure them directly. Another interesting point is that the
$\bar b\to\bar d$ EW penguin amplitude $P_{\mbox{{\scriptsize EW}}}$ 
allows in principle to fix the uncertainty $\Delta\alpha$ (see 
(\ref{deltaalpha})) arising from EW penguins in the GL 
method\cite{gl} of extracting $\alpha$ and to check whether it is e.g.\
in agreement with (\ref{deltaalpha-est}). Since EW penguins are ``rare'' 
FCNC processes that are absent at tree level within the 
SM, it may well be that ``New Physics'' contributes to them significantly 
through additionally present virtual particles in the loops. Consequently
EW penguins may give hints to physics beyond the SM.

We have just seen an example of a simple strategy to determine EW penguin 
amplitudes experimentally. In Ref.\cite{PAPI}, where more involved methods
to accomplish this task are discussed, it was pointed out that the 
central input to control EW penguins in a quantitative way is the CKM angle 
$\gamma$. Consequently determinations of this UT angle are not only 
important in respect of testing the SM description of CP violation but also 
to shed light on the physics of EW penguins. 

\runninghead{Summary and Concluding Remarks}{Summary and Concluding Remarks}
\section{Summary and Concluding Remarks}
\noindent
The $B$-meson system provides a very fertile ground for studying CP violation 
and extracting CKM phases. In this respect neutral $B_q$ decays 
($q\in\{d,s\}$) are particularly promising. The point is 
that ``mixing-induced'' 
CP-violating asymmetries are closely related to angles of the UT in some
cases. For example, the ``gold-plated'' decay $B_d\to J/\psi\,
K_{\mbox{{\scriptsize S}}}$ allows an extraction of $\sin(2\beta)$ to 
excellent accuracy because of its particular decay structure, and 
$B_d\to\pi^+\pi^-$ probes $\sin(2\alpha)$. However, hadronic uncertainties
arising from QCD penguins preclude a theoretical clean determination of
$\sin(2\alpha)$ by measuring only ${\cal A}_{\mbox{{\scriptsize 
CP}}}^{\mbox{{\scriptsize mix-ind}}}(B_d\to\pi^+\pi^-)$. Consequently more 
involved strategies are required to extract $\alpha$. Such methods are 
fortunately already available and certainly time will tell which of them 
is most promising from an experimental point of view. 

In the case of $B_s\to\rho^0K_{\mbox{{\scriptsize S}}}$, which appeared
frequently in the literature as a tool to determine $\gamma$, penguin 
contributions are expected to lead to serious problems so that a meaningful
extraction of $\gamma$ from this mode should not be possible. There
are, however, other $B_s$ decays that may allow determinations of this UT 
angle, in some cases even in a clean way. Unfortunately 
$B^0_s-\overline{B^0_s}$ mixing may be too fast to be resolved with 
present vertex technology so that these strategies are experimentally very 
challenging. 

An alternative route to extract CKM phases from $B_s$ decays and explore 
CP violation in these modes may be provided by the width difference 
of the $B_s$ system that is expected to be sizable. Interestingly the 
rapid oscillatory $\Delta M_st$ terms cancel in untagged $B_s$ data 
samples that depend therefore only on two different exponents 
$e^{-\Gamma_L^{(s)}t}$ and $e^{-\Gamma_H^{(s)}t}$. Several strategies to 
extract $\gamma$ and the Wolfenstein parameter $\eta$ from untagged 
$B_s$ decays have been proposed recently. Here time-dependent angular 
distributions for $B_s$ decays into admixtures of CP eigenstates and 
exclusive channels that are caused by $\bar b\to\bar uc\bar s$ 
($b\to c\bar us$) quark-level transitions play a key role. Such untagged 
methods are obviously much more promising in respect of efficiency, 
acceptance and purity. However, their feasibility depends crucially on 
$\Delta\Gamma_s$ and it is not clear at present whether it will turn out 
to be large enough.

Theoretical analyses of CP violation in charged $B$ decays are usually very
technical and suffer in general from large hadronic uncertainties. 
Consequently CP-violating asymmetries in charged $B$ decays are mainly
interesting in view of excluding ``superweak'' models of CP violation
in an unambiguous way. Nevertheless, if one combines branching ratios of
charged $B$ decays in a clever way, they may allow determinations of 
angles of the UT, in some cases even without hadronic uncertainties. 

To this end certain relations among decay amplitudes are used. The 
prototype of this approach are $B\to DK$ amplitude triangles that allow 
a clean determination of $\gamma$. Unfortunately one has to deal with 
experimental problems in that strategy of fixing this UT angle. Whereas 
the $B\to DK$ triangle relations are valid exactly, one may also use the 
$SU(3)$ flavor symmetry of strong interactions with certain plausible 
dynamical assumptions to derive approximate relations among non-leptonic 
$B\to\{\pi\pi,\pi K,K\overline{K}\}$ decay amplitudes which may
allow extractions of CKM phases and strong final state interaction phases
by measuring only the corresponding branching ratios. This approach
has been very popular over the recent years. It suffers, however, from
limitations due to non-factorizable $SU(3)$-breaking, QCD penguins with
internal up- and charm-quark exchanges and also EW penguins. 

Contrary to na\"\i ve expectations, EW penguins may play an important --
in some cases, e.g.\ $B_s\to\pi^0\phi$, even dominant -- role in certain
non-leptonic $B$ decays because of the large top-quark mass. The EW
penguin contributions spoil the determination of $\gamma$ using 
$B^+\to\{\pi^+\pi^0,\pi^+K^0,\pi^0K^+\}$ (and charge-conjugate)
$SU(3)$ triangle relations and require in general more involved 
geometrical constructions, e.g.\ $B\to\pi K$ quadrangles, to extract this 
UT angle which are difficult from an experimental point of view. There is, 
however, also a simple ``estimate'' of $\gamma$ using only triangles which
involve the $B^+\to\pi^+K^0$, $B^0_d\to\pi^-K^+$ and charge-conjugate
decay amplitudes. This approximate approach is more promising for 
experimentalists and may turn into a ``determination'' if the magnitude 
of the color-allowed $\bar b\to\bar s$ current-current amplitude, which is
its major input, can be determined reliably. Measuring in addition the
branching ratios for $B^\pm\to\pi^0K^\pm$ also the EW penguin amplitudes
can be determined experimentally which should allow valuable insights into
the physics of these FCNC processes.  There are more refined strategies
to control EW penguins in a quantitative way that require $\gamma$ as
an input. Consequently a determination of this UT angle is not only 
important to test the SM description of CP violation but also to shed
light on these ``rare'' FCNC processes which might give hints to ``New
Physics''. 

I hope that I could convince the reader in this article that the physics 
potential of the $B$ system in respect of CP violation and exploring 
penguins is enormous and that certainly a very exciting future of $B$ 
physics is ahead of us.

\vspace{0.3cm}
{\it Acknowledgements:\/}
I would like to thank Andrzej Buras, Isard Dunietz and Thomas Mannel for 
collaboration on some of the topics presented in this review. 

\runninghead{Bibliography} {Bibliography}

\end{document}